\RequirePackage{lineno} 
\documentclass[prc,twocolumn,superscriptaddress,showpacs,amssymb,amsmath,amsfonts,aps,nofootinbib]{revtex4-1}

\usepackage{array,mathtools,amssymb,booktabs}
\newcolumntype{C}{>{$}c<{$}}
\AtBeginDocument{
\heavyrulewidth=.08em
\lightrulewidth=.05em
\cmidrulewidth=.03em
\belowrulesep=.65ex
\belowbottomsep=0pt
\aboverulesep=.4ex
\abovetopsep=0pt
\cmidrulesep=\doublerulesep
\cmidrulekern=.5em
\defaultaddspace=.5em
}
\usepackage{graphicx}
\usepackage{dcolumn}
\newcolumntype{d}[1]{D{.}{.}{-1}} 

\usepackage{latexsym}
\usepackage{amsmath} 
\usepackage{url}
\usepackage{natbib}

\usepackage{mathtools}
\usepackage{framed}
\usepackage{diagbox}
\usepackage{multirow}
\usepackage{hyperref}
\usepackage{float}

\usepackage[margin=20mm]{geometry}

\newcolumntype{P}[1]{>{\centering\arraybackslash}p{#1}}

\newcolumntype{C}[1]{>{\centering\arraybackslash}m{#1}}

\begin{document}

\date{\today}
\title{\large Measurements of the $\gamma_{v} p \rightarrow p'
\pi^{+} \pi^{-}$
cross section with the CLAS detector  for
$0.4\enskip {\rm GeV^{2}} < Q^{2} < 1.0\enskip {\rm GeV^{2}}$ and
$1.3\enskip {\rm GeV} < W <1.825\enskip {\rm GeV}$\\}

\newcommand*{\ANL}{Argonne National Laboratory, Argonne, Illinois 60439}
\newcommand*{\ANLindex}{1}
\affiliation{\ANL}
\newcommand*{\ASU}{Arizona State University, Tempe, Arizona 85287-1504}
\newcommand*{\ASUindex}{2}
\affiliation{\ASU}
\newcommand*{\CSUDH}{California State University, Dominguez Hills, Carson, CA 90747}
\newcommand*{\CSUDHindex}{3}
\affiliation{\CSUDH}
\newcommand*{\CANISIUS}{Canisius College, Buffalo, NY}
\newcommand*{\CANISIUSindex}{4}
\affiliation{\CANISIUS}
\newcommand*{\CMU}{Carnegie Mellon University, Pittsburgh, Pennsylvania 15213}
\newcommand*{\CMUindex}{5}
\affiliation{\CMU}
\newcommand*{\CUA}{Catholic University of America, Washington, D.C. 20064}
\newcommand*{\CUAindex}{6}
\affiliation{\CUA}
\newcommand*{\SACLAY}{IRFU, CEA, Universit'e Paris-Saclay, F-91191 Gif-sur-Yvette, France}
\newcommand*{\SACLAYindex}{7}
\affiliation{\SACLAY}
\newcommand*{\CNU}{Christopher Newport University, Newport News, Virginia 23606}
\newcommand*{\CNUindex}{8}
\affiliation{\CNU}
\newcommand*{\UCONN}{University of Connecticut, Storrs, Connecticut 06269}
\newcommand*{\UCONNindex}{9}
\affiliation{\UCONN}
\newcommand*{\DUKE}{Duke University, Durham, North Carolina 27708-0305}
\newcommand*{\DUKEindex}{10}
\affiliation{\DUKE}
\newcommand*{\FU}{Fairfield University, Fairfield CT 06824}
\newcommand*{\FUindex}{11}
\affiliation{\FU}
\newcommand*{\FERRARAU}{Universita' di Ferrara , 44121 Ferrara, Italy}
\newcommand*{\FERRARAUindex}{12}
\affiliation{\FERRARAU}
\newcommand*{\FIU}{Florida International University, Miami, Florida 33199}
\newcommand*{\FIUindex}{13}
\affiliation{\FIU}
\newcommand*{\FSU}{Florida State University, Tallahassee, Florida 32306}
\newcommand*{\FSUindex}{14}
\affiliation{\FSU}
\newcommand*{\Genova}{Universit$\grave{a}$ di Genova, 16146 Genova, Italy}
\newcommand*{\Genovaindex}{15}
\affiliation{\Genova}
\newcommand*{\GWUI}{The George Washington University, Washington, DC 20052}
\newcommand*{\GWUIindex}{16}
\affiliation{\GWUI}
\newcommand*{\ISU}{Idaho State University, Pocatello, Idaho 83209}
\newcommand*{\ISUindex}{17}
\affiliation{\ISU}
\newcommand*{\INFNFE}{INFN, Sezione di Ferrara, 44100 Ferrara, Italy}
\newcommand*{\INFNFEindex}{18}
\affiliation{\INFNFE}
\newcommand*{\INFNFR}{INFN, Laboratori Nazionali di Frascati, 00044 Frascati, Italy}
\newcommand*{\INFNFRindex}{19}
\affiliation{\INFNFR}
\newcommand*{\INFNGE}{INFN, Sezione di Genova, 16146 Genova, Italy}
\newcommand*{\INFNGEindex}{20}
\affiliation{\INFNGE}
\newcommand*{\INFNRO}{INFN, Sezione di Roma Tor Vergata, 00133 Rome, Italy}
\newcommand*{\INFNROindex}{21}
\affiliation{\INFNRO}
\newcommand*{\ORSAY}{Institut de Physique Nucl\'eaire, CNRS/IN2P3 and Universit\'e Paris Sud, Orsay, France}
\newcommand*{\ORSAYindex}{22}
\affiliation{\ORSAY}
\newcommand*{\ITEP}{Institute of Theoretical and Experimental Physics, Moscow, 117259, Russia}
\newcommand*{\ITEPindex}{23}
\affiliation{\ITEP}
\newcommand*{\JMU}{James Madison University, Harrisonburg, Virginia 22807}
\newcommand*{\JMUindex}{24}
\affiliation{\JMU}
\newcommand*{\KNU}{Kyungpook National University, Daegu 41566, Republic of Korea}
\newcommand*{\KNUindex}{25}
\affiliation{\KNU}
\newcommand*{\MISS}{Mississippi State University, Mississippi State, MS 39762-5167}
\newcommand*{\MISSindex}{26}
\affiliation{\MISS}
\newcommand*{\UNH}{University of New Hampshire, Durham, New Hampshire 03824-3568}
\newcommand*{\UNHindex}{27}
\affiliation{\UNH}
\newcommand*{\NSU}{Norfolk State University, Norfolk, Virginia 23504}
\newcommand*{\NSUindex}{28}
\affiliation{\NSU}
\newcommand*{\OHIOU}{Ohio University, Athens, Ohio  45701}
\newcommand*{\OHIOUindex}{29}
\affiliation{\OHIOU}
\newcommand*{\ODU}{Old Dominion University, Norfolk, Virginia 23529}
\newcommand*{\ODUindex}{30}
\affiliation{\ODU}
\newcommand*{\RPI}{Rensselaer Polytechnic Institute, Troy, New York 12180-3590}
\newcommand*{\RPIindex}{31}
\affiliation{\RPI}
\newcommand*{\URICH}{University of Richmond, Richmond, Virginia 23173}
\newcommand*{\URICHindex}{32}
\affiliation{\URICH}
\newcommand*{\ROMAII}{Universita' di Roma Tor Vergata, 00133 Rome Italy}
\newcommand*{\ROMAIIindex}{33}
\affiliation{\ROMAII}
\newcommand*{\MSU}{Skobeltsyn Institute of Nuclear Physics, Lomonosov Moscow State University, 119234 Moscow, Russia}
\newcommand*{\MSUindex}{34}
\affiliation{\MSU}
\newcommand*{\SCAROLINA}{University of South Carolina, Columbia, South Carolina 29208}
\newcommand*{\SCAROLINAindex}{35}
\affiliation{\SCAROLINA}
\newcommand*{\TEMPLE}{Temple University,  Philadelphia, PA 19122 }
\newcommand*{\TEMPLEindex}{36}
\affiliation{\TEMPLE}
\newcommand*{\JLAB}{Thomas Jefferson National Accelerator Facility, Newport News, Virginia 23606}
\newcommand*{\JLABindex}{37}
\affiliation{\JLAB}
\newcommand*{\UTFSM}{Universidad T\'{e}cnica Federico Santa Mar\'{i}a, Casilla 110-V Valpara\'{i}so, Chile}
\newcommand*{\UTFSMindex}{38}
\affiliation{\UTFSM}
\newcommand*{\EDINBURGH}{Edinburgh University, Edinburgh EH9 3JZ, United Kingdom}
\newcommand*{\EDINBURGHindex}{39}
\affiliation{\EDINBURGH}
\newcommand*{\GLASGOW}{University of Glasgow, Glasgow G12 8QQ, United Kingdom}
\newcommand*{\GLASGOWindex}{40}
\affiliation{\GLASGOW}
\newcommand*{\VCU}{Virginia Commonwealth University, Richmond, VA 23220}
\newcommand*{\VCUindex}{41}
\affiliation{\VCU}
\newcommand*{\VT}{Virginia Tech, Blacksburg, Virginia   24061-0435}
\newcommand*{\VTindex}{42}
\affiliation{\VT}
\newcommand*{\VIRGINIA}{University of Virginia, Charlottesville, Virginia 22901}
\newcommand*{\VIRGINIAindex}{43}
\affiliation{\VIRGINIA}
\newcommand*{\WM}{College of William and Mary, Williamsburg, Virginia 23187-8795}
\newcommand*{\WMindex}{44}
\affiliation{\WM}
\newcommand*{\YEREVAN}{Yerevan Physics Institute, 375036 Yerevan, Armenia}
\newcommand*{\YEREVANindex}{45}
\affiliation{\YEREVAN}

\newcommand*{\NOWISU}{Idaho State University, Pocatello, Idaho 83209}
\newcommand*{\NOWINFNGE}{INFN, Sezione di Genova, 16146 Genova, Italy}

\author {G.V.~Fedotov} 
\affiliation{\MSU}
\affiliation{\OHIOU}
\author {Iu.A.~Skorodumina} 
\affiliation{\SCAROLINA}
\author {V.D.~Burkert} 
\affiliation{\JLAB}
\author {R.W.~Gothe} 
\affiliation{\SCAROLINA}
\author {K.~Hicks} 
\affiliation{\OHIOU}
\author {V.I.~Mokeev} 
\affiliation{\JLAB}
\affiliation{\MSU}
\author {S. Adhikari} 
\affiliation{\FIU}
\author {Whitney~R.~Armstrong} 
\affiliation{\ANL}
\author {H.~Avakian} 
\affiliation{\JLAB}
\author {J.~Ball} 
\affiliation{\SACLAY}
% ...

\collaboration{The CLAS Collaboration}
\noaffiliation

\begin{abstract}
New results on the single-differential and fully-integrated
cross sections for the process $\gamma_{v} p \rightarrow p' \pi^{+} \pi^{-}$ are presented. 
The experimental data  were collected with the CLAS detector at
Jefferson Laboratory.
Measurements were carried out in the kinematic region of the reaction invariant mass $W$ from 1.3
to 1.825 GeV and
the photon virtuality $Q^2$ from 0.4 to 1.0~GeV$^2$. The cross sections were obtained in narrow $Q^{2}$ bins (0.05 GeV$^{2}$) with the smallest statistical uncertainties achieved in double-pion electroproduction experiments to date.
The results were found to be in agreement with previously available data where they overlap.
A preliminary interpretation of the extracted cross sections, which was based on  a phenomenological meson-baryon  reaction model, revealed substantial relative contributions from  nucleon resonances.
The data offer promising prospects to improve knowledge on the  $Q^{2}$-evolution of the electrocouplings of most resonances with masses up to $\sim$1.8~GeV.
\end{abstract}

\pacs{ 11.55.Fv, 13.40.Gp, 13.60.Le, 14.20.Gk  }

\maketitle

\section{Introduction  }
\label{intro}

During the last several decades, experiments have been performed in laboratories all over the world in order to investigate exclusive reactions of meson photo- and electroproduction off proton targets.
This investigation is typically carried out through the detailed analysis of the experimental data
 with the goal of extracting various observables. Further theoretical and phenomenological interpretations of the extracted observables provide valuable information on nucleon structure and features of the strong interaction~\cite{Burkert:2016kyi,Krusche:2003ik,Aznauryan:2011qj,Skorodumina:2016pnb}.
 
 A large amount  of experimental data on exclusive meson photo- and electroproduction has been collected in Hall B at Jefferson Lab with the CLAS detector~\cite{Me03}.
The analysis of these data has already provided a lot of information on differential cross sections and different single- and double-polarization asymmetries with almost complete coverage of the final hadron phase space\footnote[1]{The numerical results on observables measured with the CLAS detector are available in the CLAS physics database~\cite{CLAS_DB}.}. Some kinematic areas, however, are still lacking this information.

This paper introduces new information on the fully-integrated and single-differential cross sections of the reaction  $\gamma_{v} p \rightarrow p' \pi^{+} \pi^{-}$ at $1.3$~GeV $< W < 1.825$~GeV and $0.4$~GeV$^{2}$ $< Q^{2} < 1.0$~GeV$^{2}$. The cross sections were extracted along the standards of the CLAS data analysis and added into the CLAS physics database~\cite{CLAS_DB}. They are also available on GitHub~\cite{Github:data}. High experimental statistics allow for narrow binning (i.e. 0.05~GeV$^{2}$ in $Q^{2}$ and 25~MeV in $W$), as well as smaller statistical uncertainties than were achieved in previous studies of double-pion electroproduction cross sections~\cite{Fedotov:2008aa,Ripani:2002ss,Isupov:2017lnd}. The conditions of the experiment and the data analysis procedure are described in Sections II~-~IV.

The kinematic region covered by the analyzed data has already been partially investigated by measurements of double-pion electroproduction cross sections~\cite{Ripani:2002ss,Fedotov:2008aa}.  
The cross sections reported in Ref.~\cite{Fedotov:2008aa}, although extracted in  $Q^{2}$ bins of the same width (0.05~GeV$^{2}$),  overlap with the present results only in the low region  $0.45$~GeV$^{2}$ $< Q^{2} < 0.6$~GeV$^{2}$ and $W$ up to $\sim$1.55~GeV. The comparison of the present results with the measurements from Ref.~\cite{Fedotov:2008aa} is given in Section~\ref{prev_data}.
The cross sections reported in Ref.~\cite{Ripani:2002ss} for $1.4$~GeV $< W < 1.825$~GeV, that have been extracted in much wider $Q^{2}$ bins $0.5$~GeV$^{2}$ $< Q^{2} < 0.8$~GeV$^{2}$ and $0.8$~GeV$^{2}$ $< Q^{2} < 1.1$~GeV$^{2}$, also partially overlap with the results reported here. However, since they have been averaged over a large $Q^{2}$ range, direct comparisons with these data are not straightforward and are not shown here.

One of the promising ways to move closer to the understanding of nucleon structure and principles of the strong interaction is the studies of nucleon excited states~\cite{Burkert:2016kyi,Krusche:2003ik,Aznauryan:2011qj,Skorodumina:2016pnb}.
The extracted cross sections are of great significance for these studies due to the essential sensitivity of the double-pion electroproduction channel to the manifestation of resonances above the $\Delta(1232)$. Most of these excited states have a considerable branching ratio to the $N\pi\pi$ final state, especially those with masses above 1.6~GeV, which are known to decay mostly by the emission of two charged pions. Beside that, the reported cross sections benefit from a narrow $Q^{2}$ binning, which is valuable for investigating the resonant structure  through establishing the $Q^{2}$-evolution of the resonance electrocouplings.

The most common way to investigate  nucleon resonances  is to perform a phenomenological analysis of the observables within a reaction model, as in the case of the double-pion exclusive channel with the JLab - Moscow State University (Russia) model JM~\cite{Mokeev:2015lda}. This model, which aims at the extraction of resonance electrocouplings and the identification of different reaction mechanisms, has proven itself as an effective tool for the analysis of the experimental cross sections~\cite{Mokeev:2008iw,Mokeev:2012vsa,Mokeev:2015lda}.

Section V introduces the JM model based preliminary interpretation of the extracted cross sections, which includes the estimation of  contributions from nucleon resonances. 
The relative resonant contributions to the cross section are found to range from 20\% to 70\% (depending on the kinematic region), which is a very promising indication that a reliable extraction of the resonance electrocouplings within the JM model will be possible.

The complete analysis of the present cross sections within the JM model, which aims to determine the evolution of the electrocouplings of most nucleon resonances  with masses up to $\sim$1.8~GeV (including the new potential candidate state
$N'(1720)3/2^+$~\cite{Mokeev:2015moa}), will be the subject of a future publication.

\section{Experimental setup}

The data reported in this paper were acquired at JLab Hall~B  with the CEBAF Large Acceptance Spectrometer (CLAS)~\cite{Me03}, which consisted of six sectors that were operated as independent detectors. Each sector included Drift Chamber (DC), a \v Cerenkov Counter (CC), a Time-Of-Flight system (TOF), and a sampling Electromagnetic Calorimeter (EC). The CLAS detector had a toroidal magnetic field that bent charged particle trajectories and therefore allowed for the determination of their momenta in the DC.
The electron beam was provided by the Continuous Electron Beam Accelerator Facility (CEBAF).
The measurements were part of the  ``e1e" run period that lasted from November 2002 until January 2003 and included several datasets with different configurations (hydrogen and deuterium targets as well as two different beam energies of 1 GeV and 2.039 GeV).

\begin{figure}[htp]
\begin{center}
\frame{\includegraphics[width=8.5cm]{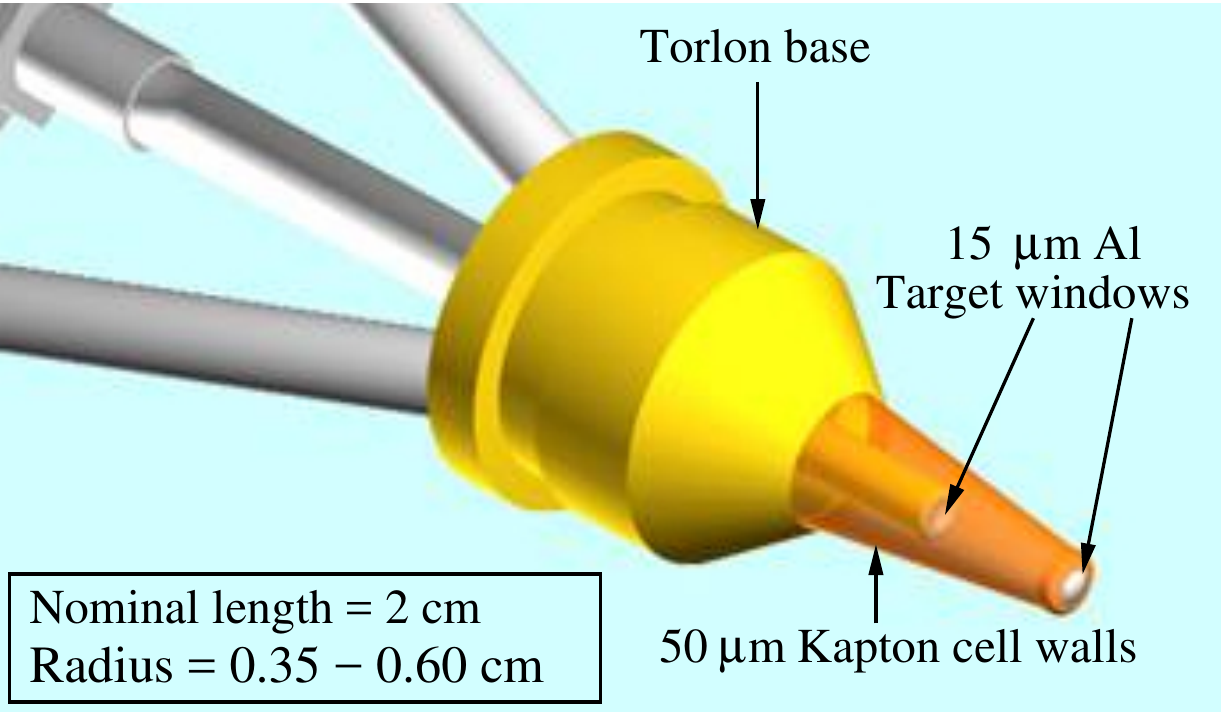}}
\vspace{-0.1cm}
\caption{(colors online) The target cell and support structure used during the CLAS ``e1e" run period.}
\label{fig:e1e_target}
\end{center}
\end{figure} 

The experimental configuration for the analyzed dataset was the following. The torus field setting was such as to bend negative particles toward the beamline (inbending configuration).
The data  were obtained with a 2-cm-long liquid hydrogen target, located at -0.4 cm along the $z$-axis (near the center of CLAS), and a 2.039 GeV electron beam.

The target was specific to the ``e1e" run period and its setup is presented in Fig.~\ref{fig:e1e_target}. In order to avoid bubble formation, the target had a special conical shape that allowed draining the bubbles away from the beam interaction region. The target cell had 15-$\mu$m-thick aluminum entrance and exit windows. In addition, an aluminum foil was located downstream of the target. This foil was made exactly the same as the entry/exit windows of the target cell and served for both  the estimation of the number of events that originated in the target windows and the precise determination of the target $z$ position along the beamline.

\begin{figure}[htp]
\begin{center}
 \includegraphics[width=8cm,keepaspectratio,clip,trim={0 0 0 0}]{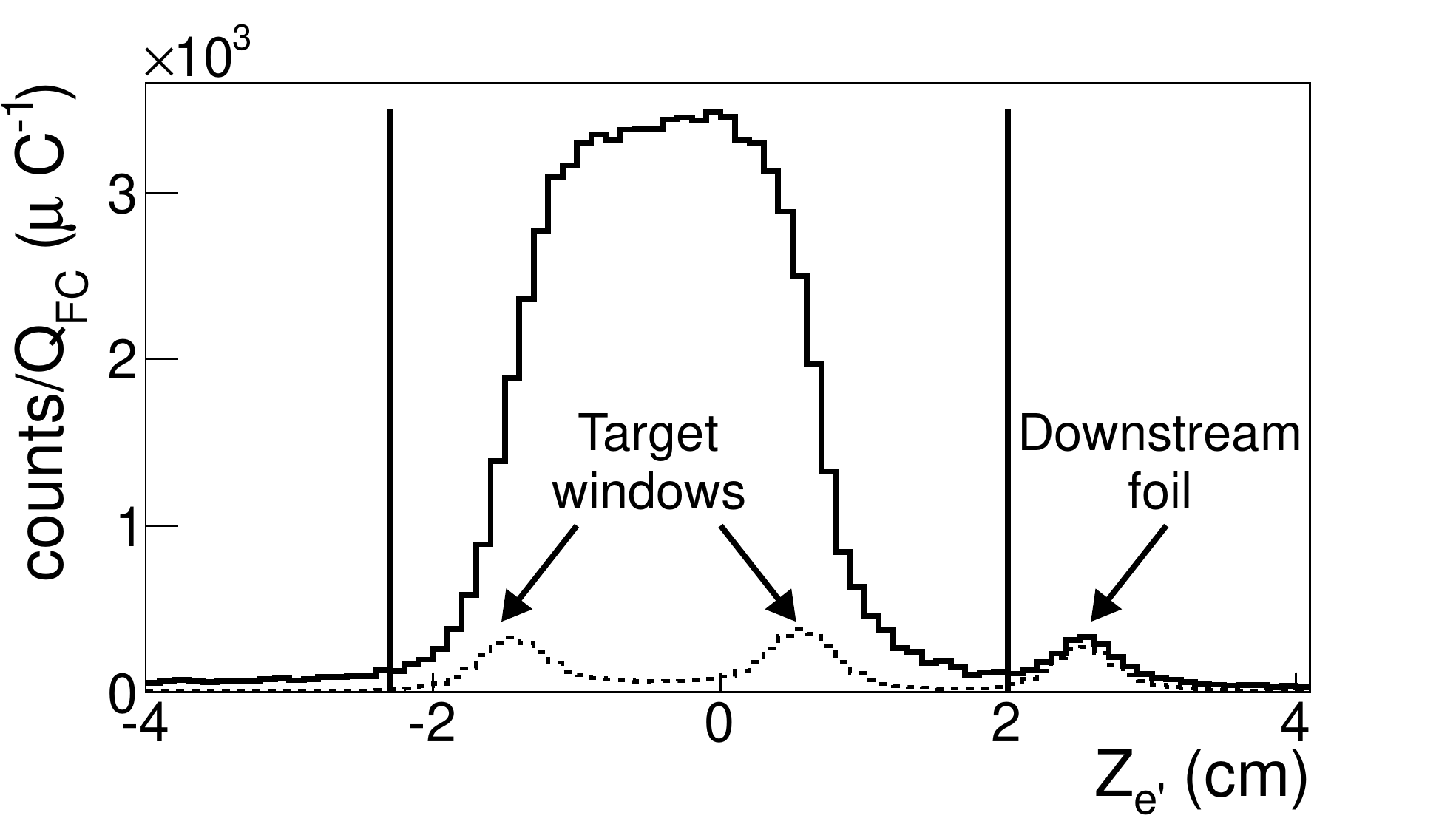}
\vspace{-0.1cm}
\caption{Distributions of the electron $z$ coordinate at the vertex for full (solid curve) and empty (dashed curve) target runs. The vertical lines show the applied cuts. Both full and empty target distributions are normalized to the corresponding charge accumulated on the Faraday Cup (FC).}
\label{fig:zvertex}
\end{center}
\end{figure} 

The dataset included runs with the target cell filled with liquid hydrogen (full) as well as runs with an empty target cell (empty). The latter served to subtract the contribution from the background events produced  by the scattering of electrons on the target windows. 
In Fig.~\ref{fig:zvertex} the distributions of electron coordinate $z$ at the interaction vertex are shown for events from both empty (dashed curve) and full (solid curve) target runs. Both distributions are normalized to the corresponding charge accumulated on the Faraday Cup (FC). The value of the vertex coordinate $z$ was corrected for the effects of beam-offset\footnote[2]{The beam offset is the deviation of the beam position from the CLAS central line $(x,y)=(0,0)$ that can lead to the inaccurate determination of the vertex position.} at the stage of data calibration. Both distributions in Fig.~\ref{fig:zvertex} demonstrate the well-separated peak around $z_{e'} = 2.4$~cm originating from the downstream aluminum foil. The distribution of events from the empty target runs also shows two other similar peaks that correspond to the windows of the target cell. In addition to the empty target event subtraction, a cut on the $z$-coordinate of the electron was applied. This cut is shown by the two vertical lines in Fig.~\ref{fig:zvertex}: events outside these lines were excluded from the analysis. 
 
\section{Exclusive reaction event selection}
\label{expt}

To identify the reaction $e p \rightarrow e' p' \pi^{+} \pi^{-}$, the scattered electron and at least two final state hadrons need to be detected, while the four-momentum of the remaining hadron can be calculated from energy-momentum conservation. The fastest particle that gives signals in all four parts of the CLAS detector (DC, CC, TOF, and EC) was chosen as the electron candidate for each event. To identify hadrons, only signals in the DC and TOF were required.

\subsection{Electron identification}

To reveal good electrons from all electron candidates,  electromagnetic calorimeter (EC) and \v Cerenkov counter (CC) responses were analyzed.

According to Ref.~\cite{Egian:007}, the overall EC resolution, as well as uncertainties in the EC output summing electronics lead to the fluctuation of the EC response near the hardware threshold. Therefore, to select only reliable EC signals, a minimal cut on the scattered electron momentum $P_{e'}$ (which is known from the DC) should be applied at the software level. As it was suggested in Ref.~\cite{Egian:007}, this cut was chosen to be $P_{e'} > 0.461$~GeV.

In the next step, a so-called sampling fraction cut was applied to  eliminate in part the pion contamination. To develop this cut, the fact that electrons and pions
had different  energy deposition patterns in the EC was used. The energy  
deposited by an electron $(E_{\text{tot}})$ is proportional to its momentum $(P_{e'})$, while a $\pi^{-}$ loses a constant
amount of energy per scintillator ($\approx 2$~MeV/cm) independently of its momentum. 
Therefore, for electrons the quantity $E_{\text{tot}}/P_{e'}$ plotted as a function of $P_{e'}$ should follow a straight line that is parallel to the $x$-axis (in reality this line has a slight slope). This line is located  around the value 1/3 on the $y$-axis, since by the EC design an electron loses about 1/3 of its energy in the active scintillators.

\begin{figure}[htp]
\begin{center}
 \includegraphics[width=7cm,keepaspectratio]{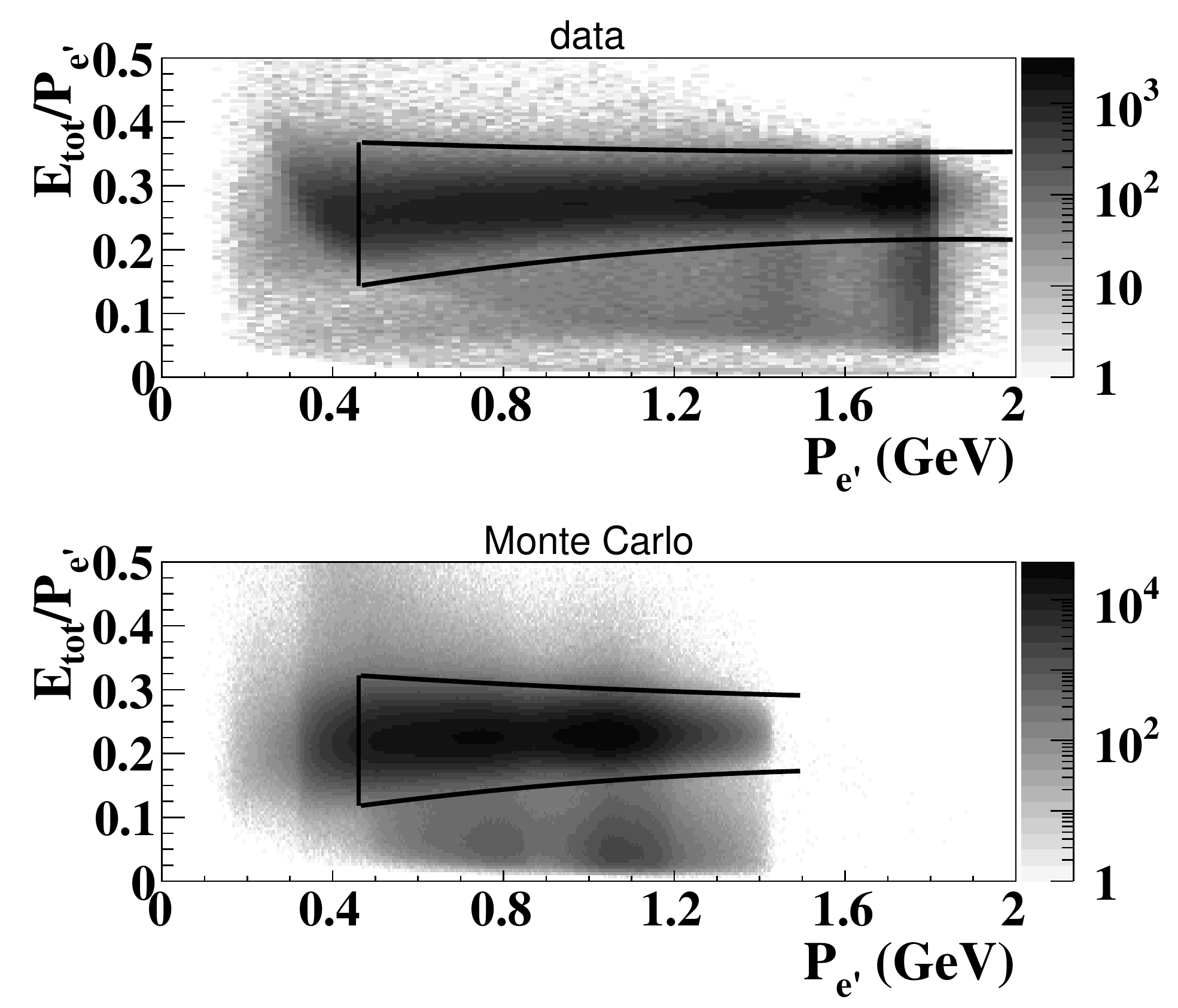}
\vspace{-0.1cm}
\caption{Sampling fraction distributions for the data (top plot) and the Monte Carlo (bottom plot). Both plots correspond to CLAS sector 1. Events between the curves were treated as good electron candidates.}
\label{fig:ec_cut}
\end{center}
\end{figure}

In Fig.~\ref{fig:ec_cut} the total energy deposited in the EC divided by the particle momentum is shown as a function of the particle momentum for the data (top plot) and the Monte Carlo (bottom plot). In this figure, a cut on the minimal scattered electron momentum is shown by the vertical line segment, while the other two curves correspond to the sampling fraction cut that was determined via a Gaussian fit to different momentum slices of the distribution. The distributions for the experimental data and the Monte Carlo simulation differ, since the former is plotted for inclusive electrons, while the latter is for simulated double pion events only.
The mean value of the simulated distribution turned out to be slightly below that of the experimental one 
due to the approximations used in the reproduction of electromagnetic showers in the Monte Carlo reconstruction procedure.

To improve the quality of electron candidate selection and $\pi^{-}/e^{-}$ separation, a \v Cerenkov counter was used.
As was shown in Ref.~\cite{Osipenko:2004}, there was a contamination in the measured CC spectrum that manifested itself as a peak at low number of photoelectrons (the so-called few photoelectron peak). The main source of this contamination was found to be the coincidence of accidental photomultiplier tube (PMT) noise with a  pion track measured in the DC~\cite{Osipenko:2004}. 

\begin{figure}[htp]
\begin{center}

\includegraphics[width=7cm,keepaspectratio]{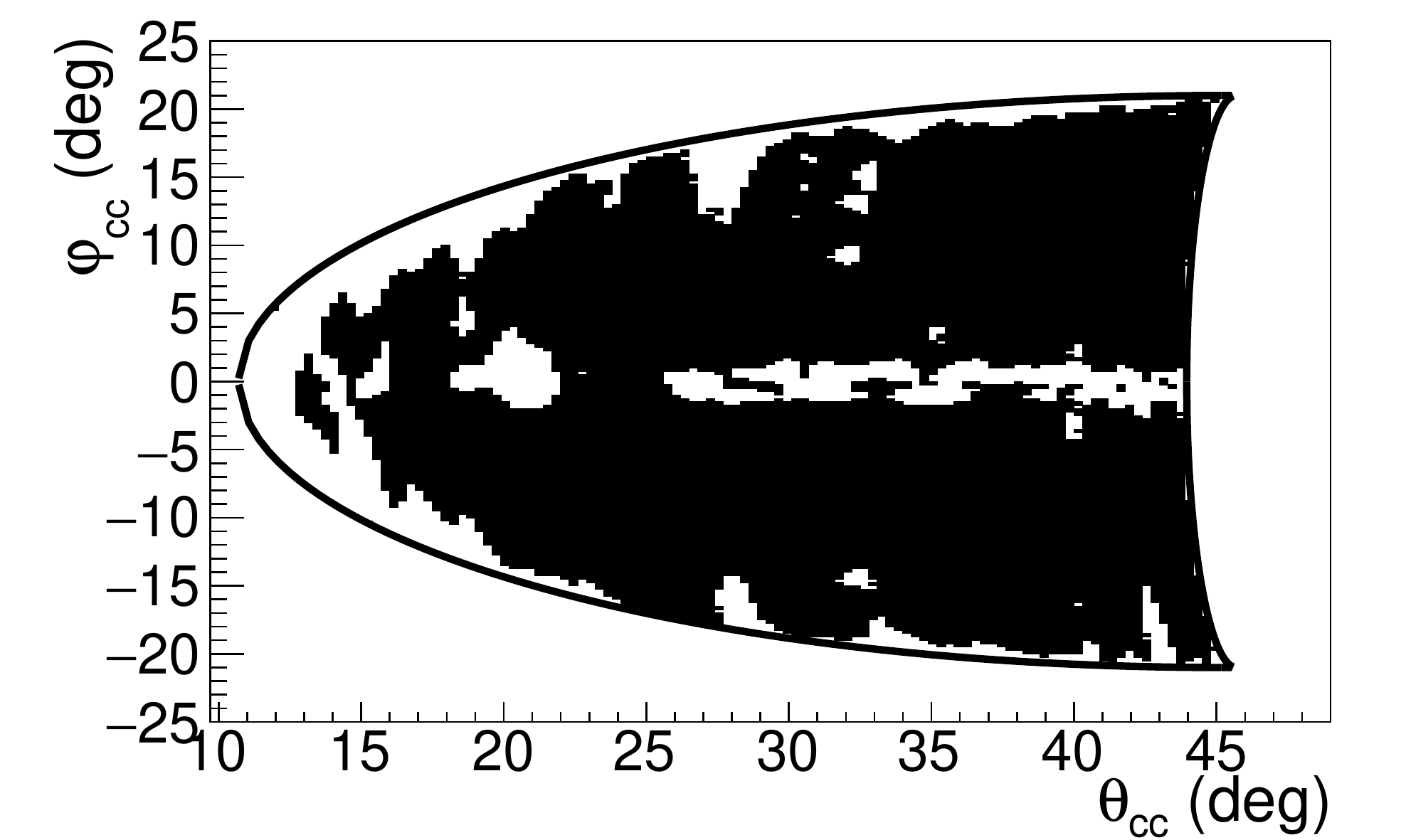}
\vspace{-0.1cm}
\caption{The CC regions with reliable detection efficiency are shown in black as a function of the polar ($\theta_{\text{cc}}$) and azimuthal ($\varphi_{\text{cc}}$) angles in the CC plane for CLAS sector 1. These regions were selected according to the criterion~\eqref{eq:ineff_zone_cut}. The curves, which are superimposed on the distribution,  show an overall fiducial cut that was applied in the CC plane.}
\label{fig:ph_vs_ph_cc}
\end{center}
\end{figure}

It turned out that the CC had some inefficient zones  that could not be simulated by the Monte Carlo technique as being too dependent on specific features of the CC design.
Signals from these zones, being depleted of photoelectrons, shifted the measured CC spectrum toward zero and therefore add up to the few photoelectron peak. Thus the inefficient zones can be differentiated from the efficient ones by a more pronounced few photoelectron peak. The following criterion for the geometrical selection of the efficient zones in the CC was used (see Ref.~\cite{Fed_an_note:2017} for details)  
\begin{equation}
\frac{N_{\text{N}_{\text{ph. el.}}>5}(\theta_{\text{cc}},\varphi_{\text{cc}})}{N_{\text{tot}}(\theta_{\text{cc}},\varphi_{\text{cc}})} > 0.8,
\label{eq:ineff_zone_cut}
\end{equation}
where the denominator corresponds to the total number of events in the particular $(\theta_{\text{cc}},\varphi_{\text{cc}})$ bin, while the numerator corresponds to the number of events with more than five photoelectrons in the same $(\theta_{\text{cc}},\varphi_{\text{cc}})$ bin. The polar $(\theta_{\text{cc}})$ and azimuthal $(\varphi_{\text{cc}})$ angles of the electron candidate are defined in the CC plane.

In Fig.~\ref{fig:ph_vs_ph_cc} the distribution of the CC regions with reliable detection efficiency, which were selected according to the criterion~\eqref{eq:ineff_zone_cut}, are shown in black as a function of $\theta_{\text{cc}}$ and $\varphi_{\text{cc}}$ for CLAS sector 1. As is seen in Fig.~\ref{fig:ph_vs_ph_cc}, there was an inefficient area in the middle of the sector (shown in white). This was expected since two CC mirrors were joined there. The curves, which are superimposed on the distribution,  show an overall fiducial cut that is applied in the CC plane. Then, within that overall cut, for both the experimental data and the Monte Carlo simulation, only electron candidates that originated from the black regions were analyzed.

\begin{figure}[htp]
\begin{center}
 \includegraphics[width=7cm,keepaspectratio]{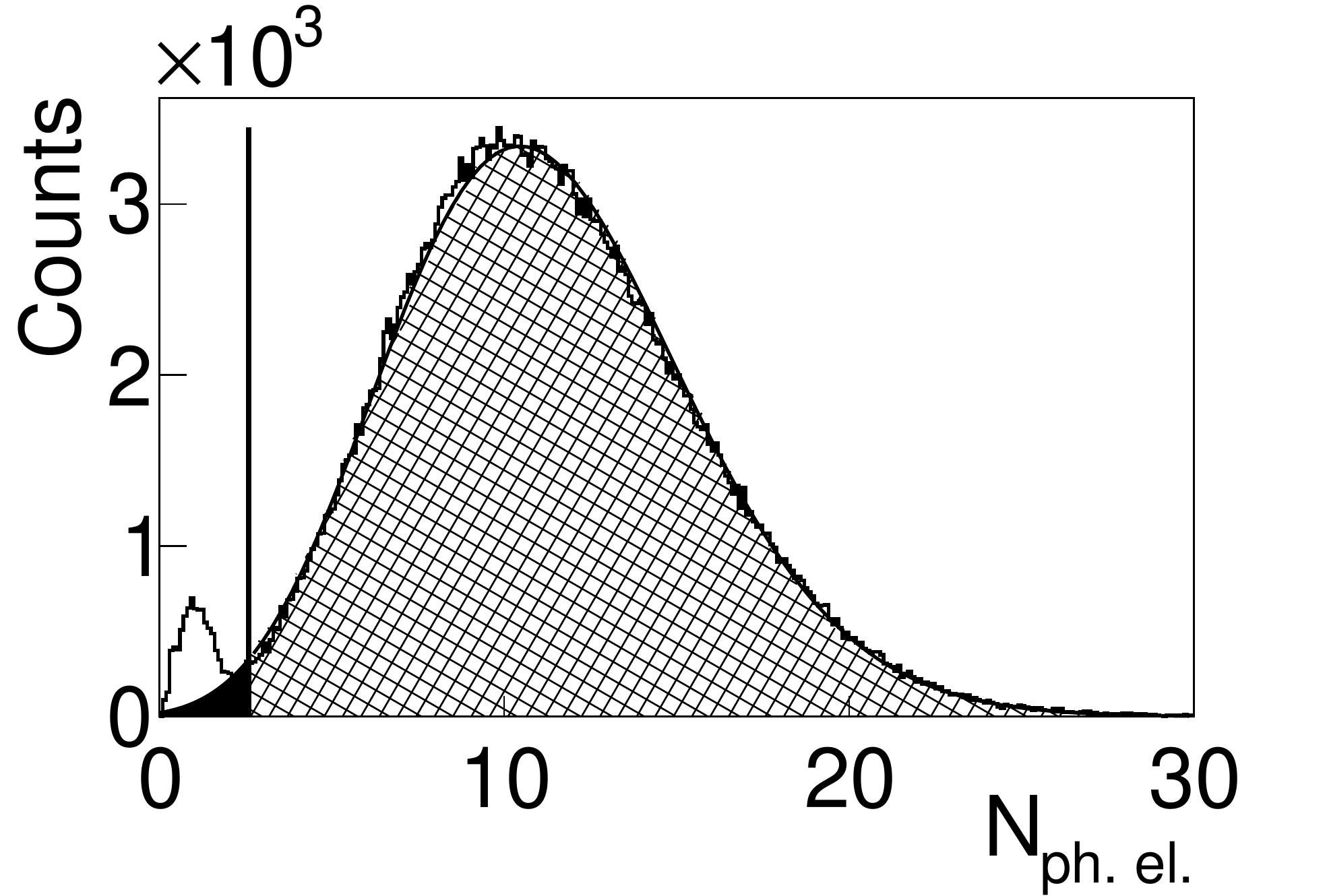}
\vspace{-0.1cm}
\caption{Number of photoelectrons for the left side PMT in segment 10 of sector 1 of the CC. The black curve shows the fit by the function given by Eq.~\eqref{eq:cc_Poisson}. The vertical line shows the applied cut. Regions that are needed to calculate the correction factor (see Eq.~\eqref{eq:cc_corr_fact}) are shown in hatch and in black. }
\label{fig:ph_el}
\end{center}
\end{figure}

Although being substantially reduced after elimination of signals from the inefficient zones, the few photoelectron peak was still present in the experimental CC spectrum as shown in Fig.~\ref{fig:ph_el}. This peak in the  photoelectron distribution was cut out for each PMT in each CC segment individually. The cut position for one particular PMT is shown by the vertical line in Fig.~\ref{fig:ph_el}.
Since there was no way  of reproducing the photoelectron spectrum by a Monte Carlo simulation, this cut was applied only to the experimental data, and good electrons lost in this way were recovered by the following procedure. The part of the distribution on the right side of the vertical line was fit by the function given by Eq.~\eqref{eq:cc_Poisson}, which is a slightly modified Poisson distribution, 

\begin{equation}
y = P_{1}\left(\frac{P_{3}^{\frac{x}{P_{2}}}}{\Gamma\left(\frac{x}{P_{2}}+1\right)}
\right)e^{-P_{3}},
\label{eq:cc_Poisson}
\end{equation}
where $P_{1}$, $P_{2}$, and $P_{3}$ are free fit parameters.

The fitting function was then continued into the region on the left side of the vertical line. In this way the two regions, shown in black and in hatch in Fig.~\ref{fig:ph_el}, were determined. Finally, the correction factors were defined by Eq.~\eqref{eq:cc_corr_fact} and applied as a weight for each event which corresponded to the particular PMT. 

\begin{equation}
F_{\text{ph. el.}} = \frac{\text{hatched area} {\,\,+\,\,} \text{black  area}}{\text{hatched area}} \textrm{.}
\label{eq:cc_corr_fact}
\end{equation}

The correction factor $F_{\text{ph. el.}}$ depended on PMT number and was typically on the level of a few percent.

\subsection{Hadron identification}

\begin{figure}[htp]
\begin{center}
 \includegraphics[width=6cm,keepaspectratio]{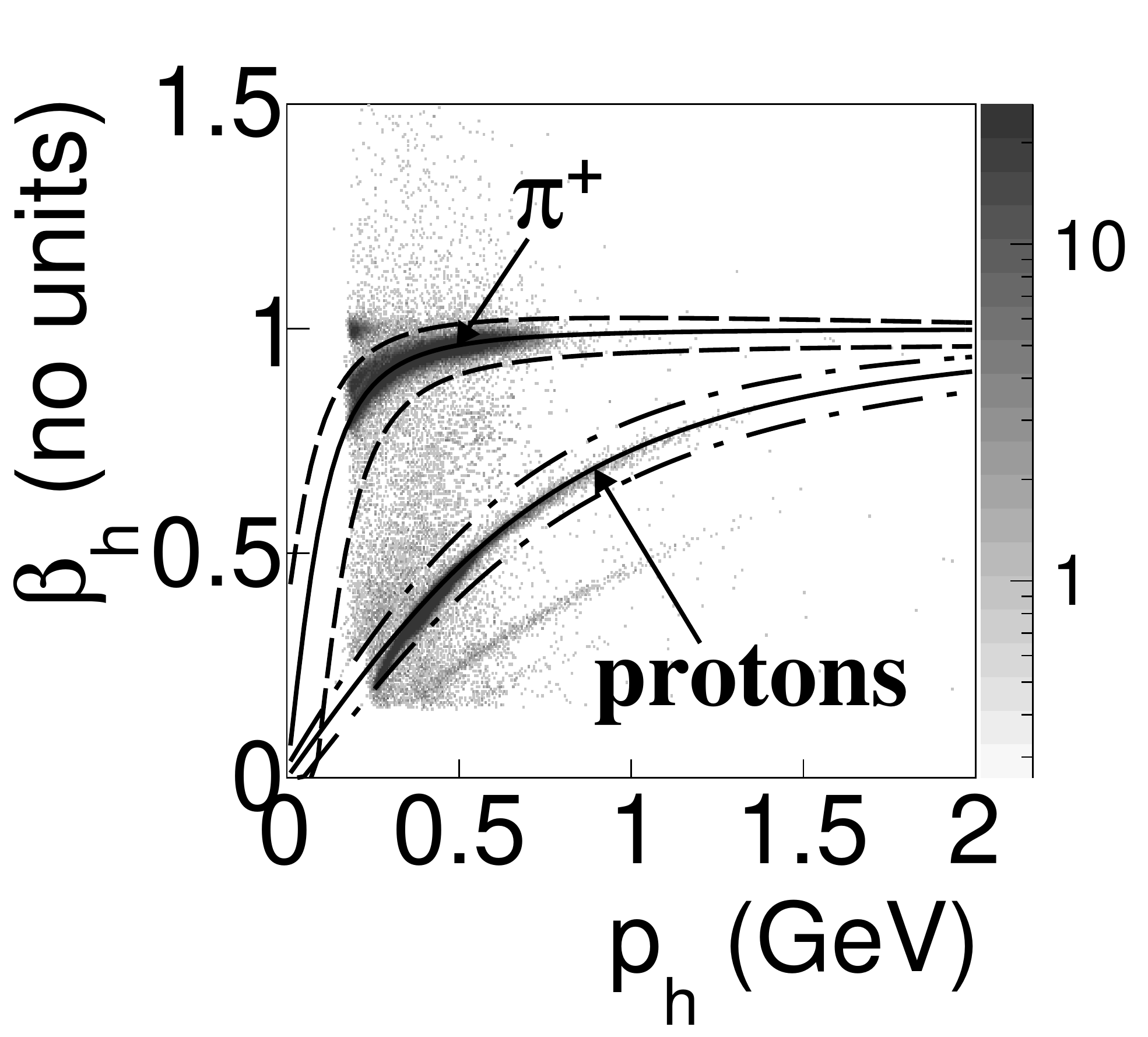}
\includegraphics[width=6cm,keepaspectratio]{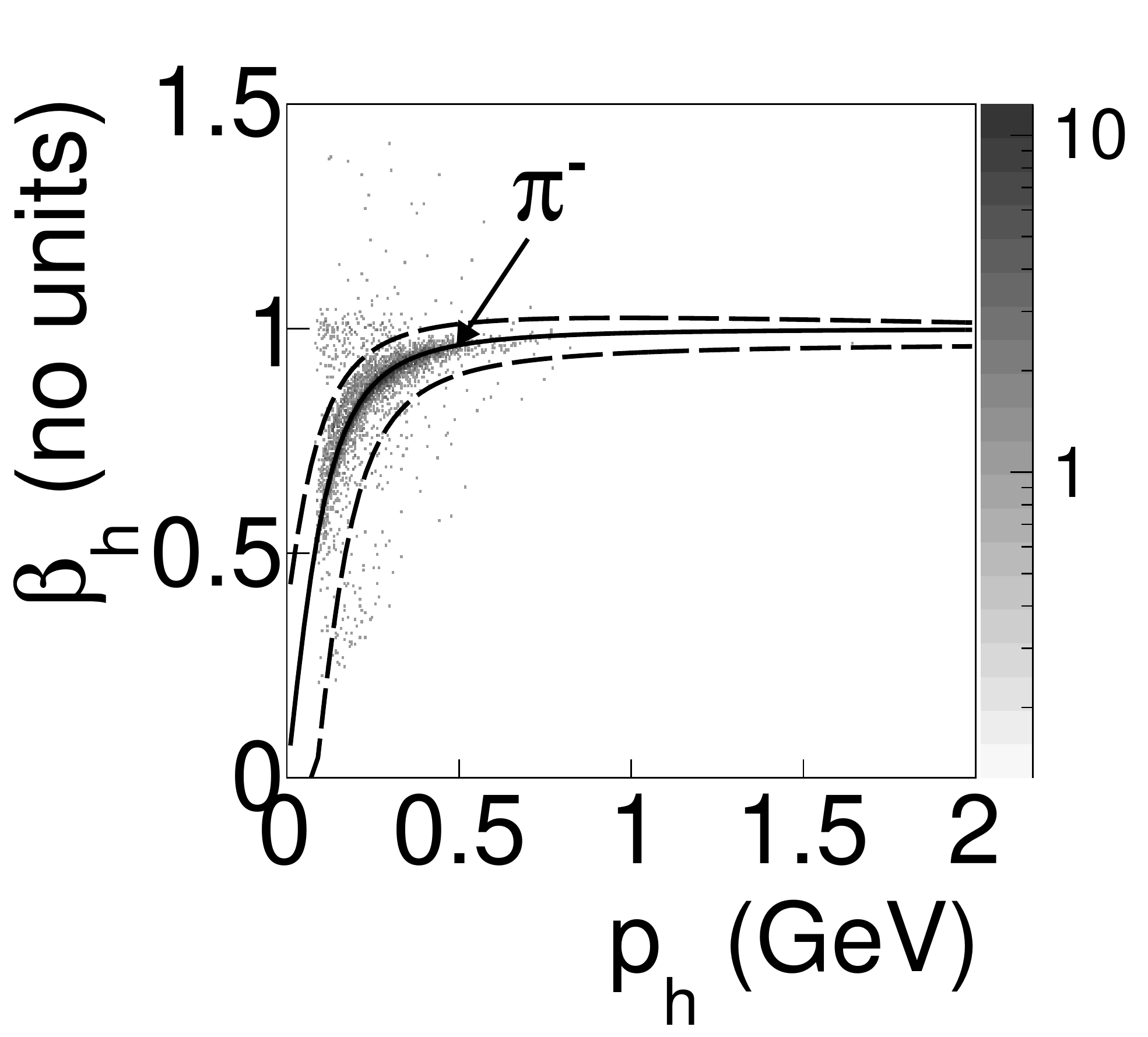} 
\vspace{-0.1cm}
\caption{$\beta_{h}$ versus momentum distributions for positively charged hadron candidates (top plot) and  negatively charged hadron candidates (bottom plot) for scintillator number 34 in CLAS sector 1. The black solid curves correspond to the nominal $\beta_{\text{n}}$ given by Eq.~\eqref{eq:hadron_hadronmass}. 
Events between the dashed and dot-dashed curves were selected as $\pi^{+}$ ($\pi^{-}$) and protons, respectively.}
\label{fig:b_vs_p}
\end{center}
\end{figure} 

The CLAS TOF system provided timing information, based on which  the velocity $(\beta_{h} = v_{h}/c)$ of the hadron candidate was calculated.  
The value of the hadron candidate momentum $(p_{h})$
was in turn provided by the DC.  The
charged hadron can be identified by a comparison of $\beta_{h}$, determined by the TOF, with $\beta_{\text{n}}$ given by:

\begin{equation}
\beta_{\text{n}}=\frac{p_{h}}{\sqrt{p_{h}^{2}+m_{h}^{2}}},
\label{eq:hadron_hadronmass}
\end{equation}
where $\beta_{\text{n}}$ is the nominal value that is calculated using the hadron candidate momentum $(p_{h})$ and an exact hadron mass assumption $m_{h}$.

The experimental event distributions $\beta_{h}$ versus $p_{h}$ were investigated for each TOF scintillator in each CLAS sector. An example of these distributions is shown in Fig.~\ref{fig:b_vs_p} for positively charged hadron candidates (top plot) and negatively charged hadron candidates (bottom plot). 
 The example is given for scintillator 34 of CLAS sector 1. In Fig.~\ref{fig:b_vs_p} the solid curves are given for $\beta_{\text{n}}$ calculated according to Eq.~\eqref{eq:hadron_hadronmass} for the corresponding hadron mass assumptions. 
The event bands of the pion and proton candidates are clearly seen around the corresponding $\beta_{\text{n}}$ curves.  
The dashed curves show the cuts that were used for pions identification, while the dot-dashed curves serve to identify protons. 

During the run, some TOF scintillator counters worked improperly and therefore their signals were considered to be unreliable and were removed from consideration in both data and simulation. 
For properly working counters, the hadron identification cuts were chosen to be the same as shown in Fig.~\ref{fig:b_vs_p}. They were applied on both experimental and reconstructed Monte Carlo events. It was found that for some scintillators  the hadron candidate bands in the experimental distributions were slightly shifted from the nominal positions. A special procedure was developed  to correct the timing information for the affected TOF counters~\cite{Fed_an_note:2017}.

\subsection{Momentum corrections}

Due to slight misalignments in the DC positions,  small inaccuracies in
the description of the torus magnetic field, and other possible reasons, the measured momentum and angle of
particles had some small systematic deviations from the real values. 
Since the effects were of an unknown origin, they could not be simulated, and therefore a special  momentum correction procedure was needed for the experimental data. 
According to Ref.~\cite{KPark:momcorr}, the evidence of the need of such corrections is most directly seen in the dependence of the elastic peak position on the azimuthal angle of the scattered electron. It is shown in Ref.~\cite{KPark:momcorr} that the elastic peak position turns out to be shifted from the proton mass value and this shift depends on CLAS sector. 

The significance of the above effect depends on the beam energy. It was found that in this dataset, with the beam energy of 2.039 GeV,  a small shift ($\sim$ 3 MeV) in the elastic peak position took place, while Ref.~\cite{KPark:momcorr} demonstrated that in case of  5.754 GeV beam energy, this shift reached 20 MeV. Moreover, Ref.~\cite{KPark:momcorr} also showed that this effect became discernible only if the particle momentum was sufficiently high (e.g. for pions the correction was needed only if their momentum was higher than 2 GeV). Here, due to the small beam energy and the fact that in double-pion kinematics hadrons carry only a small portion of the total momentum, the correction is needed only for electrons, while deviations in hadron momenta can be neglected.

\begin{figure}[htp]
\begin{center}
 \includegraphics[width=6cm,keepaspectratio]{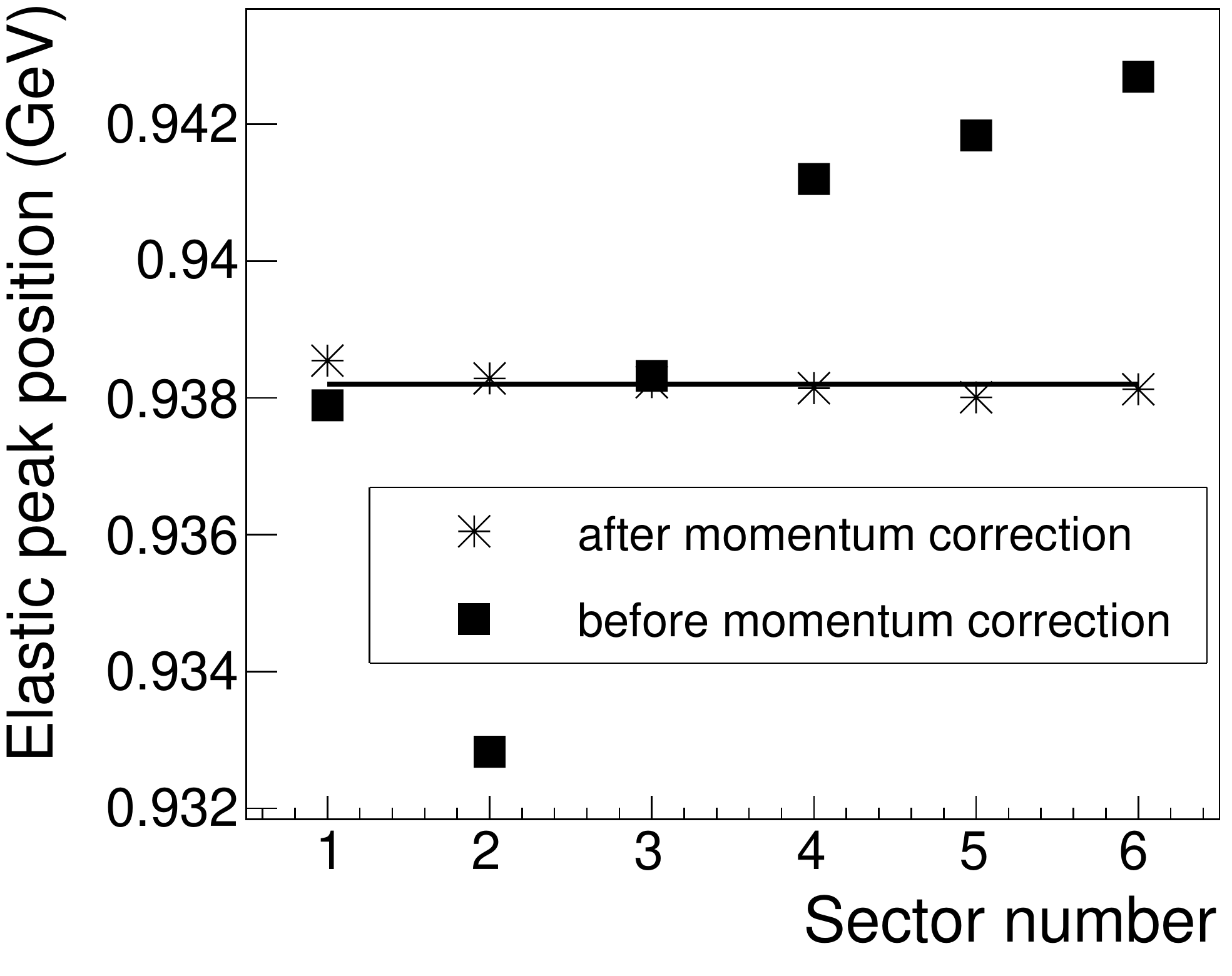}
\vspace{-0.1cm}
\caption{Elastic peak position for six CLAS sectors before (squares) and after (stars) the electron momentum correction. The horizontal line shows the proton mass. }
\label{fig:elast_pic_position}
\end{center}
\end{figure}

The electron momentum corrections used for this dataset were developed according to Ref.~\cite{KPark:momcorr} for each CLAS sector individually and included an electron momentum magnitude correction, as well as an electron polar angle correction. Although the corrections were established using elastic events, they were applied for all electron candidates in the dataset.  The influence of these corrections on the elastic peak position is shown in Fig.~\ref{fig:elast_pic_position}. The corrections bring the position of the elastic peak closer to the proton mass for all six CLAS sectors.

The  above effects do not lead to substantial distortions of the hadron momenta. However, hadrons lose a part of their energy due to their interaction with detector and target media, hence their measured momentum appears to be lower than the actual value. 
Simulation of the CLAS detector  correctly propagates hadrons through the media and, therefore, the effect of the hadron energy loss is included into the efficiency and does not impact the extracted cross section value.
However, in order to avoid shifts in the distributions of some kinematic quantities (e.g. missing masses) from their expected values, 
an energy loss correction was applied to the proton momentum magnitude,
 since the low-energy protons were affected the most by energy loss in the materials. The simulation of the CLAS detector was used to establish the correction function, which then was applied for both experimental and reconstructed Monte Carlo events.

\subsection{Other cuts}

\subsubsection{Fiducial cuts}

The active detection solid angle of the CLAS detector
 was smaller than $4\pi$~\cite{Me03} as the areas covered by the torus field coils
 were not 
equipped with any detection system, thus forming gaps in the azimuthal angle coverage. In addition, the detection area was also limited in polar angle from 8$^{o}$ up to 45$^{o}$ for electrons and up to 140$^{o}$ for other charged particles.
The edges of the detection area, being affected by rescattering from the
coils, field distortions, and similar effects should be excluded from consideration by applying specific (fiducial) cuts on the kinematic variables
(momentum and angles) of each particle. These cuts were applied for both real events and Monte Carlo reconstructed events.

\begin{figure}[htp]
\begin{center}
\includegraphics[width=7cm,keepaspectratio]{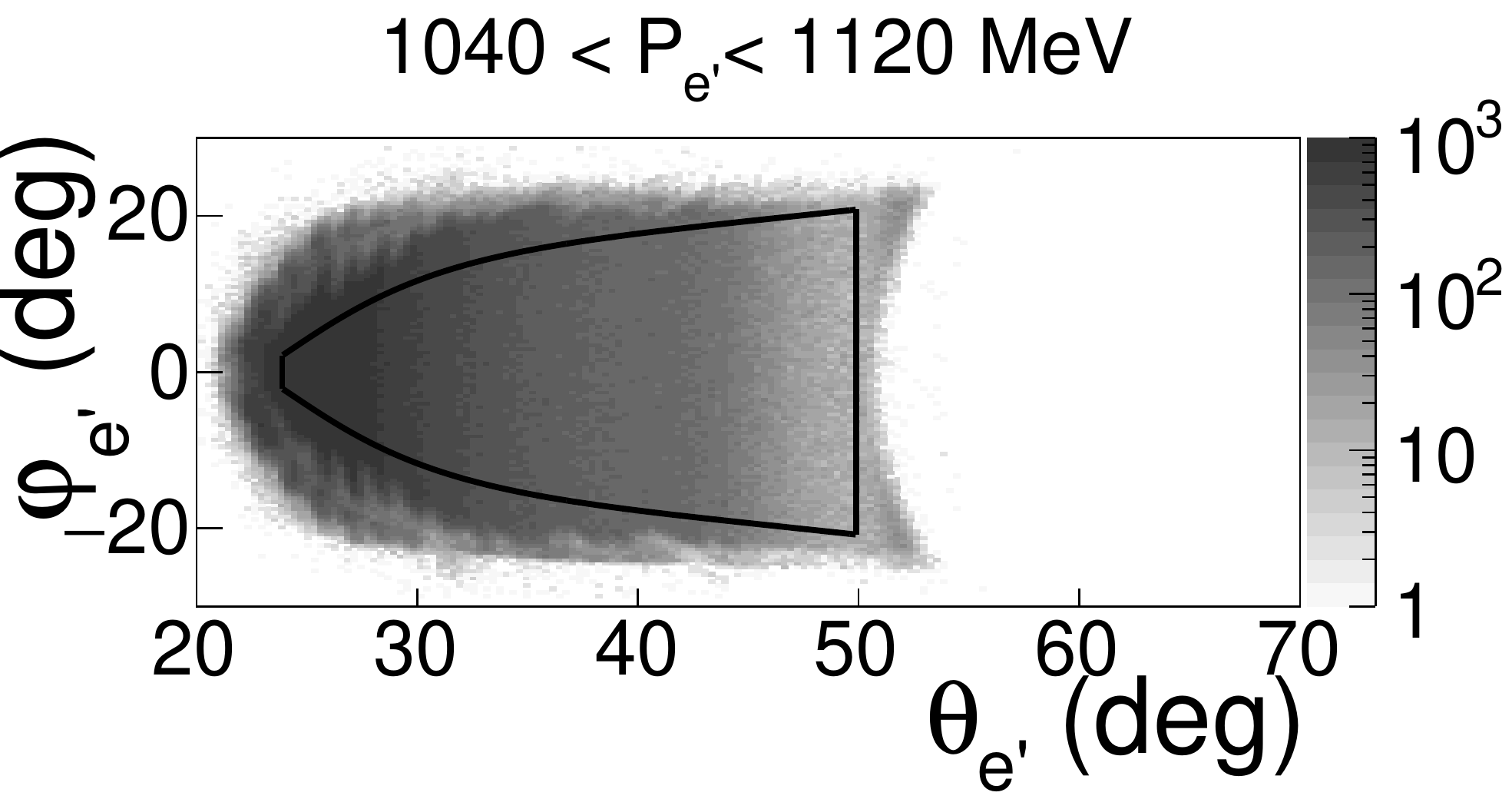}
\includegraphics[width=7cm,keepaspectratio]{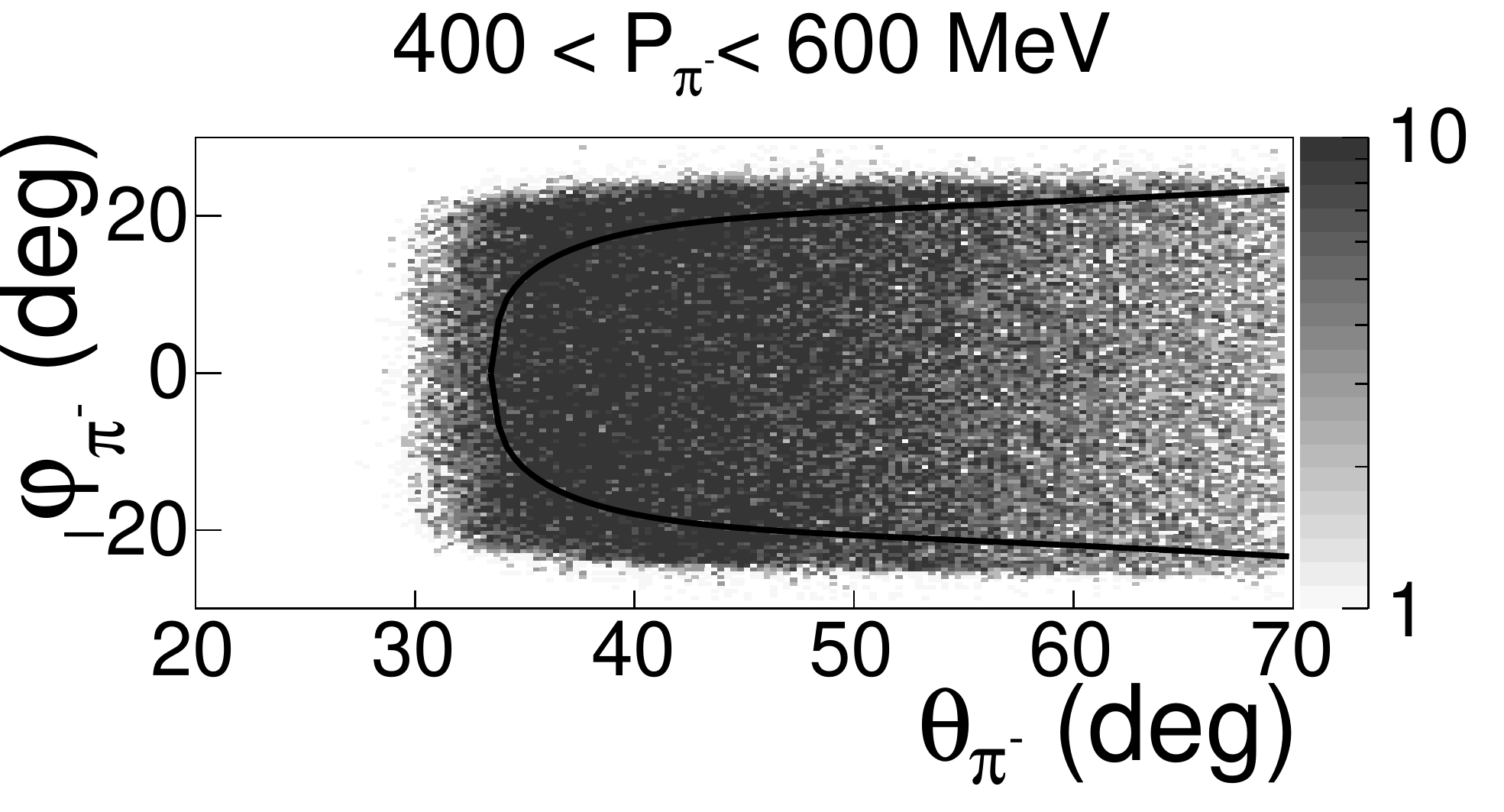} 
\vspace{-0.1cm}
\caption{Fiducial cuts for negatively charged particles. The top plot shows the $\varphi$ versus $\theta$ distribution for electrons, while the bottom plot corresponds to that for $\pi^{-}$. Both distributions are given for sector 1 of CLAS and the range over momentum specified in the plots.  The solid black curves show the applied fiducial cuts. }
\label{fig:fid_cuts_neg}
\end{center}
\end{figure} 

The ``e1e" run period used a torus magnetic field configuration that forced negatively charged particles to be inbending. For these particles, sector independent, symmetrical, and momentum dependent cuts were applied. 
Fig.~\ref{fig:fid_cuts_neg} shows the number of detected electrons (top plot) and $\pi^{-}$ (bottom plot) as a function of the angles $\varphi$ and $\theta$  for CLAS sector 1 in a specific momentum slice. The angles $\varphi$ and $\theta$ were taken at the interaction vertex. The solid black curves correspond to the applied fiducial cuts that select the regions with a relatively flat particle density along the azimuthal angle.

\begin{figure}[htp]
\begin{center}
 \includegraphics[width=7cm,keepaspectratio]{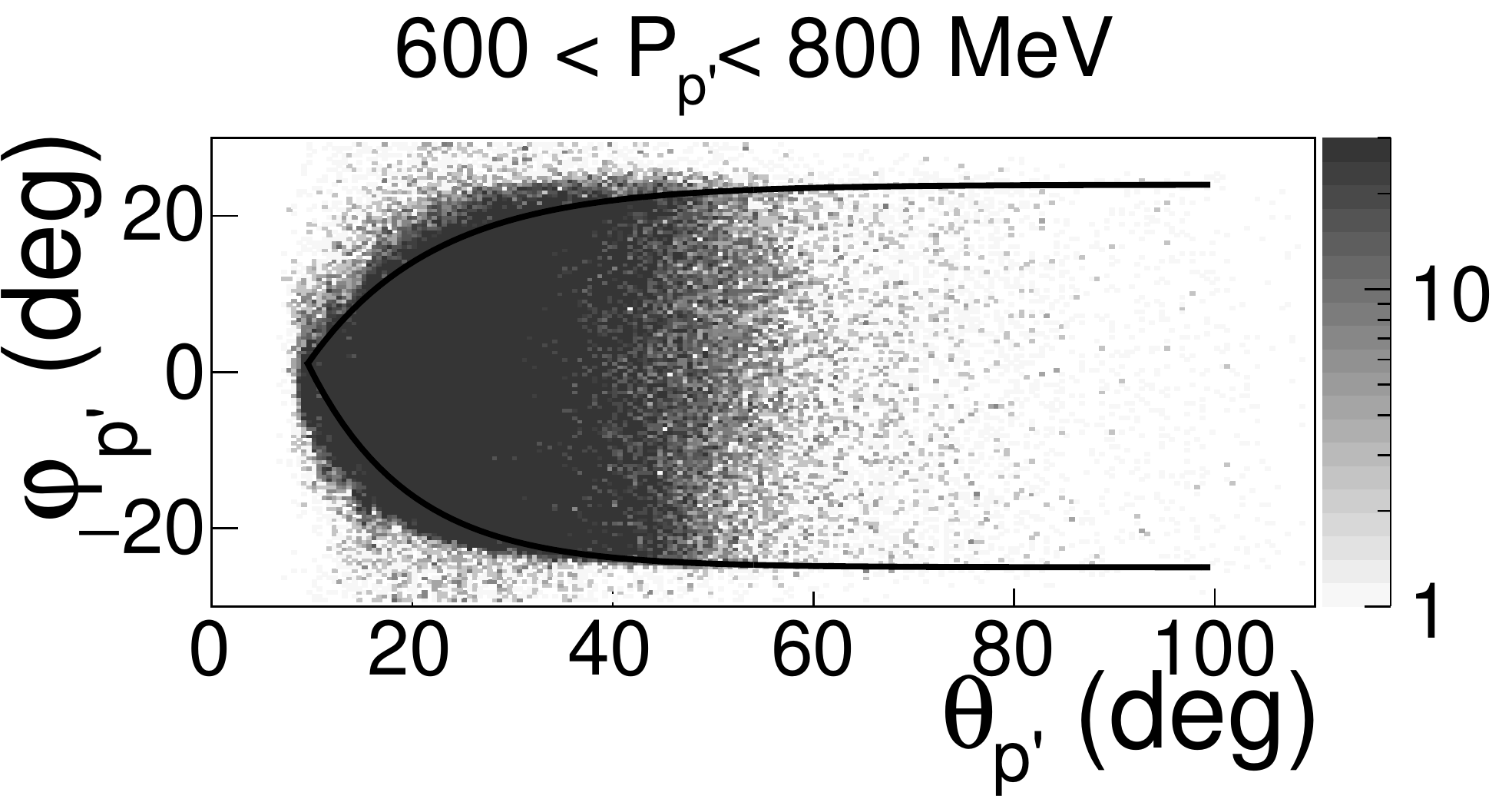}
\includegraphics[width=7cm,keepaspectratio]{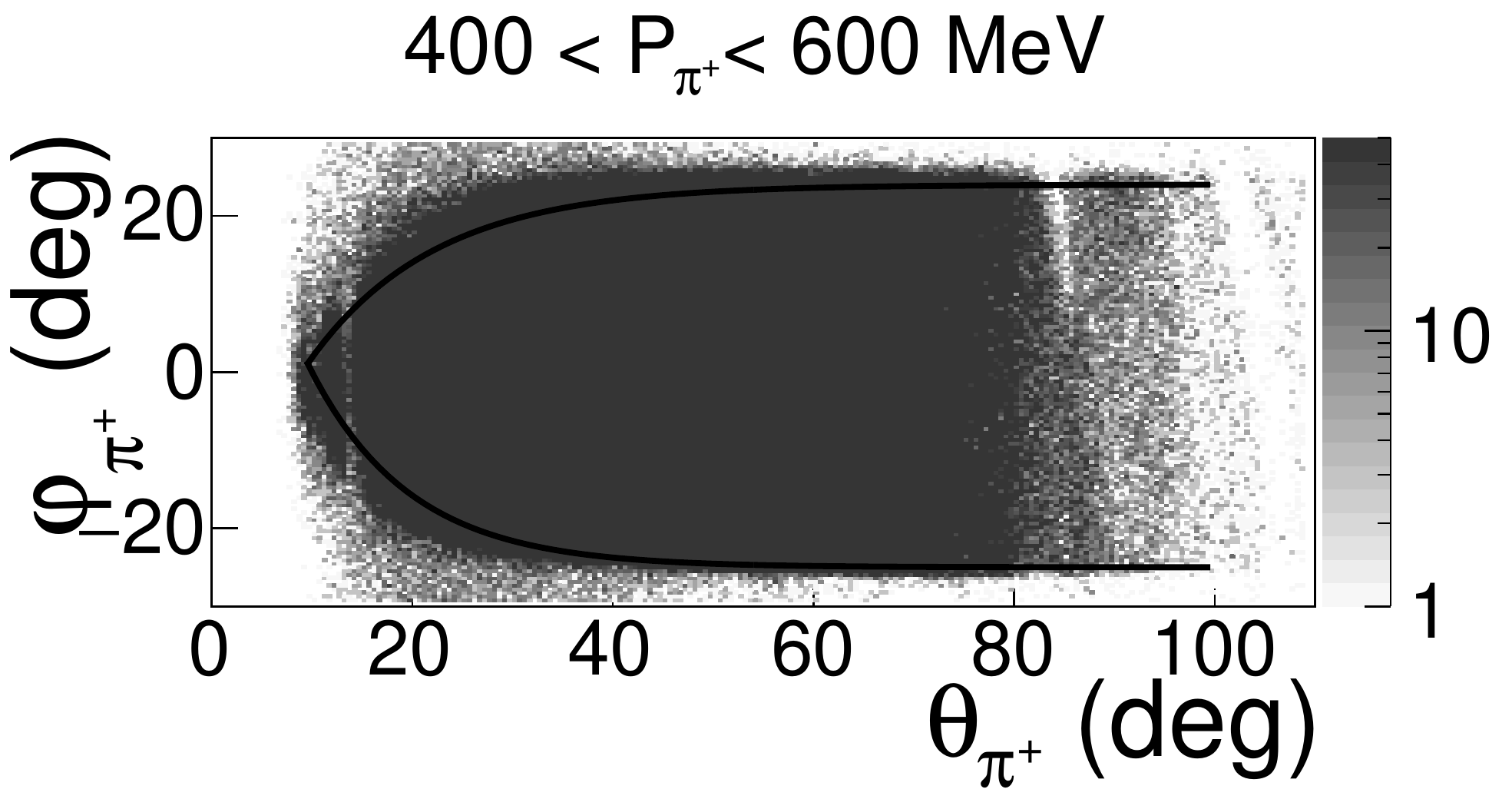} 
\vspace{-0.1cm}
\caption{Fiducial cuts for positively charged particles. The top plot shows the $\varphi$ versus $\theta$ distribution for protons, while the bottom plot corresponds to that for $\pi^{+}$. Both distributions are given for sector 1 of CLAS and the range over momentum specified in the plots.  The solid black curves show the applied fiducial cuts. }
\label{fig:fid_cuts_pos}
\end{center}
\end{figure}

For positively charged particles, which were outbending in the ``e1e" run period, momentum independent and slightly asymmetrical fiducial cuts are the best choice. These cuts were established in the same way as for negatively charged particles, i.e.
by selecting the areas with a relatively flat particle density along the $\varphi$ angle. In Fig.~\ref{fig:fid_cuts_pos} these cuts are shown by the black curves that are superimposed on the $\varphi$ versus $\theta$ event distributions for protons (top plot) and $\pi^{+}$ (bottom plot). All angles are given at the interaction vertex.

\begin{figure}[htp]
\begin{center}
 \includegraphics[width=8cm,keepaspectratio]{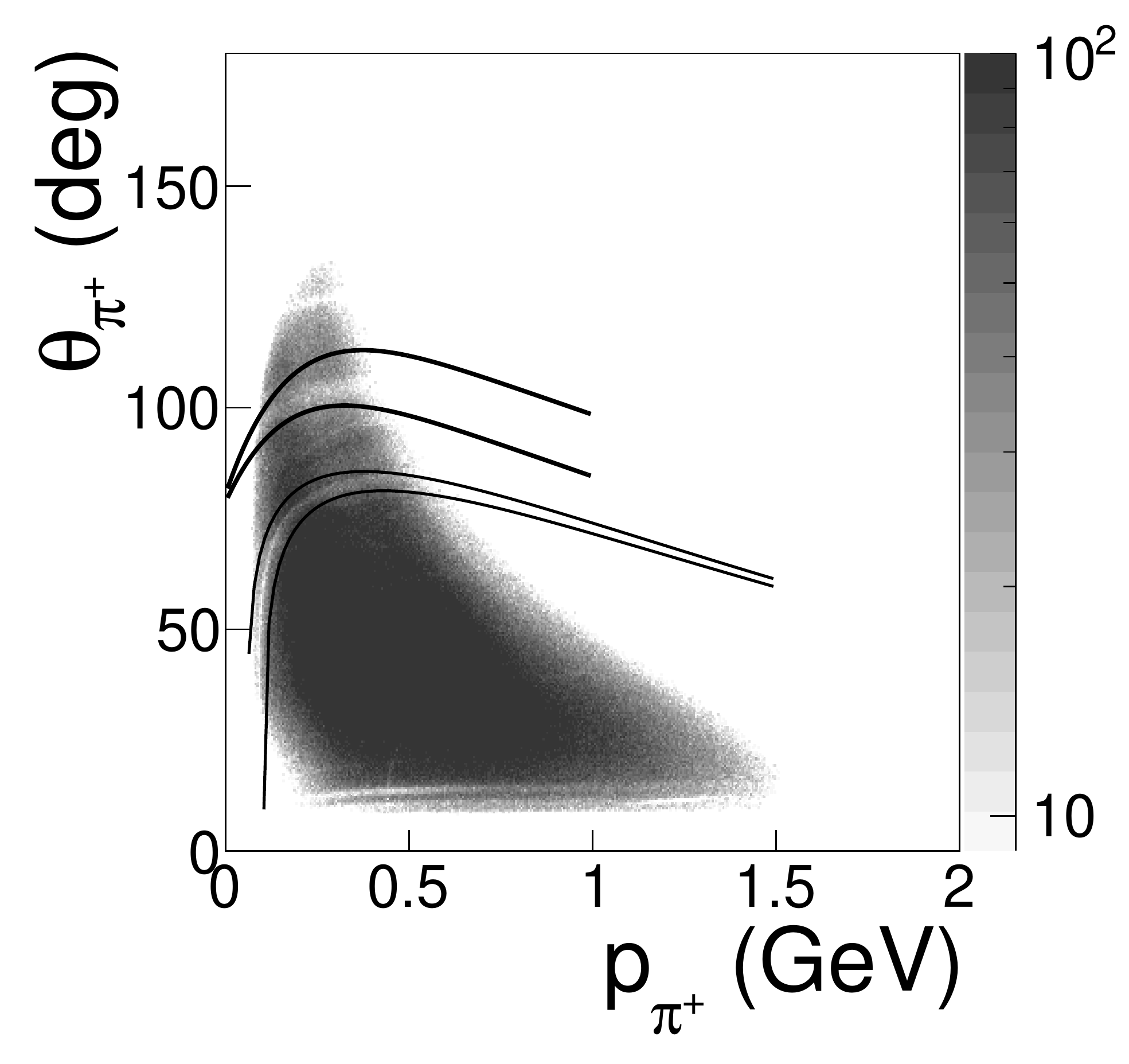} 
\vspace{-0.1cm}
\caption{$\theta$ versus momentum distribution for $\pi^{+}$ in CLAS sector 1. The angle $\theta$ was taken at the point of the interaction. The black curves show the applied fiducial cuts.}
\label{fig:fid_cuts_th_vs_p}
\end{center}
\end{figure}

Some additional inefficient areas, not related to the CLAS geometrical acceptance, were revealed in this dataset. These areas were typically caused by the DC and TOF system inefficiencies (dead wires or PMTs).  To exclude them from consideration, additional fiducial cuts on the $\theta$ versus momentum distributions were applied, where $\theta$ was taken at the point of the  interaction.  These cuts were different for each CLAS sector. An example of the cut for a $\pi^{+}$ in sector 1 of CLAS is shown by the black curves in Fig.~\ref{fig:fid_cuts_th_vs_p}.

\subsubsection{Data quality check}

\begin{figure}[htp]
\begin{center}
 \includegraphics[width=8cm,keepaspectratio]{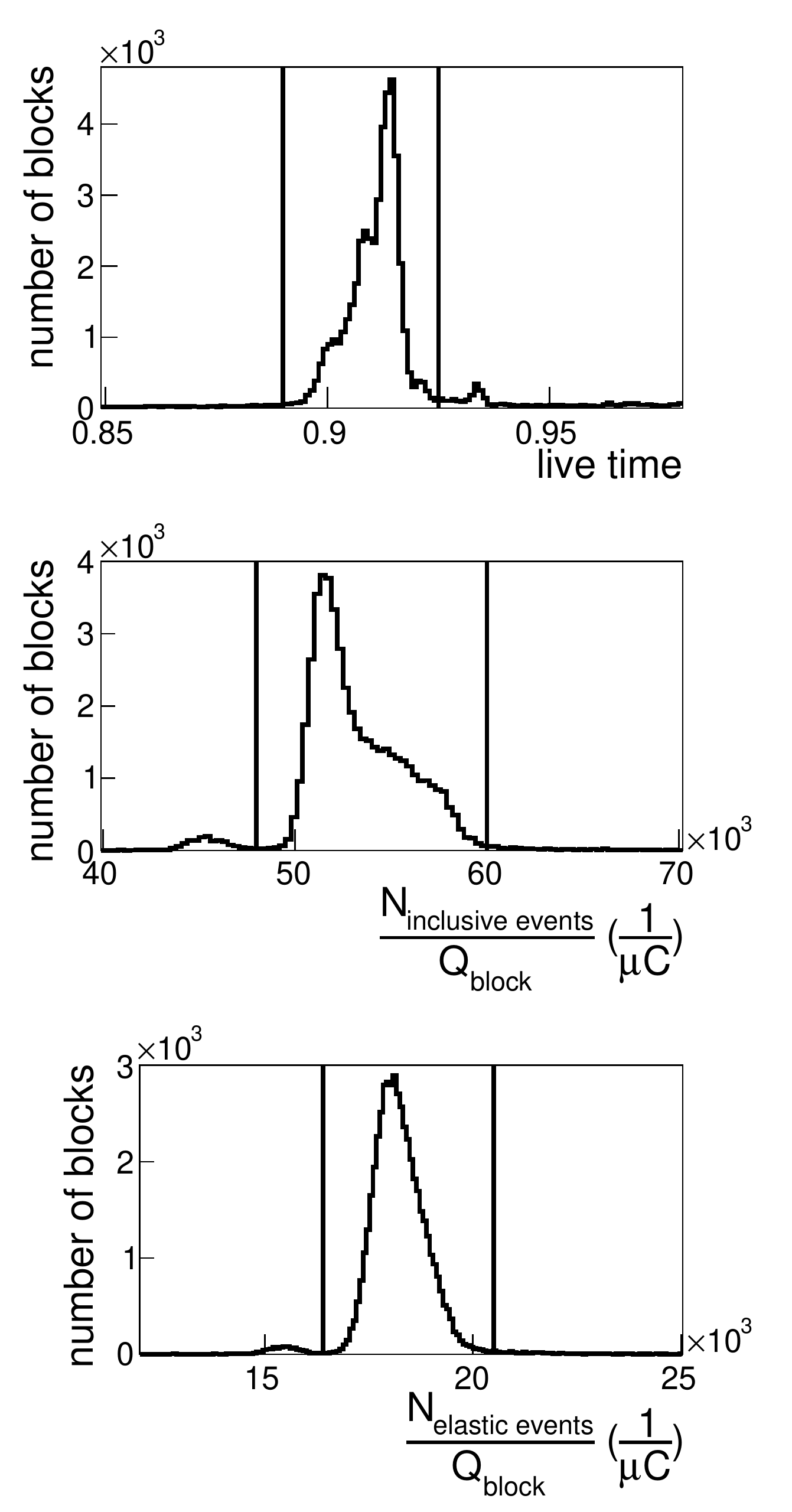} 
\vspace{-0.1cm}
\caption{Data quality check plots. The number of blocks as a function of the DAQ live time (top plot), and  the yields of inclusive (middle plot) and elastic (bottom plot) events  normalized to FC charge are shown. The vertical black  lines show the applied cuts.}
\label{fig:qcheck}
\end{center}
\end{figure}

During a long experimental run, variations of the experimental conditions, e.g. fluctuations in the target density or changes in the response of parts of the detector, can lead to fluctuations in event yields.
 Only the parts of the run with relatively stable event rates should be considered.
Therefore cuts on Data Acquisition (DAQ) live time and number of events per Faraday Cup (FC) charge need to be established. 

The FC charge was updated with a given frequency, hence the whole run time could be divided into blocks. Each block corresponded to the portion of time between two FC charge readouts. The block number ranged from one to a certain maximum number over the run time.

The DAQ live time is the portion of time within the block during which the DAQ was able to accumulate events. A significant deviation of the live time from the average value indicates event rate alteration. 

In Fig.~\ref{fig:qcheck}, the number of blocks is shown as functions of the DAQ live time and  the yields of inclusive and elastic events normalized to FC charge (from top to bottom). 
The blocks between the vertical black  lines in Fig.~\ref{fig:qcheck} were taken into consideration.

\subsubsection{Exclusivity cut}

For picking out the reaction $e p \rightarrow e' p' \pi^{+} \pi^{-} $, it is sufficient to register two final state hadrons along with the scattered electron. The four-momentum of the remaining unregistered hadron can be recovered using energy-momentum conservation (the ``missing mass" technique). Thus one can distinguish between four different event topologies depending on the specific combination of registered final hadrons ($X$ is the undetected part): 

\begin{enumerate}
\item $e p \rightarrow e' p' \pi^{+} X$\text{,}
\item $e p \rightarrow e' p' \pi^{-} X$\text{,}
\item $e p \rightarrow e' \pi^{+} \pi^{-} X$\text{, and}
\item $e p \rightarrow e' p \pi^{+} \pi^{-} X$.
\end{enumerate}

\begin{figure}[htp!]
\begin{center}
 \includegraphics[width=7cm,keepaspectratio]{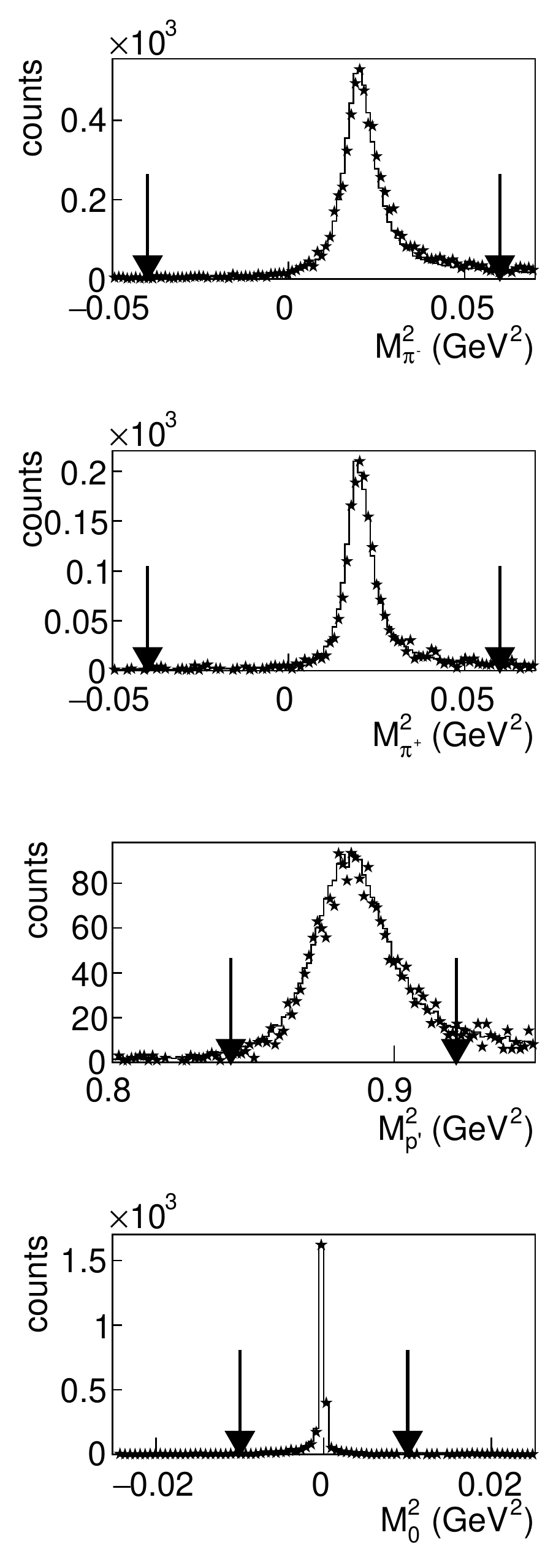} 
\vspace{-0.1cm}
\caption{
Missing mass squared ($M_{X}^{2}$) distributions for the four event topologies for $1.675$~GeV $< W < 1.7$~GeV and $0.45$~GeV$^{2} < Q^{2} < 0.5$~GeV$^{2}$ in comparison with the Monte Carlo. The stars show the experimental data, while the curves are from the simulation.
The plots show the topologies 1 to 4 from top to bottom. The arrows show the applied exclusivity cuts. 
Each distribution is normalized to the corresponding integral.}
\label{fig:miss_mass}
\end{center}
\end{figure}

Due to the experimental conditions, topology 1 with a $\pi^{-}$ missing contains about 50\% of the total statistics, while the remaining half of the total is relatively equally distributed among the other topologies that require a $\pi^{-}$ detection. This uneven distribution of the statistics between the topologies originates from the fact that CLAS does not cover the polar angle range $0\,^{\circ}\mathrm{} < \theta_{\text{lab}} < 8\,^{\circ}\mathrm{}$ ~\cite{Me03}. 
The presence of this forward acceptance hole does not affect much the registration of the positive particles ($p$ and $\pi^{+}$), since their trajectories are bent by the magnetic field away from the hole, whereas the negative particles ($e$ and $\pi^{-}$) are inbending so that their trajectories are bent in the forward direction. Electrons, having generally a high momentum, undergo small track curvature, and the presence of the forward hole leads for them only to
a constraint on the minimal achievable $Q^2$. However, for negative pions the situation is dramatic: being heavier and slower they are bent dominantly into the forward detector hole and, therefore, most of them cannot be detected. This leads to the fact that the $\pi^{-}$ missing topology contains the dominant part of the statistics.

The topologies were defined so that they did not overlap. For example, the topology $e p \rightarrow e' p' \pi^{+} X$ required the presence of $e'$, $p'$ and $\pi^{+}$ candidates and the absence of $\pi^{-}$ candidates, avoiding in this way double counting. In most of the CLAS papers on double-pion electroproduction~\cite{Isupov:2017lnd,Ripani:2002ss,Fedotov:2008aa}, only topologies 1 and 4 were used. However, in this study all four topologies were used in combination. 
This approach allowed not only an increase of the analyzed statistics (about 50\%), but also to populate events in a broader part of the reaction phase space, since the topologies had non-identical kinematic coverage.

For the case when one of the final hadrons was not detected, the missing mass $M_{X}$ for the reaction $e p \rightarrow e' h_1 h_2 X$ is determined by

\begin{equation}
M_{X}^{2} = (P_{e} + P_{p} -P_{e'} - P_{h_1} - P_{h_2})^{2},
\end{equation} 
where $P_{h_1}$ and $P_{h_2}$ are the four-momenta of the registered final state hadrons, $P_{e}$ and $P_{p}$ the four-momenta of the initial state electron and proton, and $P_{e'}$ the four-momentum of the scattered electron.

For topology 4, the missing mass $M_{X}$ for the reaction $e p \rightarrow e' p' \pi^+ \pi^- X$ is given by

\begin{equation}
M_{X}^{2} = (P_{e} + P_{p} -P_{e'} - P_{\pi^+} - P_{\pi^-} - P_{p'})^{2},
\end{equation} 
where $P_{e}$, $P_{p}$, $P_{e'}$, $P_{\pi^+}$, $P_{\pi^-}$,  and $P_{p'}$  are the four-momenta of the initial and final state particles.

The distributions of the missing mass squared ($M_{X}^{2}$) for various topologies are shown in Fig.~\ref{fig:miss_mass} for $1.675$~GeV $< W < 1.7$~GeV in comparison with the Monte Carlo. The stars show the experimental data, while the curves are from the simulation.
The plots in Fig.~\ref{fig:miss_mass} represent the topologies 1 to 4 from top to bottom. The arrows show the applied exclusivity cuts. 
Each distribution in Fig.~\ref{fig:miss_mass} is normalized to the corresponding integral.

Fig.~\ref{fig:miss_mass} demonstrates good agreement between the experimental and the Monte Carlo distributions,  since the simulation included both radiative effects and a background from other exclusive channels. 
The former was taken into account according to the inclusive approach~\cite{Mo:1968cg}. 
The main source of the exclusive background was found to be the reaction $e p \rightarrow e' p' \pi^{+} \pi^{-} \pi^{0}$. The events for that reaction were simulated along  with the double-pion events, 
considering the ratio of three-pion/double-pion cross sections
 taken from Ref.~\cite{Wu:2005wf}. The simulation of double-pion events was carried out based on the JM05 version of double-pion production model~\cite{Ripani:2000va,Aznauryan:2005tp,Mokeev:2005re}, while for three-pion events a phase space distribution was assumed. 
 
For the purpose of the cross section calculations, experimental events from all four topologies were summed up in each multi-dimensional bin. With respect to the simulation, the reconstructed Monte Carlo events were also subject to the same summation.

\section{Cross section calculation}

\subsection{Kinematic variables}
\label{sec_kin_var}

Once the selection of the double-pion events has been carried out, the four-momenta of the final state hadrons are known (either detected or calculated as missing) and defined in the lab frame that corresponds to the system where the target proton is at rest and the axis orientation is the following: $z_{\text{lab}}$ -- along the beam, $y_{\text{lab}}$ -- pointing upwards with respect to the Hall floor, and $x_{\text{lab}}$ -- along $[\vec y_{\text{lab}} \times \vec z_{\text{lab}}]$.

The cross sections were obtained in the single-photon exchange approximation in the center of mass frame of the {\em virtual photon -- initial proton} system (c.m.s.).   
The c.m.s. is uniquely defined as the system where the initial proton and the virtual photon exchanged in the scattering move towards each other with the axis $z_{\text{cms}}$ along the photon and the net momentum equal to zero. The axis $x_{\text{cms}}$ is situated in the electron scattering plane, while $y_{\text{cms}}$ is along $[\vec z_{\text{cms}} \times \vec x_{\text{cms}}]$.

To transform the lab system to the c.m.s., two rotations and one
boost should be performed~\cite{Fed_an_note:2017}.
The first rotation situates the axis $x$ in the electron scattering plane.
The second one aligns the axis $z$ with the virtual photon direction. 
Then the boost along $z$ is performed.

The kinematic variables that describe the final hadronic state are calculated from the four-momenta of the final hadrons in the c.m.s.~\cite{Isupov:2017lnd,Fedotov:2008aa}.
The three-body final state is 
unambiguously determined by five kinematic
variables. 
Beside that, the variables $W$ and $Q^{2}$ are needed to describe the initial state.
 
There are several ways to choose the five variables for the description of the final hadronic state. In this study the following generalized set of
variables is used~\cite{Fed_an_note:2017,Byckling:1971vca,Isupov:2017lnd,Fedotov:2008aa,Mokeev:2015lda}.

\begin{itemize}
\item invariant mass of the first pair of 
hadrons $M_{h_{1}h_{2}}$;
\item invariant mass of the second pair of
hadrons $M_{h_{2}h_{3}}$;
\item the first hadron solid angle $\Omega_{h_{1}} = (\theta_{h_{1}}, \varphi_{h_{1}})$;
\item the angle $\alpha_{h_{1}}$ between the two planes (i) defined by the three-momenta of
the virtual photon (or initial proton) and the first final state hadron and (ii) defined by the three-momenta of all final state hadrons (see Appendix~\ref{app_a}).
\end{itemize}

The cross sections  were obtained in three sets of variables depending on
various assignments for the first, second, and
third final hadrons:

\begin{enumerate}
\item $\boldsymbol{first\; - p',\; second \; - \pi^{+},\; third - \pi^{-}}$: \\ $M_{p'\pi^{+}}$, $M_{\pi^{+}\pi^{-}}$, $\theta_{p'}$, $\varphi_{p'}$,  $\alpha_{p'}$ (or $\alpha_{(pp')(\pi^{+}\pi^{-})}$),
\item $\boldsymbol{first\; - \pi^{-},\; second \; - \pi^{+},\; third - p'}$: $M_{\pi^{-}\pi^{+}}$, $M_{\pi^{+}p'}$, $\theta_{\pi^{-}}$, $\varphi_{\pi^{-}}$, $\alpha_{\pi^{-}}$ (or $\alpha_{(p\pi^{-})(p'\pi^{+})}$ ), and
\item $\boldsymbol{first\; - \pi^{+},\; second \; - \pi^{-},\; third - p'}$: $M_{\pi^{+}\pi^{-}}$, $M_{\pi^{-}p'}$, $\theta_{\pi^{+}}$, $\varphi_{\pi^{+}}$, $\alpha_{\pi^{+}}$ (or $\alpha_{(p\pi^{+})(p'\pi^{-})}$ ).
\end{enumerate}

\subsection{Binning and kinematic coverage}

\begin{figure}[htp]
\begin{center}
 \includegraphics[width=8.5cm,keepaspectratio,trim=0mm 4mm 0mm 0mm,clip]{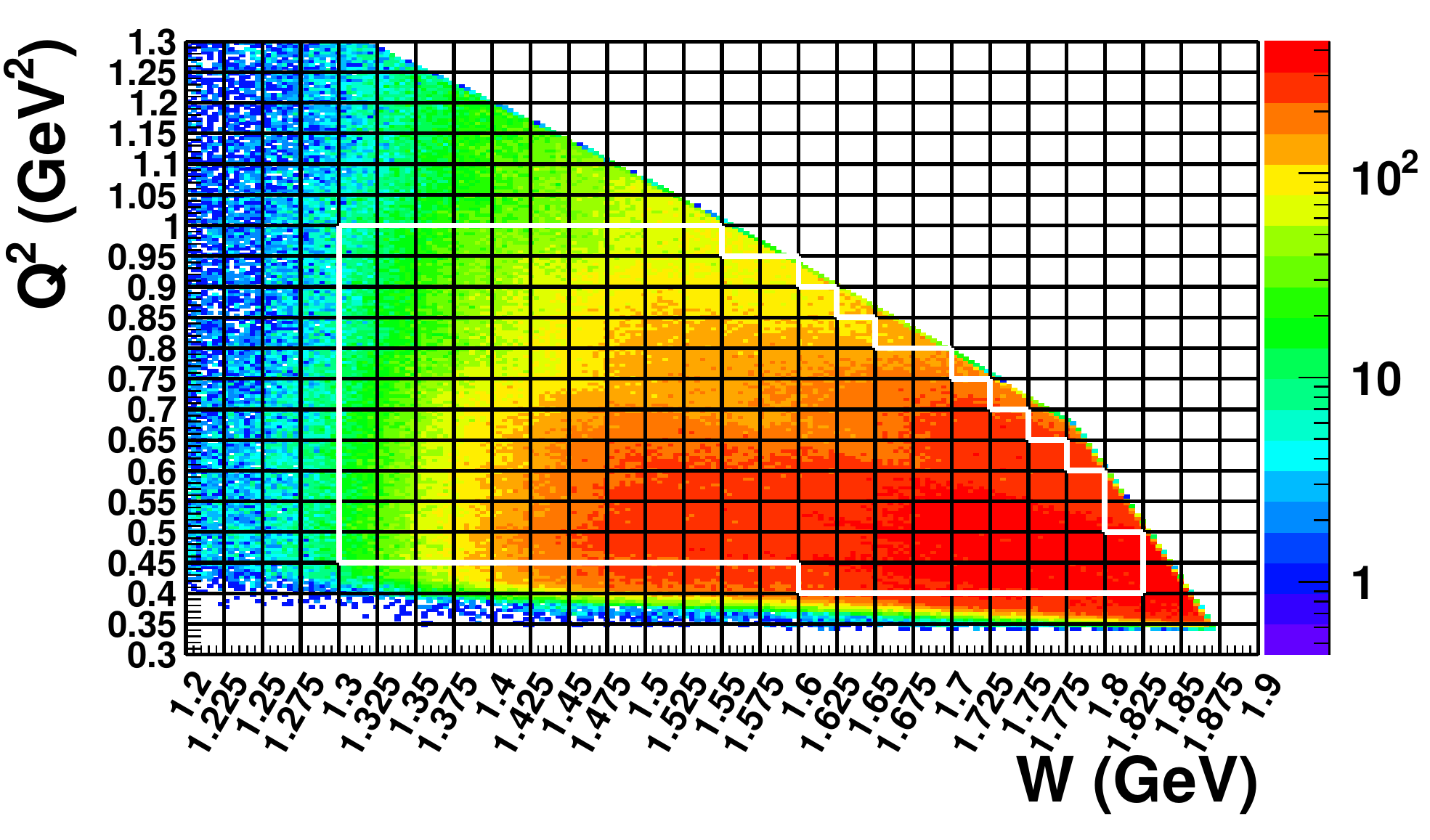} 
\caption{(colors online) $Q^2$ versus $W$ distribution populated with selected double-pion events. The cross section was calculated in 2D cells within the white boundaries.}
\label{fig:q2vsw}
\end{center}
\end{figure}

The kinematic coverage in the initial state variables is shown by the $Q^{2}$ versus $W$ distribution in Fig. \ref{fig:q2vsw}. The  distribution represents the number of exclusive double-pion events left after the cuts and corrections described above. 
The white boundary limits the analyzed kinematic area, where the double-pion cross sections were extracted, and encompasses about 1.2 million events. The black grid demonstrates the chosen binning in the initial state variables.

The binning in the hadronic variables is listed in Table~\ref{tab:summary_bins}. It was chosen to maintain reasonable statistical uncertainties of the single-differential cross sections for all $W$ and $Q^{2}$ bins. The binning choice also takes into account the cross section drop near the double-pion production threshold at $\approx 1.22$~GeV, as well as the broadening of the reaction phase space with increasing $W$.

\begin{table*}[htp]
\centering 
\caption{\small Number of bins for each hadronic variable \label{tab:summary_bins}}
\normalsize
  \begin{tabular}{lm{4cm}P{2cm}P{2cm}P{2cm}P{2cm}}
    \toprule
    & & \multicolumn{4}{c}{W range (GeV)} \\
    \multicolumn{2}{c}{\centering Hadronic variable }  & $1.3 - 1.35$ & $1.35 - 1.4$ & $1.4-1.45$ & $>1.45$ \\
    \cmidrule(l{5pt}r{15pt}){1-2} \cmidrule(l{5pt}r{5pt}){3-6}
    M        & Invariant mass       &   8  & 10 & 12 & 12  \\
    $\theta$ & Polar angle          &   6  & 8  & 10 & 10  \\
    $\varphi$   & Azimuthal angle      &   5  & 5  & 5  & 8   \\
    $\alpha$ & Angle between planes &   5  & 6  & 8  & 8   \\
    \bottomrule
  \end{tabular}
\end{table*}

Special attention is required for the binning in the invariant masses. 
The upper and lower boundaries of the invariant mass distributions depend on the hadron masses and $W$ as: 

\begin{equation}
\begin{aligned}
M_{\text{lower}} & = m_{h_1} + m_{h_2}~\text{and} \\
M_{\text{upper}}(W) & = W - m_{h_3}, \label{eq:inv_mass_boundary}
\end{aligned}  
\end{equation}
where  $m_{h_1}$, $m_{h_2}$, and $m_{h_3}$ are the masses of the final hadrons.

Since the cross section is calculated in a bin 
$W_{\text{left}} < W < W_{\text{right}}$, the boundary of $M_{\text{upper}}$ is not distinct. 
For the purpose of binning in mass, the value of $M_{\text{upper}}$ was 
calculated using $W_{\text{center}}$, at the center of the $W$ bin.
As a result, some events with $W > W_{\text{center}}$ turned out to be located beyond $M_{\text{upper}}$. 
Hence it was decided to use a specific arrangement of mass bins with
the bin width $\Delta M$ determined as:

\begin{equation}
\begin{aligned}
\Delta M = \frac{M_{\text{upper}}(W_{\text{center}})-M_{\text{lower}}}{N_{\text{bins}}-1}, \label{eq:bin_width}
\end{aligned}  
\end{equation} 
where $N_{\text{bins}}$ is the number of the bins specified in the first row of Table~\ref{tab:summary_bins}.

\begin{figure}[htp]
\begin{center}
\includegraphics[width=8.6cm]{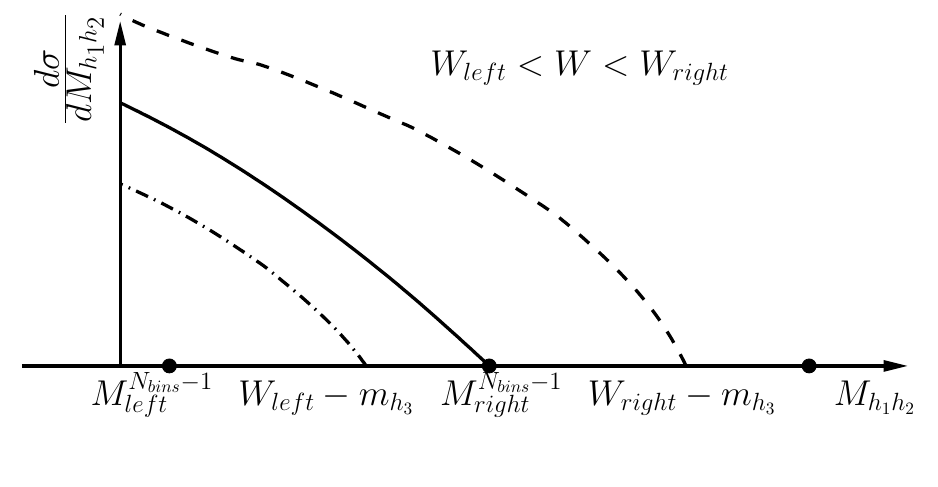}\vspace{-0.5cm}
\caption{\small Schematic representation of the invariant mass distributions ending in 
 $M_{\text{upper}}$ calculated according to Eq.~\eqref{eq:inv_mass_boundary} for three 
choices of $W$ at $W_{\text{left}}$ (dot-dashed), $W_{\text{center}}$ (solid) and 
$W_{\text{right}}$ (dashed).
The black points at $M_{\text{left}}^{N_{\text{bins}}-1}$ and $M_{\text{right}}^{N_{\text{bins}}-1}$ show the left and right boundaries of the next to last bin, respectively.
\label{fig:mass_corr}}
\end{center}
\end{figure}

The chosen arrangement of bins forces the last bin to be situated completely out of the boundaries given by Eq.~\eqref{eq:inv_mass_boundary} using $W_{\text{center}}$. The cross section 
for this extra bin was very small, but it was kept so that no 
events were lost. When integrating the cross section over the 
mass distribution, these events in the extra bin were included, 
but a cross section for this bin is not reported.

The cross section in the next to last bin (labeled as bin number 
$N_{\text{bins}}-1$) should be treated carefully. This is best illustrated in Fig.~\ref{fig:mass_corr}, which shows schematically the 
distribution of events in mass, ending in $M_{\text{upper}}$ for three 
choices of $W$ at $W_{\text{left}}$ (dot-dashed), $W_{\text{center}}$ (solid) and 
$W_{\text{right}}$ (dashed).
The black points at $M_{\text{left}}^{N_{\text{bins}}-1}$ and $M_{\text{right}}^{N_{\text{bins}}-1}$ show the left and right boundaries of the next to last bin, respectively. In the next to last bin events with $W < W_{\text{center}}$ are 
distributed over a range, which is less than $\Delta M$ defined by Eq.~\eqref{eq:bin_width}.
However, when extracting the cross sections, the event yield was divided by the full bin width $\Delta M$, thus leading to an underestimation of the cross section.

The correction for this effect was made using the TWOPEG double-pion event generator~\cite{Skorodum:EG}, because the statistics of the experimental data were not sufficient for this purpose. 
The correction factor to the cross section in the next to last bin is the ratio of the simulated cross sections calculated with fixed $\Delta M$ defined by Eq.~\eqref{eq:bin_width} and with $\widetilde{\Delta M} = W - m_{h_{3}} - M_{\text{left}}^{N_{\text{bins}}-1}$, which was different for each generated event. This factor provided the correction to the cross section in the next to last bin that varied from 5\% to 10\%.

In addition to the above procedure, one more binning issue should be considered.
The cross section extracted within the bin in any kinematic variable was assigned to its central point. In the areas with non-linear cross section behavior, the finite bin size caused the distortion of the cross section value due to its averaging within the bin. To cure this effect, a binning correction was applied that included a cubical spline approximation for the cross section shape~\cite{Fed_an_note:2017}. The typical value of the correction was $\sim$~1\% rising up to 4\% for some data-points at low $W$.

\subsection{Cross section formula}
\label{cr_sect_formula}

In the single-photon exchange approximation, the virtual photoproduction cross section $\sigma_{\text{v}}$ (which is the focus of this paper) is connected with the experimental electron scattering cross section $\sigma_{e}$ via: 

\begin{equation}
\begin{aligned}
\frac{\textrm{d}^{5}\sigma_{\text{v}}}{\textrm{d}^{5}\tau} & = \frac{1}{\Gamma_{\text{v}}}
\frac{\textrm{d}^{7}\sigma_{e}}{\textrm{d}W\textrm{d}Q^{2}\textrm{d}^{5}\tau} \textrm{ ,} \\
\textrm{d}^{5}\tau & = \textrm{d}M_{h_{1}h_{2}}\textrm{d}M_{h_{2}h_{3}}\textrm{d}\Omega_{h_{1}}
\textrm{d}\alpha_{h_{1}} \textrm{.}
\label{fulldiff}
\end {aligned}
\end{equation}

Here $\textrm{d}^{5}\tau$ is the differential of the five independent variables of the final $\pi^{+}\pi^{-}p$ state that were described in Sec.~\ref{sec_kin_var}, $\Gamma_{\text{v}}$ is the 
virtual photon flux given by

\begin{equation}
\Gamma_{\text{v}}(W,Q^{2}) =
\frac{\alpha}{4\pi}\frac{1}{E_{\text{beam}}^{2}m_{p}^{2}}\frac{W(W^{2}-m_{p}^{2})}
{(1-\varepsilon_{\text{T}})Q^{2}} \textrm{ ,}
\label{flux}
\end{equation}
where $\alpha$ is the fine structure constant $\left(1/137\right)$, $m_{p}$ is the proton
mass, $E_{\text{beam}} = 2.039$~GeV is the energy of the incoming electron beam, and $\varepsilon_{\text{T}}$ is the virtual photon transverse polarization, given by

\begin{equation}
\varepsilon_{\text{T}} = \left( 1 + 2\left( 1 +
\frac{\nu^{2}}{Q^{2}} \right)
\tan^{2}\left(\frac{\theta_{e'}}{2}\right) \right)^{-1} \textrm{.}
\label{polarization}
\end{equation}

Here $\nu = E_\text{{beam}} - E_{e'}$ is the virtual photon energy, while $E_{e'}$ and
$\theta_{e'}$ are the energy and the polar angle of the scattered electron in the
lab frame, respectively.

The experimental electron scattering cross section $\sigma_{e}$ introduced in Eq.~\eqref{fulldiff} was calculated as

\begin{equation}
\frac{\textrm{d}^{7}\sigma_{e}}{\textrm{d}W\textrm{d}Q^{2}\textrm{d}^{5}\tau} = \frac{1}{\mathcal{E} \cdot R} 
\frac{\left( \frac{N_{\text{full}}}{Q_{\text{full}}}-\frac{N_{\text{empty}}}{Q_{\text{empty}}} \right)}{
\Delta W \Delta Q^{2} \Delta^{5} \tau \left( \frac{l \rho N_{A}}{q_{e}M_{H}} \right)} \textrm{ ,}
\label{expcrossect}
\end{equation}
where $N_{\text{full}}$ and $N_{\text{empty}}$ are the numbers of selected double-pion events inside the
seven-dimensional bin for runs with hydrogen and
empty target, respectively. 
Each event was weighted with the corresponding photoelectron correction factor given by Eq.~\eqref{eq:cc_corr_fact}.
Also  $Q_{\text{full}}= 5999.64$~$\mu$C and $Q_{\text{empty}} = 334.603$~$\mu$C are the  values of the charge accumulated on the Faraday Cup for runs with hydrogen and empty target, respectively, and $q_{e} =1.6 \cdot 10^{-19}$ C is the elementary charge, $\rho = 0.0708$  g/cm$^{3}$ is the density of liquid hydrogen at a temperature of 20~K,
$l = 2$ cm is the length of the target, $M_{H} = 1.00794$ g/mol is the molar density of
the natural mixture of hydrogen, and  $N_{A} =6.02 \cdot 10^{23}$ mol$^{-1}$ is Avogadro's
number.

In Eq.~\eqref{expcrossect}  $\mathcal{E} = \mathcal{E}(\Delta W, \Delta Q^{2}, \Delta^{5} \tau)$ is the detector efficiency for the seven-dimensional bin coming from the  Monte Carlo simulation and $R = R(\Delta W, \Delta Q^{2})$ is the
radiative correction factor described in Sec.~\ref{rad_corr}.

The electron scattering cross section in the left hand side of Eq.~\eqref{expcrossect} was assumed to be obtained in the center of the finite seven-dimensional kinematic bin $\Delta W \Delta Q^{2} \Delta^{5} \tau$.

The limited
statistics of the experiment did not allow estimates
of the five-fold differential cross section $\sigma_{\text{v}}$ with a reasonable
accuracy. Therefore, being obtained on the multi-dimensional grid, the cross section $\sigma_{\text{v}}$ was then integrated over at least four hadron variables. Hence only the sets of the single-differential and fully-integrated cross sections are presented as a result here.

For each bin in $W$ and $Q^2$, the following cross sections were obtained:

\begin{equation}
\begin{aligned}
\frac{\textrm{d}\sigma_{\text{v}}}{\textrm{d}M_{h_{1}h_{2}}} & =
\int\frac{\textrm{d}^{5}\sigma_{\text{v}}}{\textrm{d}^{5}\tau}\textrm{d}M_{h_{2}h_{3}}\textrm{d}\Omega_{h_{1}}\textrm{d}\alpha_{h_{1}}\text{,}  \\
\frac{\textrm{d}\sigma_{\text{v}}}{\textrm{d}M_{h_{2}h_{3}}} & =
\int\frac{\textrm{d}^{5}\sigma_{\text{v}}}{\textrm{d}^{5}\tau}\textrm{d}M_{h_{1}h_{2}}\textrm{d}\Omega_{h_{1}}\textrm{d}\alpha_{h_{1}}\text{,} \\
\frac{\textrm{d}\sigma_{\text{v}}}{\textrm{d}(-cos\theta_{h_{1}})} & =
\int\frac{\textrm{d}^{5}\sigma_{\text{v}}}{\textrm{d}^{5}\tau}\textrm{d}M_{h_{1}h_{2}}\textrm{d}M_{h_{2}h_{3}}\textrm{d}\varphi_{h_{1}}\textrm{d}\alpha_{h_{1}}\text{,}  \\
\frac{\textrm{d}\sigma_{\text{v}}}{\textrm{d}\alpha_{h_{1}}} & =
\int\frac{\textrm{d}^{5}\sigma_{\text{v}}}{\textrm{d}^{5}\tau} \textrm{d}M_{h_{1}h_{2}}\textrm{d}M_{h_{2}h_{3}}\textrm{d}\Omega_{h_{1}},~\text{and} \\
\sigma^{int}_{\text{v}}(W,Q^{2}) & =
\int\frac{\textrm{d}^{5}\sigma_{\text{v}}}{\textrm{d}^{5}\tau}\textrm{d}M_{h_{1}h_{2}}\textrm{d}M_{h_{2}h_{3}}\textrm{d}\Omega_{h_{1}}\textrm{d}\alpha_{h_{1}}.
\end{aligned}
\label{inegr5diff}
\end{equation}

Since the cross sections were obtained on the
five-dimensional kinematic grid,
integrals in Eq.~\eqref{inegr5diff} were calculated numerically 
on that grid.

\subsection{Efficiency evaluation}
\label{eff_err}

For the Monte Carlo simulation the GENEV event generator~\cite{Genev:Golova} developed by 
Genova group was used. This event generator uses the JM05 model~\cite{Mokeev:2005re} for the investigated double-pion channel,
while for the background channel $e p \rightarrow e'p'\pi^{+}\pi^{-}\pi^{0}$, which was generated along with the main one, GENEV assumes a phase space distribution for all kinematic variables. The simulation accounts for radiative effects according to the approach described in Ref.~\cite{Mo:1968cg}.

\begin{figure}[htp]
\begin{center}
\includegraphics[width=8.5cm,clip,trim={0 1mm 0 0}]{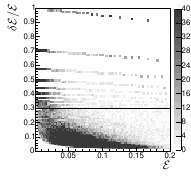}
\caption{\small The number of five-dimensional cells plotted as a function of the relative efficiency uncertainty versus efficiency. The example is given for one particular bin in $W$ and $Q^2$ ($1.625$~GeV $< W < 1.65$~GeV and $0.5$~GeV$^{2} < Q^{2} < 0.55$~GeV$^{2}$).} \label{fig:eff_err}
\end{center}
\end{figure}

The generated events were passed through the GEANT based detector simulation and reconstruction procedures. The efficiency factor $\mathcal{E}$ from Eq.~\eqref{expcrossect} was then calculated in each $\Delta W\Delta Q^{2}\Delta^{5} \tau$ bin as:

\begin{equation}
\mathcal{E}(\Delta W, \Delta Q^{2}, \Delta^{5} \tau) = \frac{N_{\text{rec}}}{N_{\text{gen}}},
\label{efficiency}
\end{equation}
where $N_{\text{gen}}$ is the number of generated double-pion events (without any cuts) inside the multi-dimensional bin, while $N_{\text{rec}}$ is the number of reconstructed either double- or three-pion events that survived in the bin after event selection. This definition of the efficiency factor $\mathcal{E}$ accounted for the three-pion background that was negligible at $W < 1.6$ GeV  and
increased up to a few percent at $W \approx 1.8$ GeV. The averaged (over all analyzed multi-dimensional cells) value of the efficiency was found to be about 11\%.

\begin{figure*}[htp]
\begin{center}
\includegraphics[width=13cm,keepaspectratio]{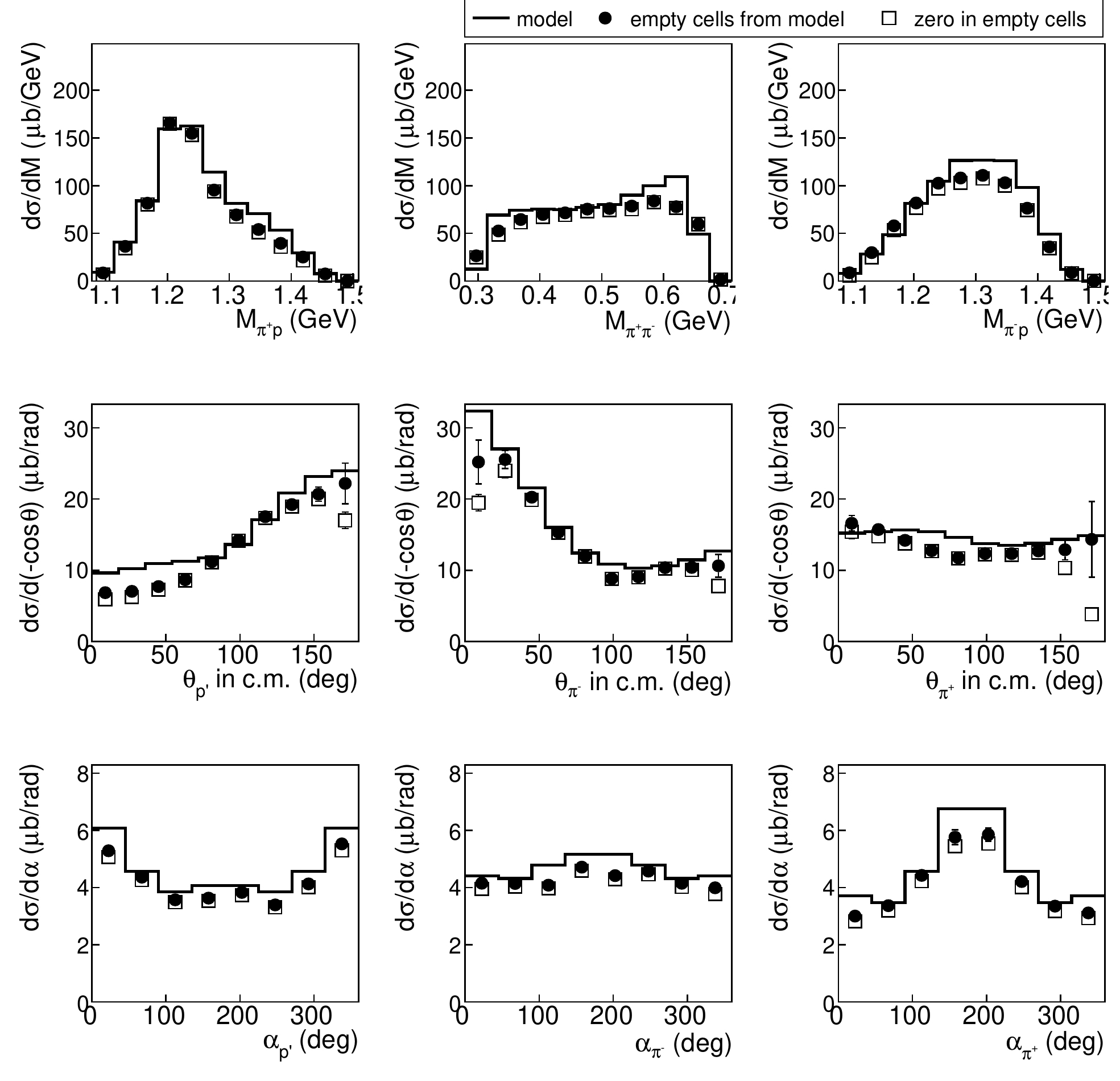}
\caption{\small The extracted single-differential cross sections  for the cases when the contribution from the empty cells was ignored (empty squares) and when it was taken into account (black circles). The former are reported with the uncertainty $\delta_{\text{stat}}^{\text{tot}}$ given by Eq.~\eqref{errortot} (it is smaller than the symbol size), while the latter are with the uncertainty $\delta_{\text{stat,mod}}^{\text{tot}}$ given by Eq.~\eqref{eq:error_stat_mod}.
The curves show the TWOPEG cross sections that were used as a model assumption for the empty cell contribution. All distributions are given for one particular bin in $W$ and $Q^2$ ($W = 1.6125$~GeV, $Q^2 = 0.475$~GeV$^2$).} \label{fig:topologies}
\end{center}
\end{figure*}

Due to the blind areas in the geometrical coverage of the CLAS detector, some kinematic bins of the double-pion production phase space turned out to have zero acceptance. 
In such bins, which are usually called empty cells,  the cross section cannot be experimentally defined. The empty cells contribute to the integrals in Eq.~\eqref{inegr5diff} along with the other kinematic bins.  Ignoring the contribution from the empty cells leads to a systematic cross section underestimation and, therefore, some model assumptions for the cross section in these cells are needed. This situation causes a slight model dependence of the final result. 

A special procedure was developed in order to take into account the contributions from the empty cells to the integrals in Eq.~\eqref{inegr5diff}.
The map of the empty cells was determined using the Monte Carlo simulation. A cell was treated as empty, if it contained generated events ($N_{\text{gen}} > 0$), but did not contain any reconstructed events ($N_{\text{rec}} = 0$). 

Additionally, the efficiency in some kinematic bins could not be reliably determined due to boundary effects, bin to bin event migration, and limited Monte Carlo statistics.  
Such cells were excluded from consideration and also treated as empty cells.
They can be differentiated from the cells with reliable efficiency by a larger relative efficiency uncertainty $\frac{\delta \mathcal{E}}{\mathcal{E}}$ (absolute efficiency uncertainty $\delta \mathcal{E}$ is defined in Sect.~\ref{stat_unc}).
In order to determine the criterion for the cell exclusion, the distribution shown in Fig.~\ref{fig:eff_err} was produced for each bin in $W$ and $Q^{2}$.
This figure gives the uncertainty $\frac{\delta \mathcal{E}}{\mathcal{E}}$   versus efficiency $\mathcal{E}$, showing the number of multi-dimensional cells. As is seen in  Fig.~\ref{fig:eff_err}, cells with relative efficiency uncertainty greater than 30\% are clustered along the horizontal stripes. This clustering originates from the fact that efficiency was obtained by the division of two integer numbers and reveals the bins with small statistics of the reconstructed events. Moreover, these horizontal stripes contain many cells with unreliable extremely small efficiency. Therefore,  the multi-dimensional bins that are located above the horizontal  line in Fig.~\ref{fig:eff_err} were excluded from consideration and treated as empty cells.

Once the map of the empty cells was determined, the cross section produced by the TWOPEG event  generator~\cite{Skorodum:EG} was used as a model assumption for these kinematic bins. This event generator employs the double-pion cross sections from the recent version of the JM15 model fit to the data~\cite{Ripani:2002ss,Mokeev:2012vsa,Fedotov:2008aa,Golovach:note}, as well as the data~\cite{Wu:2005wf,ABBHHM:1968aa} itself and up to now provides the best cross section estimation in the investigated kinematic region.
Ref.~\cite{Skorodum:EG} describes in detail the approach used in TWOPEG in order to estimate the cross sections.

Fig.~\ref{fig:topologies} introduces the single-differential cross sections given by Eq.~\eqref{inegr5diff} extracted for three sets of the kinematic variables described in Sect.~\ref{sec_kin_var}.  The empty squares correspond to the case when the contribution from the empty cells was ignored, and the  black circles are for the case when that was taken into account in the way described above. The black curves represent the TWOPEG cross sections that were used as a model assumption. The figure demonstrates a reasonably small contribution from the empty cells (and therefore a small model dependence of the results) that was achieved using all four available reaction topologies in combination. Only the edge points in the $\theta$ distributions reveal pronounced empty cell contributions  due to the negligible/zero CLAS acceptance in the corresponding directions. 
To account for the model dependence, the part of the single-differential cross section that came from the empty cells was assigned a 50\% relative uncertainty. The corresponding absolute uncertainty $\delta_{\text{model}}$ was combined with the total statistical uncertainty, as was done in Refs.~\cite{Isupov:2017lnd,Golovach:note}.

\subsection{Radiative correction}
\label{rad_corr}

The radiative correction to the extracted cross sections was performed using the TWOPEG event generator for the double-pion electroproduction~\cite{Skorodum:EG}, which accounts for the radiative effects by means of the well-known approach of Ref.~\cite{Mo:1968cg}. 
This approach has successfully proven itself as an efficient tool to calculate inclusive radiative cross section from the non-radiative one.
In Ref.~\cite{Mo:1968cg} the approach is applied to the inclusive case, while in TWOPEG, the double-pion integrated cross sections are used instead. The radiative photons are supposed to be emitted collinearly either to the direction of the initial or scattered electron (the so-called ``peaking approximation").

In Refs.~\cite{Mo:1968cg,Skorodum:EG} the calculation of the radiative cross section is split into two parts. The ``soft" part assumes the energy of the  emitted radiative photon to be less than a certain minimal value (10 MeV), while the ``hard" part is for the photons with an energy greater than that value. 
The ``soft" part is evaluated
explicitly, while for the calculation of the ``hard" part, an inclusive hadronic
tensor is assumed. 
The latter assumption is however considered adequate, since approaches
that are capable of describing radiative processes
in exclusive double-pion electroproduction are not yet available.

\begin{figure}[htp]
\begin{center}
\includegraphics[width=6cm]{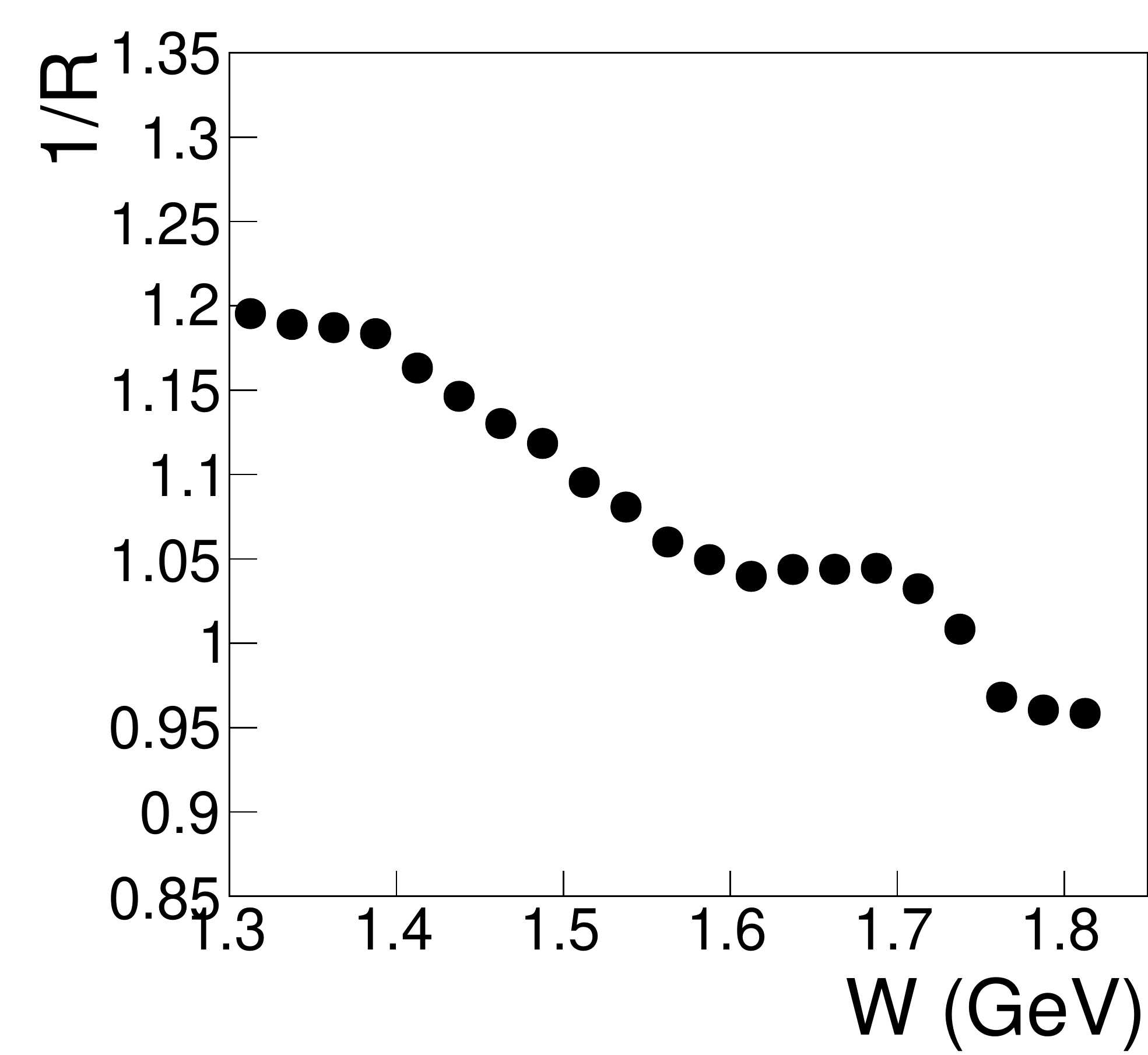}
\caption{\small The quantity $1/R$ (see Eq.~\eqref{eq_radcorrfact})
as a function of $W$ averaged over all considered $Q^{2}$ bins.} \label{radcorrfact}
\end{center}
\end{figure}

The radiative
correction factor $R$ in Eq.~(\ref{expcrossect})
was determined in the following way.
The double-pion events either with or without radiative effects were generated with TWOPEG, then the ratio given by Eq.~\eqref{eq_radcorrfact} was taken in each $\Delta W \Delta Q^{2}$ bin. 

\begin{equation}
\label{eq_radcorrfact}
R(\Delta W,\Delta Q^{2}) = \frac{N_{\text{rad}}^{\text{2D}}}{N_{\text{norad}}^{\text{2D}}} \textrm{ ,}
\end{equation}
where $N_{\text{rad}}^{\text{2D}}$ and $N_{\text{norad}}^{\text{2D}}$ are the
numbers of generated events in each $\Delta W \Delta Q^{2}$ bin
with and without radiative effects, respectively. Neither $N_{\text{rad}}^{\text{2D}}$ nor $N_{\text{norad}}^{\text{2D}}$ are subject to any cuts.

This approach gives the correction factor $R$ only as a function of $W$ and $Q^{2}$, disregarding its dependence on the hadronic variables.
However, 
the need to integrate the cross section at least over four hadronic variables (see Eq.~\eqref{inegr5diff}) considerably reduces the influence of
the final state hadron kinematics on the
radiative correction factor, thus justifying 
the applicability of the procedure~\cite{Mo:1968cg,Skorodum:EG}.

 The quantity $1/R$, which is averaged over all considered $Q^{2}$ bins, is plotted in  Fig.~\ref{radcorrfact} as a function of $W$. The dependence of the correction factor on $Q^{2}$ was found to be negligible.  The uncertainties associated with the statistics of the generated events are very small and therefore  not seen in Fig.~\ref{radcorrfact}.

\subsection{Statistical uncertainties}
\label{stat_unc}

The limited statistics of both the experimental data and the Monte Carlo simulation  are two sources of statistical fluctuations of the extracted cross sections.
The cut on the efficiency uncertainty described in Sec.~\ref{eff_err} was chosen in a way that the latter source gives a minor contribution to the total statistical uncertainty.

The absolute statistical  uncertainty to the five-fold differential
virtual photoproduction cross section caused by the statistics of the experimental data was calculated as

 \begin{equation}
\delta_{\text{stat}}^{\text{exp}}(\Delta^{5} \tau) = \frac{1}{\mathcal{E}} \frac{1}{R}
\frac{1}{ \Gamma_{\text{v}} }
\frac{\sqrt{\left( \frac{N_{\text{full}}}{Q_{\text{full}}^{2}}+\frac{N_{\text{empty}}}{Q_{\text{empty}}^{2}} \right) } }{
\Delta W \Delta Q^{2} \Delta^{5} \tau \left( \frac{l \rho N_{A}}{q_{e}M_{H}} \right)}.
\label{staterrors}
\end{equation}

The absolute uncertainty to the cross section due to the limited Monte Carlo statistics was  estimated as

\begin{equation}
\delta_{\text{stat}}^{\text{MC}}(\Delta^{5} \tau) = \frac{\textrm{d}^{5}\sigma_{\text{v}}}{\textrm{d}^{5}\tau} \left( \frac{\delta \mathcal{E}}{\mathcal{E}} \right),
\label{montecarloerror}
\end{equation}
where $\mathcal{E}$ is the efficiency inside the multi-dimensional bin defined by Eq.~\eqref{efficiency}, while $\delta \mathcal{E}$ is its absolute statistical uncertainty.

Due to the fact that $N_{\text{gen}}$ and $N_{\text{rec}}$ in Eq.~\eqref{efficiency} are not independent, the usual method of partial derivatives is not applicable in order to calculate $\delta \mathcal{E}$.
Therefore the special approach described in Ref.~\cite{Laforge:1996ts} was used for this purpose.
Neglecting the event migration between the bins, this approach gives the following expression for  
 the absolute statistical uncertainty of the efficiency,
 
\begin{equation}
\delta \mathcal{E} =
\sqrt{\frac{(N_{\text{gen}}-N_{\text{rec}})N_{\text{rec}}}{N_{\text{gen}}^{3}}}.
\label{efferror}
\end{equation}

The two parts of the statistical uncertainty given by Eqs.~\eqref{staterrors} and \eqref{montecarloerror} were combined quadratically into the total absolute statistical uncertainty to the cross section in the multi-dimensional bin:

\begin{equation}
\delta_{\text{stat}}^{\text{tot}}(\Delta^{5} \tau) =
\sqrt{\left (\delta_{\text{stat}}^{\text{exp}} \right )^{2} + \left (\delta_{\text{stat}}^{\text{MC}}\right )^{2}}.
\label{errortot}
\end{equation}

The uncertainties $\delta_{\text{stat}}^{\text{tot}}$ for the extracted  single-differential cross sections were obtained from the uncertainties  $\delta_{\text{stat}}^{\text{tot}}(\Delta^{5} \tau)$ of the five-fold differential cross sections according to the standard error propagation rules.

Finally for the single-differential cross sections, the total statistical uncertainty $\delta_{\text{stat}}^{\text{tot}}$ was  combined with the uncertainty $\delta_{\text{model}}$, which accounted for the model dependence of the results that came from the empty cell contribution (see Sect.~\ref{eff_err}):

\begin{equation}
\delta_{\text{stat,mod}}^{\text{tot}} =
\sqrt{\left (\delta_{\text{stat}}^{\text{tot}} \right )^{2} + \left (\delta_{\text{model}}\right )^{2}}.
\label{eq:error_stat_mod}
\end{equation}

\begin{figure*}[htp!]
\begin{center}
\includegraphics[width=17cm,,clip,trim={0 0 0 0}]{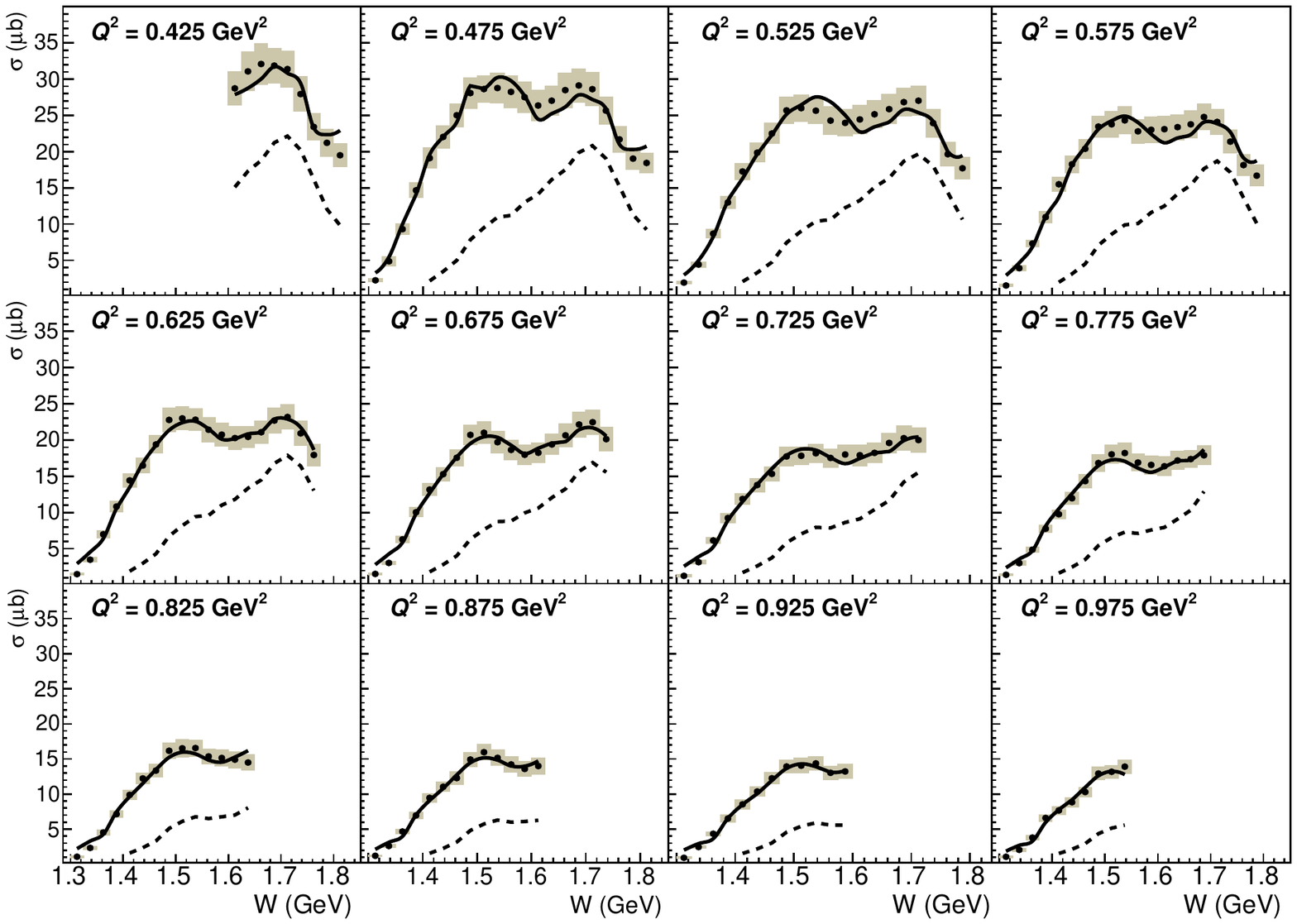}
\caption{\small  The $W$-dependencies of the integrated cross sections (symbols) in various bins in $Q^{2}$. The gray shadowed area  for each point is the total cross section uncertainty, which is the uncertainty $\delta^{\text{tot}}_{\text{stat,mod}}$ given by Eq.~\eqref{eq:error_stat_mod} summed up in quadrature with the total systematic uncertainty. The error bars that correspond to the uncertainty $\delta^{\text{tot}}_{\text{stat,mod}}$ only, are smaller than the symbol size.  The solid curves are the cross section prediction obtained from TWOPEG~\cite{Skorodum:EG}, while the dashed curves correspond to the resonant contribution estimated within the unitarized Breit-Wigner ansatz of the JM model~\cite{Mokeev:2012vsa,Mokeev:2015lda} (see text for more details).}
\label{fig:sys_err}

\end{center}
\end{figure*}

\begin{figure*}[htp!]
\begin{center}
\includegraphics[width=15cm]{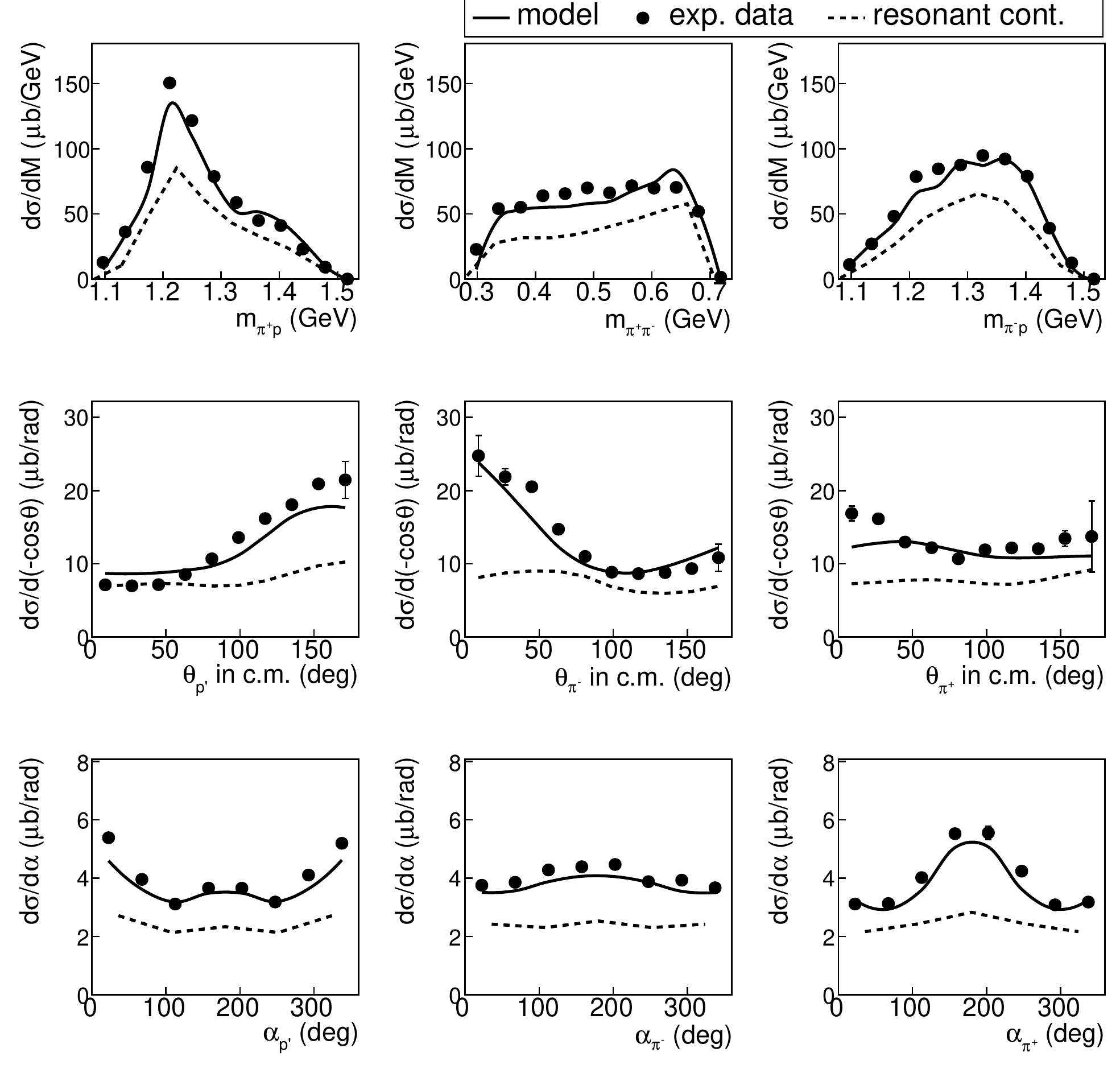}
\caption{\small 
The extracted single-differential cross sections (symbols) for one particular bin in $W$ and $Q^2$ ($W = 1.6375$~GeV, $Q^2 = 0.525$~GeV$^2$). The error bars correspond to the uncertainty $\delta_{\text{stat,mod}}^{\text{tot}}$ given by Eq.~\eqref{eq:error_stat_mod}. The solid curves are for the cross section prediction obtained from TWOPEG~\cite{Skorodum:EG}, while the dashed curves correspond to the resonant contribution estimated within the unitarized Breit-Wigner ansatz of the JM model~\cite{Mokeev:2012vsa,Mokeev:2015lda}(see text for more details).}
\label{fig:res_cont}

\end{center}
\end{figure*}

\subsection{Systematic uncertainties}

The systematic uncertainties of the obtained results dominate the statistical ones and originate from several sources.

The presence of the elastic events
in the dataset helped with the normalization verification of the extracted cross sections. For this purpose the elastic cross section was extracted and compared with the parameterization~\cite{Bosted:1994tm}, and a 3\% fluctuation was found. Therefore this value was included into the systematic uncertainty  of the extracted double-pion cross sections as a global factor. This factor takes into account  inaccuracies in the luminosity calculation (due to miscalibrations of the Faraday Cup, target
density instabilities, etc.) as well as errors in the electron registration and identification.

In order to study the systematic uncertainties, the double-pion cross sections were obtained using an alternative method of the topology combination. In contrast with the main method, where events from all four topologies were summed up in each multi-dimensional bin, the alternative one considers only those events that come from the topology with the maximal efficiency in the bin. 
 The difference between the cross sections obtained in these two ways was interpreted as a systematic uncertainty.
Since various topologies correspond to different detected final hadrons,  this uncertainty includes the errors due to the hadron identification. 
This uncertainty was calculated for each bin in  $W$ and $Q^{2}$ and found to be of the order of 2\%.

According to Sect.~\ref{sec_kin_var}, the double-pion cross sections were extracted in three sets of the  kinematic variables. The difference between the cross sections obtained by integration over these three kinematic grids was interpreted as a systematic uncertainty. This  uncertainty was computed for each bin in $W$ and $Q^{2}$ and was typically of the order of 5\%.
For the final results, the integrated cross sections averaged over these three grids are reported.

As a common practice with CLAS~\cite{Fedotov:2008aa,Isupov:2017lnd},  an extra 5\% global uncertainty was assigned to the cross section due to the inclusive radiative correction procedure (see Sect.~\ref{rad_corr}).

The uncertainties due to the sources mentioned above were summed up in quadrature to obtain the total systematic uncertainty for the integrated double-pion cross sections. The relative systematic uncertainty in each $W$ and $Q^{2}$ bin can be propagated as a global factor to the corresponding single-differential cross sections, which are reported with the uncertainty $\delta_{\text{stat,mod}}^{\text{tot}}$ only (see Eq.~\eqref{eq:error_stat_mod}).

\section{Comparison with the model and previously available data}

In Fig.~\ref{fig:sys_err} the $W$-dependencies of the extracted integrated cross sections of the reaction $\gamma_{v} p \rightarrow p' \pi^{+} \pi^{-}$ are shown by the black circles for twelve bins in $Q^{2}$. The gray shadowed areas correspond to the total cross section uncertainty, which is the uncertainty $\delta^{\text{tot}}_{\text{stat,mod}}$ given by Eq.~\eqref{eq:error_stat_mod} summed up in quadrature with the total systematic uncertainty. The error bars that correspond to the uncertainty $\delta^{\text{tot}}_{\text{stat,mod}}$ only, are smaller than the symbol size.

For each $(W,Q^{2})$ point shown in Fig.~\ref{fig:sys_err},
nine single-differential cross sections (see Eq.~\eqref{inegr5diff}) are  reported. An example of these cross sections is presented in Fig.~\ref{fig:res_cont} for the particular point  $W = 1.6375$~GeV and $Q^{2} = 0.525$~GeV$^{2}$, where the black symbols are for the single-differential cross sections, while the error  bars show the uncertainty $\delta^{\text{tot}}_{\text{stat,mod}}$. 

The whole set of the extracted cross sections is available in the CLAS physics database~\cite{CLAS_DB} and also on GitHub~\cite{Github:data}.

The extracted cross sections benefit from the minimal statistical uncertainty and the minimal model dependence among the previous studies of double-pion electroproduction cross sections~\cite{Fedotov:2008aa,Isupov:2017lnd,Ripani:2002ss}. This was achieved due to the high experimental statistics and the fact that four reaction topologies were analyzed in combination.

\subsection{Comparison with the model }

A preliminary interpretation of the extracted cross sections was based on the meson-baryon reaction model JM, which is currently the only available approach for phenomenological analysis of the double-pion electroproduction cross sections. This model aims at extracting the resonance electrocouplings as well as
establishing the contributions from different reaction subchannels and has proven itself as an effective tool for the analysis of the experimental cross sections~\cite{Mokeev:2008iw,Mokeev:2012vsa,Mokeev:2015lda}.

The preliminary interpretation of the results included the estimations of the full double-pion cross sections (integrated and single-differential),  as well as their resonant parts. The former is shown in Fig.~\ref{fig:sys_err} and Fig.~\ref{fig:res_cont} by the solid curves, while the latter by the dashed curves.

For this study a fit of the obtained results within the JM model was not performed, therefore an estimation of the full double-pion cross sections was obtained using the JM model based TWOPEG~\cite{Skorodum:EG} event generator.
This generator employs the five-fold differential structure functions from the recent version of the JM model fit to all existing CLAS results on double-pion photo- and electroproduction~\cite{Ripani:2002ss,Mokeev:2012vsa,Fedotov:2008aa,Golovach:note}. 
In the kinematic areas already covered by the CLAS data, TWOPEG performs the interpolation of the model structure functions and successfully reproduces the available integrated and single-differential cross sections. 
In the areas not yet covered by the CLAS data, special
extrapolation procedures have been applied that included additional world data on the integrated photoproduction cross sections~\cite{Wu:2005wf,ABBHHM:1968aa}. This event generator gives the absolute cross section values (see Ref.~\cite{Skorodum:EG} for details) that can be treated as a cross section prediction.
To perform a comparison with the reported cross sections, TWOPEG predictions  were adjusted to them using their experimentally established $Q^{2}$-dependence.
The quality of the description of the experimental results achieved in this way is shown in  Fig.~\ref{fig:sys_err} and Fig.~\ref{fig:res_cont} by the solid curves for the integrated and single-differential cross sections, respectively.

\begin{figure}[htp!]
\begin{center}
\includegraphics[width=8cm]{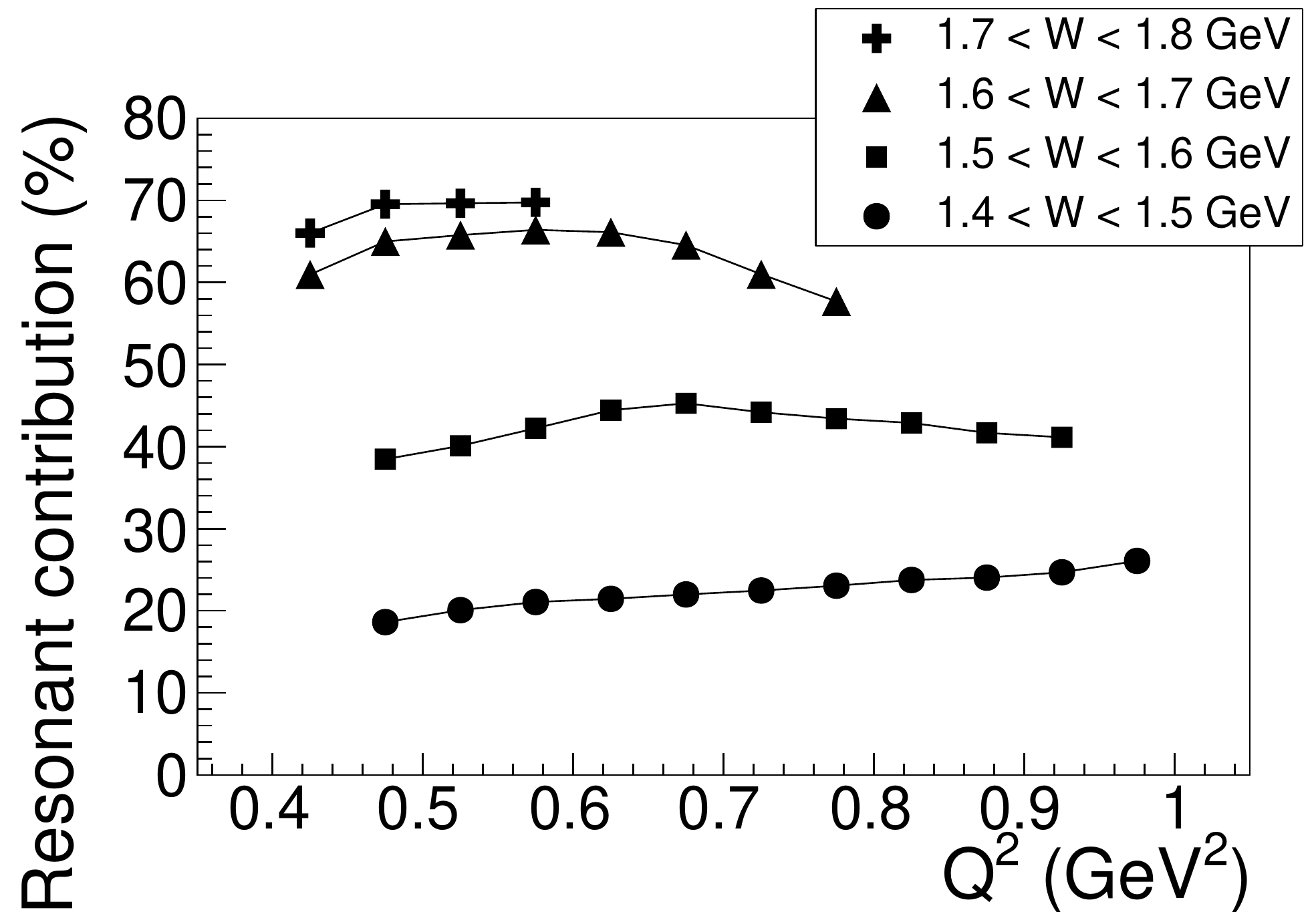}
\caption{\small Estimated relative resonant contribution to the integrated double-pion cross section  as a function of $Q^{2}$ (see text for details).  The different symbols connected with lines correspond to different $W$ ranges.} \label{fig:relative_res_cont}
\end{center}
\end{figure}

\begin{figure*}[htp!]
\begin{center}
\includegraphics[width=17.2cm]{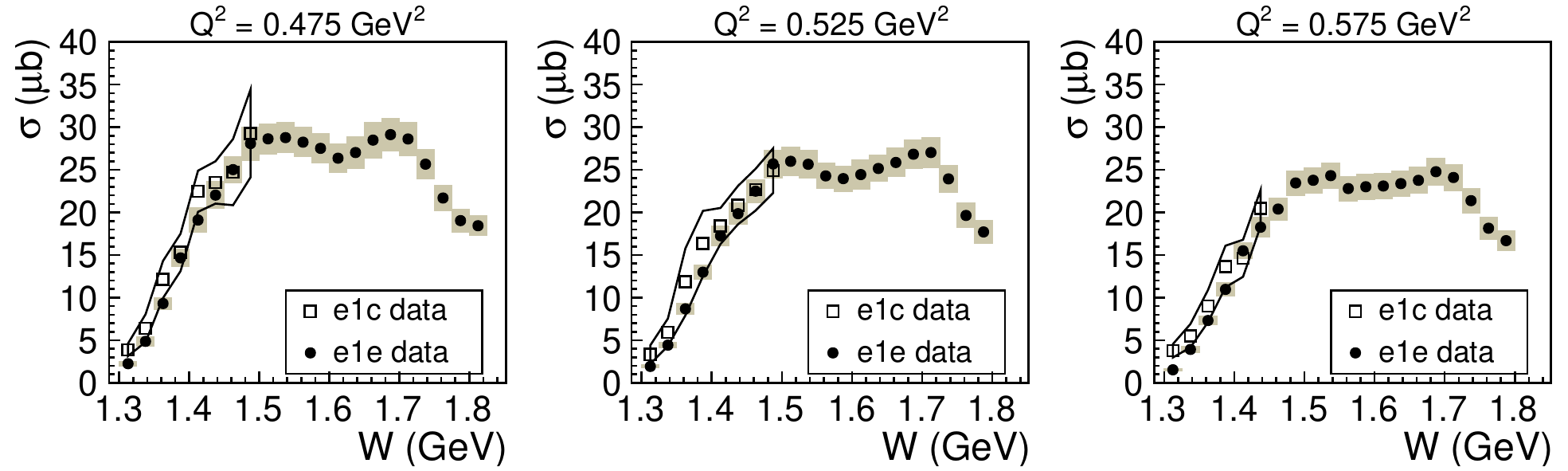}
\caption{\small The $W$-dependencies of the extracted  cross sections (black circles) in comparison with the available data~\cite{Fedotov:2008aa} (open squares) for three points in $Q^{2}$. The total cross section uncertainty (which includes both systematic and statistical uncertainties) is shown by the gray shadowed area for the new results (``e1e"), while for the results from Ref.~\cite{Fedotov:2008aa} (``e1c''), it is shown by the bands.}
\label{fig:e1e_e1c}
\end{center}
\end{figure*}

The resonant contribution to the full cross section was estimated using the unitarized Breit-Wigner ansatz of the JM model~\cite{Mokeev:2012vsa}. 
The model considered that, in the investigated $W$ range, the dominant part of the resonant contribution to the cross section is formed by the following nine resonances: P$_{11}$(1440), D$_{13}$(1520), S$_{11}$(1535), S$_{31}$(1620), S$_{11}$(1650), F$_{15}$(1680), D$_{33}$(1700), P$_{13}$(1720), and P$^{'}_{13}$(1720)\footnote[3]{In the updated PDG format N(1440)1/2$^{+}$, N(1520)3/2$^{-}$, N(1535)1/2$^{-}$, $\Delta$(1620)1/2$^{-}$, N(1650)1/2$^{-}$, N(1680)5/2$^{+}$, $\Delta$(1700)3/2$^{-}$, N(1720)$3/2^{+}$, and N$'$(1720)$3/2^{+}$, respectively.}, where P$^{'}_{13}$(1720) is a new potential candidate state~\cite{Mokeev:2015moa}. The electrocouplings of these nine states in the investigated $Q^{2}$ range were evaluated using the functions of their $Q^{2}$-dependences taken from the study~\cite{Isupov:2017lnd}. These functions were obtained as a polynomial fit of the available data on the resonance electrocouplings including those at the photon point~\cite{Mok_page,Dugger:2009pn,Aznauryan:2009mx,Mokeev:2012vsa,Mokeev:2015lda,Aznauryan:2009mx,Dalton:2008aa,Denizli:2007tq,Thompson:2000by,Armstrong:1998wg,Burkert:2002zz,Park:2014yea,Tiator:2011pw,Agashe:2014kda}. Ref.~\cite{Isupov:2017lnd} describes in detail the fit procedure. Due to the scarce data on electrocouplings close to the photon point  and the fact that the $S_{1/2}$ does not exist at the photon point, the fit for the $S_{1/2}$ electrocoupling of the resonances S$_{31}$(1620), F$_{15}(1680)$, and P$^{'}_{13}$(1720) is unreliable at $Q^{2} \lesssim 0.6$~GeV$^{2}$. Therefore, for these three states at $Q^{2} \lesssim 0.6$~GeV$^{2}$ the constant value of the $S_{1/2}$ taken at the last available $Q^{2}$ point was used.

Additionally, the states P$_{33}$(1600), D$_{15}$(1675), D$_{13}$(1700)\footnote[4]{$\Delta$(1600)3/2$^{+}$, N(1675)5/2$^{-}$, N(1700)3/2$^{-}$, respectively.}, although giving an insignificant contribution comparing with the nine resonances mentioned above, were included into the calculations with the values of their electrocouplings taken from the studies~\cite{Mokeev:2015lda,Mokeev:2005re} for $Q^{2} =0.65$~GeV$^{2}$. In order to partially take into account a contribution from the tails of the high-lying states, the resonances F$_{35}$(1905) and F$_{37}$(1950)\footnote[5]{$\Delta$(1905)5/2$^{+}$ and $\Delta$(1950)7/2$^{+}$, respectively.}
were also introduced into the model with the values of their electrocouplings taken from the study~\cite{Mokeev:2005re} for $Q^{2} =0.65$~GeV$^{2}$. 
These two states give from 2\% to 20\% of the total resonant contribution as $W$ grows from 1.7~GeV to 1.8~GeV.
For all resonance states 
the unitarized Breit-Wigner ansatz~\cite{Mokeev:2012vsa} was used and the hadronic decay widths to the $\pi\Delta$ and $\rho p$ final
states were taken from Ref.~\cite{Mokeev:2015lda}.

The estimation for the resonant part of the cross section is shown by the dashed curves in Fig.~\ref{fig:sys_err} and Fig.~\ref{fig:res_cont} for the integrated and single-differential cross sections, respectively. The relative resonant contribution to the integrated cross section is shown in Fig.~\ref{fig:relative_res_cont} as a function of $Q^{2}$ for various ranges in $W$. It was obtained as the ratio of the evaluated resonant part to the TWOPEG estimation for the full cross section. Fig.~\ref{fig:relative_res_cont} demonstrates the growth of the relative resonant contribution with increasing $W$, consistent with previous studies~\cite{Mokeev:2008iw,Mokeev:2015lda}. For small $W \sim 1.45$~GeV, this contribution stays on a level of 20\%, while at higher $W \sim 1.75$~GeV it reaches 70\%. The resonant contribution at $W \sim 1.75$~GeV is somewhat underestimated, since the resonances with masses above 1.8 GeV were not fully taken into account in this estimation.

The estimated resonant part of the cross section depends on the assumption for the $Q^{2}$ behavior of the resonance electrocouplings. Since a fit within the JM model was not performed, the uncertainty for this estimation can hardly be evaluated explicitly. A recent JM model fit of the data~\cite{Isupov:2017lnd} gives an uncertainty for the resonant part of about 6\%.

\subsection{Previously available data}
\label{prev_data}

In Fig.~\ref{fig:e1e_e1c}, the extracted integrated double-pion cross sections  are compared with the available data~\cite{Fedotov:2008aa}. 
The cross sections~\cite{Fedotov:2008aa} were obtained with a 1.515~GeV electron beam energy, which is different from that of the data reported here.
This introduces a small systematic distortion into the comparison
 caused by a beam energy dependence of the longitudinal cross section part. The kinematic coverages of these two datasets overlap only in three bins in $Q^{2}$. Meanwhile, the cross sections presented here should be treated as more reliable, since they were extracted with a more advanced technique -- i.e., the combination of all four available topologies was used instead of only two in Ref.~\cite{Fedotov:2008aa}, the map of the empty cells was better determined using  the cut on the efficiency uncertainty, the contribution from the empty cells was accounted for by the advanced method using TWOPEG~\cite{Skorodum:EG}, and furthermore, finer binning in the hadronic variables was achieved. Nevertheless, Fig.~\ref{fig:e1e_e1c} demonstrates reasonable agreement between these two sets of the cross sections within the total uncertainties.

\section{CONCLUSIONS AND OUTLOOK}

In this paper, new results on the integrated and single-differential cross sections of the reaction $\gamma_{v} p \rightarrow p' \pi^{+} \pi^{-}$  at $W$ from 1.3~GeV to 1.825~GeV and $Q^{2}$ from 0.4~GeV$^{2}$ to 1~GeV$^{2}$ are reported. The results are a significant improvement over previously available data~\cite{Fedotov:2008aa,Ripani:2002ss} in this kinematic region due to the extension in the $W$ coverage and due to the increased statistics, thereby achieving a finer binning in $Q^2$ (0.05 GeV$^2$). The whole set of the obtained cross sections is available in the CLAS physics database~\cite{CLAS_DB} and also on GitHub~\cite{Github:data}.

The kinematic coverage of the extracted cross sections overlaps with that of the previously available results~\cite{Fedotov:2008aa} in three $Q^{2}$ points 0.475, 0.525, and 0.575~GeV$^2$ for $W$ from 1.3 to $\approx 1.5$~GeV. In this region of overlap, the two cross section sets were found to be in agreement, as Fig.~\ref{fig:e1e_e1c} demonstrates. 
The double-pion cross sections reported in Ref.~\cite{Ripani:2002ss} also partially overlap with the results presented here, but since they were obtained  in much wider $Q^{2}$ bins, a comparison with them is not straightforward.

The cross section extraction procedure has some improvements in comparison with previous studies~\cite{Fedotov:2008aa,Ripani:2002ss,Isupov:2017lnd}.  An original method of revealing cells with unreliable efficiency via a cut on the relative efficiency uncertainty was applied. The cross sections in kinematic cells with zero acceptance were estimated using a recently developed event generator TWOPEG~\cite{Skorodum:EG}. All available reaction topologies were combined together to minimize statistical uncertainties as well as the contribution from kinematic cells with zero acceptance, in this way achieving a very modest model dependence of the obtained cross sections.

The obtained cross sections are compared with the predictions of the JM model based TWOPEG event generator, which currently provides the best double-pion cross section estimation in the investigated kinematic region. The comparisons presented in Fig.~\ref{fig:sys_err} and Fig.~\ref{fig:res_cont} show reasonably good agreement between the TWOPEG estimations (solid curves) and the experimental cross sections (symbols). The resonant contributions  to the cross section (dashed curves in Fig.~\ref{fig:sys_err} and Fig.~\ref{fig:res_cont}) were evaluated using the unitarized Breit-Wigner ansatz of the JM model, which includes all well established resonances in amplitude form. This estimation shows a sizable resonant contribution (see Fig.~\ref{fig:relative_res_cont}) that indicates the possibility of reliable extraction of the resonance electrocouplings.

The experimental results presented here will be further analyzed within the framework of the reaction model JM~\cite{Mokeev:2008iw,Mokeev:2012vsa,Mokeev:2015lda}. 
This analysis will eventually allow a determination of the $Q^2$-evolution of the electrocouplings of most nucleon resonances with masses up to $\sim$1.8~GeV for photon virtualities $Q^2$ from 0.425~GeV$^2$ to 0.975~GeV$^2$. For those resonances with mass greater than 1.6 GeV, which decay preferentially to the $p \pi^{+} \pi^{-}$ final state, this information will be obtained for the first time. These efforts are underway and the
results will be presented in a future publication on the subject.

\begin{acknowledgments}

The authors thank the technical staff at Jefferson Lab
and at all the participating institutions for their invaluable
contributions to the success of the experiment.
We are grateful for the help and support of the University of South Carolina (USC), the
Skobeltsyn Institute of Nuclear Physics (SINP) and the
Physics Departments at Moscow State University (MSU,
Moscow) and Ohio University (OU). 
This work was supported in part by the U.S. Department
of Energy (DOE) and National Science Foundation
(NSF), the Chilean Comisi\'on Nacional de Investigaci\'on
Cient\'ifica y Tecnol\'ogica (CONICYT), the Italian Istituto
Nazionale di Fisica Nucleare (INFN), the French Centre
National de la Recherche Scientifique (CNRS), the
French Commissariat \'a l'Energie Atomique (CEA), the
Skobeltsyn Institute of Nuclear Physics (SINP) and the
Physics Departments at Moscow State University (MSU,
Moscow) and Ohio University (OU), the Scottish Universities
Physics Alliance (SUPA), the National Research
Foundation of Korea (NRF), the UK Science and Technology
Facilities Council (STFC). Jefferson Science Associates
(JSA) operates the Thomas Jefferson National
Accelerator Facility for the United States Department of
Energy under contract DE-AC05-06OR23177.

\end{acknowledgments}

\clearpage
\section*{Appendix A: The definition of the angle $\alpha$}
\label{app_a}

The calculation of the angle $\alpha_{\pi^{-}}$ from the second set of hadron variables mentioned in Sec.~\ref{sec_kin_var} is given below. The angles $\alpha_{p'}$ and $\alpha_{\pi^{+}}$ from the other sets of variables are calculated analogously~\cite{Fed_an_note:2017}.

\begin{figure}[htp]
\begin{center}
\includegraphics[width=8cm,trim=0mm 14mm 0mm 0mm,clip]{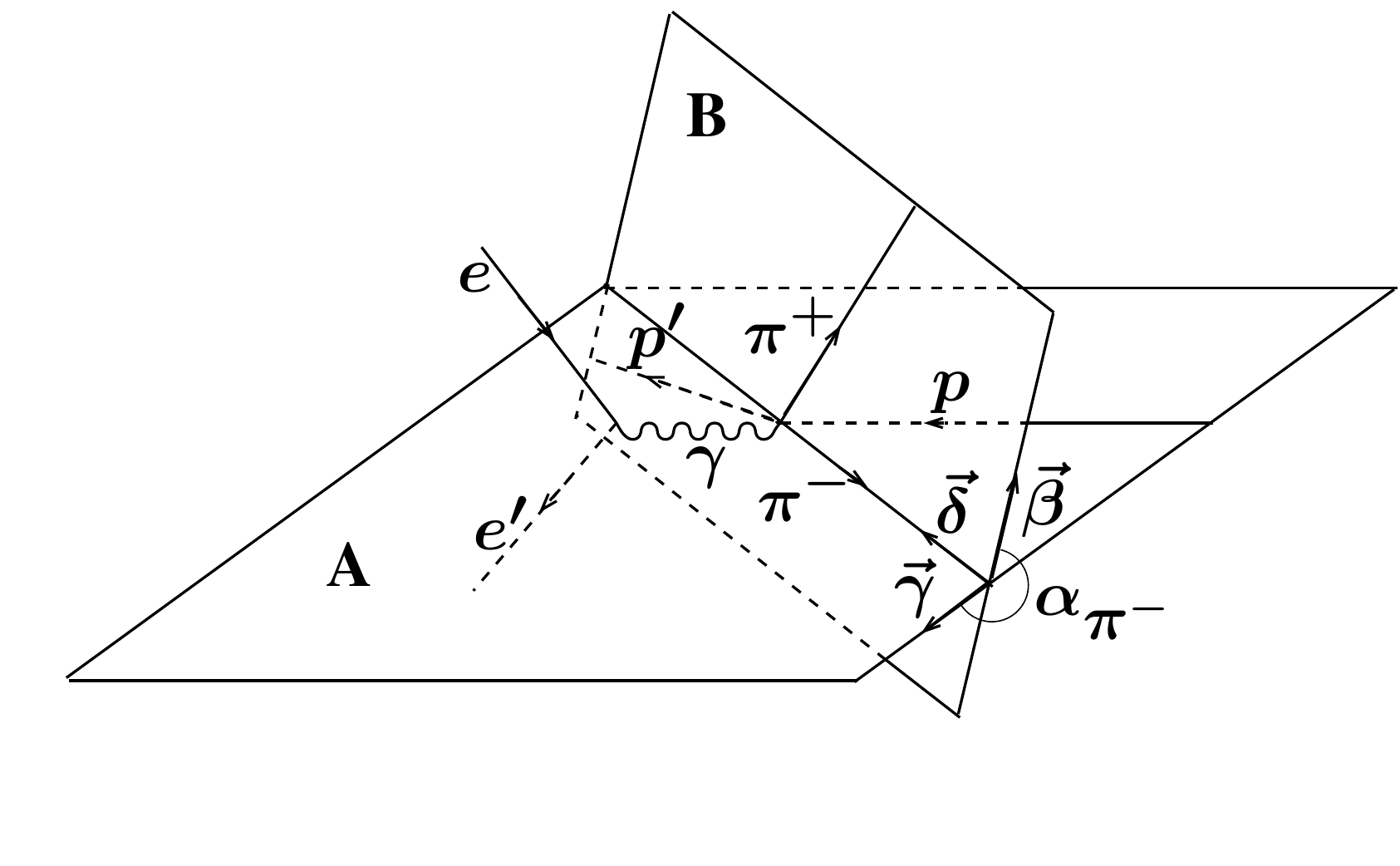}
\caption{\small Definition of the angle $\alpha_{\pi^{-}}$. The plane B is defined by the three-momenta of all final state hadrons, while the plane A is defined by  the three-momenta of the $\pi^{-}$ and initial proton. The definitions of the auxiliary vectors $\vec \beta$, $\vec \gamma$, $\vec \delta$ are given in the text.} \label{fig:cr_sec_kinematic2}
\end{center}
\end{figure}

The angle $\alpha_{\pi^{-}}$ is the angle between the two planes A and B (see Fig.~\ref{fig:cr_sec_kinematic2}).
The plane A is defined by
the initial proton and $\pi^{-}$, while the plane B is defined by the momenta of
all final state hadrons. Note that the three-momenta of the $\pi^{+}$,
$\pi^{-}$, $p'$ are in the same plane, since in the c.m.s.
their total three-momentum has to be equal to zero.

To calculate the angle $\alpha_{\pi^{-}}$, firstly two
auxiliary vectors $\vec \gamma$  and
$\vec \beta$ should be determined. The vector $\vec \gamma$ is the unit vector perpendicular to the three-momentum
$\vec P_{\pi^{-}}$, directed toward the vector $(-\vec n_{z})$ and situated in the plane A. $\vec
n_{z}$ is the unit vector directed along the $z$-axis.
The vector $\vec \beta$ is the unit vector perpendicular to the three-momentum of the $\pi^{-}$, 
directed toward the three-momentum of the $\pi^{+}$ and situated in the plane B. 
The angle between the two planes  $\alpha_{\pi^{-}}$ can be calculated as
 
\begin{equation}
\alpha_{\pi^{-}} = \arccos(\vec \gamma \cdot \vec \beta),
\label{eq:cr_sec_anglealpha}
\end{equation}
where $\arccos$ is a function that runs between zero and
$\pi$, while the angle $\alpha_{\pi^{-}}$ may vary between zero and
$2\pi$. To determine the $\alpha$ angle in the
range between $\pi$ and $2\pi$, 
the relative direction between the $\pi^{-}$ three-momentum and the vector product $\vec \delta = [ \vec \gamma \times \vec \beta ]$ of the auxiliary vectors $\vec
\gamma$ and $\vec \beta$ should be taken into account.
If the vector $\vec \delta$ is collinear to the three-momentum of the $\pi^{-}$, the angle $\alpha_{\pi^{-}}$ is determined
by Eq.~\eqref{eq:cr_sec_anglealpha}, and in the case of anti-collinearity by
\begin{equation}
\alpha_{\pi^{-}} = 2\pi - \arccos(\vec \gamma \cdot \vec \beta).
\label{eq:cr_sec_anglealpha_var}
\end{equation}

The defined above vector $\vec \gamma$ can be expressed as
\begin{eqnarray}
\vec \gamma = a_{\alpha}(-\vec n_{z}) + b_{\alpha}\vec n_{P_{\pi^{-}}} & \text{with} \nonumber \\
a_{\alpha} = \sqrt{\frac{1}{1 - (\vec n_{P_{\pi^{-}}} \cdot (-\vec n_{z} ) )^{2}}} & \text{and} \label{alphavec}\\
b_{\alpha} = - (\vec n_{P_{\pi^{-}}} \cdot (-\vec n_{z} ) ) a_{\alpha} \textrm{ ,} \nonumber
\end{eqnarray} 
where $\vec n_{P_{\pi^{-}}}$ is the unit vector directed along the three-momentum of the $\pi^{-}$ (see Fig.~\ref{fig:cr_sec_kinematic2}).

Taking the scalar products $(\vec \gamma \cdot \vec
n_{P_{\pi^{-}}})$ and $(\vec \gamma \cdot \vec  \gamma)$,
it is straightforward to verify, that $\vec \gamma$ is the unit vector perpendicular to the three-momentum of the $\pi^{-}$.

The vector $\vec \beta$ can be obtained as
\begin{eqnarray}
\vec \beta = a_{\beta}\vec n_{P_{\pi^{+}}} + b_{\beta}\vec n_{P_{\pi^{-}}} & \text{with} \nonumber \\
a_{\beta} = \sqrt{\frac{1}{1 - (\vec n_{P_{\pi^{+}}} \cdot \vec n_{P_{\pi^{-}}})^{2}}} & \text{and} \label{betavec}\\
b_{\beta} = - (\vec n_{P_{\pi^{+}}} \cdot \vec n_{P_{\pi^{-}}}) a_{\beta} \textrm{ ,} \nonumber
\end{eqnarray} 
where $\vec n_{P_{\pi^{+}}}$ is the unit vector directed along the three-momentum of the $\pi^{+}$.

Again taking the scalar products $(\vec \beta \cdot \vec
n_{P_{\pi^{-}}})$ and $(\vec \beta \cdot \vec  \beta)$,
it is straightforward to see that $\vec \beta$ is
the unit vector perpendicular to the 
three-momentum of the $\pi^{-}$.

Further detailed information about the kinematics of the reactions with three-particle final states can be found in Ref.~\cite{Byckling:1971vca}.

\bibliography{e1e_paper}{}

\begin{thebibliography}{41}%
\makeatletter
\providecommand \@ifxundefined [1]{%
 \@ifx{#1\undefined}
}%
\providecommand \@ifnum [1]{%
 \ifnum #1\expandafter \@firstoftwo
 \else \expandafter \@secondoftwo
 \fi
}%
\providecommand \@ifx [1]{%
 \ifx #1\expandafter \@firstoftwo
 \else \expandafter \@secondoftwo
 \fi
}%
\providecommand \natexlab [1]{#1}%
\providecommand \enquote  [1]{``#1''}%
\providecommand \bibnamefont  [1]{#1}%
\providecommand \bibfnamefont [1]{#1}%
\providecommand \citenamefont [1]{#1}%
\providecommand \href@noop [0]{\@secondoftwo}%
\providecommand \href [0]{\begingroup \@sanitize@url \@href}%
\providecommand \@href[1]{\@@startlink{#1}\@@href}%
\providecommand \@@href[1]{\endgroup#1\@@endlink}%
\providecommand \@sanitize@url [0]{\catcode `\\12\catcode `\$12\catcode
  `\&12\catcode `\#12\catcode `\^12\catcode `\_12\catcode `\%12\relax}%
\providecommand \@@startlink[1]{}%
\providecommand \@@endlink[0]{}%
\providecommand \url  [0]{\begingroup\@sanitize@url \@url }%
\providecommand \@url [1]{\endgroup\@href {#1}{\urlprefix }}%
\providecommand \urlprefix  [0]{URL }%
\providecommand \Eprint [0]{\href }%
\providecommand \doibase [0]{http://dx.doi.org/}%
\providecommand \selectlanguage [0]{\@gobble}%
\providecommand \bibinfo  [0]{\@secondoftwo}%
\providecommand \bibfield  [0]{\@secondoftwo}%
\providecommand \translation [1]{[#1]}%
\providecommand \BibitemOpen [0]{}%
\providecommand \bibitemStop [0]{}%
\providecommand \bibitemNoStop [0]{.\EOS\space}%
\providecommand \EOS [0]{\spacefactor3000\relax}%
\providecommand \BibitemShut  [1]{\csname bibitem#1\endcsname}%
\let\auto@bib@innerbib\@empty
\bibitem [{\citenamefont {Burkert}(2017)}]{Burkert:2016kyi}%
  \BibitemOpen
  \bibfield  {author} {\bibinfo {author} {\bibfnamefont {V.~D.}\ \bibnamefont
  {Burkert}} (\bibinfo {collaboration} {CLAS Collaboration}),\ }\href {\doibase
  10.1051/epjconf/201713401001} {\bibfield  {journal} {\bibinfo  {journal} {EPJ
  Web Conf.}\ }\textbf {\bibinfo {volume} {134}},\ \bibinfo {pages} {01001}
  (\bibinfo {year} {2017})},\ \Eprint {http://arxiv.org/abs/1610.00400}
  {arXiv:1610.00400 [nucl-ex]} \BibitemShut {NoStop}%
\bibitem [{\citenamefont {Krusche}\ and\ \citenamefont
  {Schadmand}(2003)}]{Krusche:2003ik}%
  \BibitemOpen
  \bibfield  {author} {\bibinfo {author} {\bibfnamefont {B.}~\bibnamefont
  {Krusche}}\ and\ \bibinfo {author} {\bibfnamefont {S.}~\bibnamefont
  {Schadmand}},\ }\href {\doibase 10.1016/S0146-6410(03)90005-6} {\bibfield
  {journal} {\bibinfo  {journal} {Prog. Part. Nucl. Phys.}\ }\textbf {\bibinfo
  {volume} {51}},\ \bibinfo {pages} {399} (\bibinfo {year} {2003})},\ \Eprint
  {http://arxiv.org/abs/nucl-ex/0306023} {arXiv:nucl-ex/0306023 [nucl-ex]}
  \BibitemShut {NoStop}%
\bibitem [{\citenamefont {Aznauryan}\ and\ \citenamefont
  {Burkert}(2012)}]{Aznauryan:2011qj}%
  \BibitemOpen
  \bibfield  {author} {\bibinfo {author} {\bibfnamefont {I.~G.}\ \bibnamefont
  {Aznauryan}}\ and\ \bibinfo {author} {\bibfnamefont {V.~D.}\ \bibnamefont
  {Burkert}},\ }\href {\doibase 10.1016/j.ppnp.2011.08.001} {\bibfield
  {journal} {\bibinfo  {journal} {Prog. Part. Nucl. Phys.}\ }\textbf {\bibinfo
  {volume} {67}},\ \bibinfo {pages} {1} (\bibinfo {year} {2012})},\ \Eprint
  {http://arxiv.org/abs/1109.1720} {arXiv:1109.1720 [hep-ph]} \BibitemShut
  {NoStop}%
\bibitem [{\citenamefont {Skorodumina}\ \emph {et~al.}(2015)\citenamefont
  {Skorodumina}, \citenamefont {Burkert}, \citenamefont {Golovach},
  \citenamefont {Gothe}, \citenamefont {Isupov}, \citenamefont {Ishkhanov},
  \citenamefont {Mokeev},\ and\ \citenamefont {Fedotov}}]{Skorodumina:2016pnb}%
  \BibitemOpen
  \bibfield  {author} {\bibinfo {author} {\bibfnamefont {{\relax Iu}.~A.}\
  \bibnamefont {Skorodumina}}, \bibinfo {author} {\bibfnamefont {V.~D.}\
  \bibnamefont {Burkert}}, \bibinfo {author} {\bibfnamefont {E.~N.}\
  \bibnamefont {Golovach}}, \bibinfo {author} {\bibfnamefont {R.~W.}\
  \bibnamefont {Gothe}}, \bibinfo {author} {\bibfnamefont {E.~L.}\ \bibnamefont
  {Isupov}}, \bibinfo {author} {\bibfnamefont {B.~S.}\ \bibnamefont
  {Ishkhanov}}, \bibinfo {author} {\bibfnamefont {V.~I.}\ \bibnamefont
  {Mokeev}}, \ and\ \bibinfo {author} {\bibfnamefont {G.~V.}\ \bibnamefont
  {Fedotov}},\ }\href {\doibase 10.3103/S002713491506017X} {\bibfield
  {journal} {\bibinfo  {journal} {Moscow Univ. Phys. Bull.}\ }\textbf {\bibinfo
  {volume} {70}},\ \bibinfo {pages} {429} (\bibinfo {year} {2015})},\ \bibinfo
  {note} {[Vestn. Mosk. Univ.,no.6,3(2015)]}\BibitemShut {NoStop}%
\bibitem [{\citenamefont {Mecking}\ \emph {et~al.}(2003)\citenamefont {Mecking}
  \emph {et~al.}}]{Me03}%
  \BibitemOpen
  \bibfield  {author} {\bibinfo {author} {\bibfnamefont {B.~A.}\ \bibnamefont
  {Mecking}} \emph {et~al.} (\bibinfo {collaboration} {CLAS Collaboration}),\
  }\href {\doibase http://dx.doi.org/10.1016/S0168-9002(03)01001-5} {\bibfield
  {journal} {\bibinfo  {journal} {Nucl. Instr. and Meth.}\ }\textbf {\bibinfo
  {volume} {503}},\ \bibinfo {pages} {513 } (\bibinfo {year}
  {2003})}\BibitemShut {NoStop}%
\bibitem [{CLA()}]{CLAS_DB}%
  \BibitemOpen
  \href@noop {} {\enquote {\bibinfo {title} {{CLAS Physics Database}},}\
  }\bibinfo {howpublished}
  {\url{http://clas.sinp.msu.ru/cgi-bin/jlab/db.cgi}}\BibitemShut {NoStop}%
\bibitem [{Git()}]{Github:data}%
  \BibitemOpen
  \href@noop {} {}\bibinfo {howpublished}
  {\url{https://github.com/gleb811/two\_pi\_exp\_data/tree/master/Fedotov\_data\_Q2\_0425\_0975}}\BibitemShut
  {NoStop}%
\bibitem [{\citenamefont {Fedotov}\ \emph {et~al.}(2009)\citenamefont {Fedotov}
  \emph {et~al.}}]{Fedotov:2008aa}%
  \BibitemOpen
  \bibfield  {author} {\bibinfo {author} {\bibfnamefont {G.~V.}\ \bibnamefont
  {Fedotov}} \emph {et~al.} (\bibinfo {collaboration} {CLAS Collaboration}),\
  }\href {\doibase 10.1103/PhysRevC.79.015204} {\bibfield  {journal} {\bibinfo
  {journal} {Phys. Rev.}\ }\textbf {\bibinfo {volume} {C79}},\ \bibinfo {pages}
  {015204} (\bibinfo {year} {2009})},\ \Eprint {http://arxiv.org/abs/0809.1562}
  {arXiv:0809.1562 [nucl-ex]} \BibitemShut {NoStop}%
\bibitem [{\citenamefont {Ripani}\ \emph {et~al.}(2003)\citenamefont {Ripani}
  \emph {et~al.}}]{Ripani:2002ss}%
  \BibitemOpen
  \bibfield  {author} {\bibinfo {author} {\bibfnamefont {M.}~\bibnamefont
  {Ripani}} \emph {et~al.} (\bibinfo {collaboration} {CLAS Collaboration}),\
  }\href {\doibase 10.1103/PhysRevLett.91.022002} {\bibfield  {journal}
  {\bibinfo  {journal} {Phys. Rev. Lett.}\ }\textbf {\bibinfo {volume} {91}},\
  \bibinfo {pages} {022002} (\bibinfo {year} {2003})},\ \Eprint
  {http://arxiv.org/abs/hep-ex/0210054} {arXiv:hep-ex/0210054 [hep-ex]}
  \BibitemShut {NoStop}%
\bibitem [{\citenamefont {Isupov}\ \emph {et~al.}(2017)\citenamefont {Isupov}
  \emph {et~al.}}]{Isupov:2017lnd}%
  \BibitemOpen
  \bibfield  {author} {\bibinfo {author} {\bibfnamefont {E.~L.}\ \bibnamefont
  {Isupov}} \emph {et~al.} (\bibinfo {collaboration} {CLAS Collaboration}),\
  }\href {\doibase 10.1103/PhysRevC.96.025209} {\bibfield  {journal} {\bibinfo
  {journal} {Phys. Rev.}\ }\textbf {\bibinfo {volume} {C96}},\ \bibinfo {pages}
  {025209} (\bibinfo {year} {2017})},\ \Eprint
  {http://arxiv.org/abs/1705.01901} {arXiv:1705.01901 [nucl-ex]} \BibitemShut
  {NoStop}%
\bibitem [{\citenamefont {Mokeev}\ \emph
  {et~al.}(2016{\natexlab{a}})\citenamefont {Mokeev}, \citenamefont {Burkert},
  \citenamefont {Carman}, \citenamefont {Elouadrhiri}, \citenamefont {Fedotov},
  \citenamefont {Golovatch}, \citenamefont {Gothe}, \citenamefont {Hicks},
  \citenamefont {Ishkhanov}, \citenamefont {Isupov},\ and\ \citenamefont
  {Skorodumina}}]{Mokeev:2015lda}%
  \BibitemOpen
  \bibfield  {author} {\bibinfo {author} {\bibfnamefont {V.~I.}\ \bibnamefont
  {Mokeev}}, \bibinfo {author} {\bibfnamefont {V.~D.}\ \bibnamefont {Burkert}},
  \bibinfo {author} {\bibfnamefont {D.~S.}\ \bibnamefont {Carman}}, \bibinfo
  {author} {\bibfnamefont {L.}~\bibnamefont {Elouadrhiri}}, \bibinfo {author}
  {\bibfnamefont {G.~V.}\ \bibnamefont {Fedotov}}, \bibinfo {author}
  {\bibfnamefont {E.~N.}\ \bibnamefont {Golovatch}}, \bibinfo {author}
  {\bibfnamefont {R.~W.}\ \bibnamefont {Gothe}}, \bibinfo {author}
  {\bibfnamefont {K.}~\bibnamefont {Hicks}}, \bibinfo {author} {\bibfnamefont
  {B.~S.}\ \bibnamefont {Ishkhanov}}, \bibinfo {author} {\bibfnamefont {E.~L.}\
  \bibnamefont {Isupov}}, \ and\ \bibinfo {author} {\bibfnamefont {{\relax
  Iu}.~A.}\ \bibnamefont {Skorodumina}},\ }\href {\doibase
  10.1103/PhysRevC.93.025206} {\bibfield  {journal} {\bibinfo  {journal} {Phys.
  Rev.}\ }\textbf {\bibinfo {volume} {C93}},\ \bibinfo {pages} {025206}
  (\bibinfo {year} {2016}{\natexlab{a}})},\ \Eprint
  {http://arxiv.org/abs/1509.05460} {arXiv:1509.05460 [nucl-ex]} \BibitemShut
  {NoStop}%
\bibitem [{\citenamefont {Mokeev}\ \emph {et~al.}(2009)\citenamefont {Mokeev},
  \citenamefont {Burkert}, \citenamefont {Lee}, \citenamefont {Elouadrhiri},
  \citenamefont {Fedotov},\ and\ \citenamefont {Ishkhanov}}]{Mokeev:2008iw}%
  \BibitemOpen
  \bibfield  {author} {\bibinfo {author} {\bibfnamefont {V.~I.}\ \bibnamefont
  {Mokeev}}, \bibinfo {author} {\bibfnamefont {V.~D.}\ \bibnamefont {Burkert}},
  \bibinfo {author} {\bibfnamefont {T.-S.~H.}\ \bibnamefont {Lee}}, \bibinfo
  {author} {\bibfnamefont {L.}~\bibnamefont {Elouadrhiri}}, \bibinfo {author}
  {\bibfnamefont {G.~V.}\ \bibnamefont {Fedotov}}, \ and\ \bibinfo {author}
  {\bibfnamefont {B.~S.}\ \bibnamefont {Ishkhanov}},\ }\href {\doibase
  10.1103/PhysRevC.80.045212} {\bibfield  {journal} {\bibinfo  {journal} {Phys.
  Rev.}\ }\textbf {\bibinfo {volume} {C80}},\ \bibinfo {pages} {045212}
  (\bibinfo {year} {2009})},\ \Eprint {http://arxiv.org/abs/0809.4158}
  {arXiv:0809.4158 [hep-ph]} \BibitemShut {NoStop}%
\bibitem [{\citenamefont {Mokeev}\ \emph {et~al.}(2012)\citenamefont {Mokeev}
  \emph {et~al.}}]{Mokeev:2012vsa}%
  \BibitemOpen
  \bibfield  {author} {\bibinfo {author} {\bibfnamefont {V.~I.}\ \bibnamefont
  {Mokeev}} \emph {et~al.} (\bibinfo {collaboration} {CLAS Collaboration}),\
  }\href {\doibase 10.1103/PhysRevC.86.035203} {\bibfield  {journal} {\bibinfo
  {journal} {Phys. Rev.}\ }\textbf {\bibinfo {volume} {C86}},\ \bibinfo {pages}
  {035203} (\bibinfo {year} {2012})},\ \Eprint {http://arxiv.org/abs/1205.3948}
  {arXiv:1205.3948 [nucl-ex]} \BibitemShut {NoStop}%
\bibitem [{\citenamefont {Mokeev}\ \emph
  {et~al.}(2016{\natexlab{b}})\citenamefont {Mokeev}, \citenamefont
  {Aznauryan}, \citenamefont {Burkert},\ and\ \citenamefont
  {Gothe}}]{Mokeev:2015moa}%
  \BibitemOpen
  \bibfield  {author} {\bibinfo {author} {\bibfnamefont {V.~I.}\ \bibnamefont
  {Mokeev}}, \bibinfo {author} {\bibfnamefont {I.~G.}\ \bibnamefont
  {Aznauryan}}, \bibinfo {author} {\bibfnamefont {V.~D.}\ \bibnamefont
  {Burkert}}, \ and\ \bibinfo {author} {\bibfnamefont {R.~W.}\ \bibnamefont
  {Gothe}},\ }\href {\doibase 10.1051/epjconf/201611301013} {\bibfield
  {journal} {\bibinfo  {journal} {EPJ Web Conf.}\ }\textbf {\bibinfo {volume}
  {113}},\ \bibinfo {pages} {01013} (\bibinfo {year} {2016}{\natexlab{b}})},\
  \Eprint {http://arxiv.org/abs/1508.04088} {arXiv:1508.04088 [nucl-ex]}
  \BibitemShut {NoStop}%
\bibitem [{\citenamefont {Egiyan}()}]{Egian:007}%
  \BibitemOpen
  \bibfield  {author} {\bibinfo {author} {\bibfnamefont {K.}~\bibnamefont
  {Egiyan}},\ }\href@noop {} {}\bibinfo {howpublished} {CLAS-NOTE-99-007,
  \url{https://www.jlab.org/Hall-B/notes/clas_notes99/ec_thresh.ps}}\BibitemShut
  {NoStop}%
\bibitem [{\citenamefont {Osipenko}\ \emph {et~al.}(2004)\citenamefont
  {Osipenko}, \citenamefont {Vlassov},\ and\ \citenamefont
  {Taiuti}}]{Osipenko:2004}%
  \BibitemOpen
  \bibfield  {author} {\bibinfo {author} {\bibfnamefont {M.}~\bibnamefont
  {Osipenko}}, \bibinfo {author} {\bibfnamefont {A.}~\bibnamefont {Vlassov}}, \
  and\ \bibinfo {author} {\bibfnamefont {M.}~\bibnamefont {Taiuti}},\
  }\href@noop {} {}\bibinfo {howpublished} {CLAS-NOTE-2004-020,
  \url{https://www.jlab.org/Hall-B/notes/clas_notes04/2004-020.pdf}} (\bibinfo
  {year} {2004})\BibitemShut {NoStop}%
\bibitem [{\citenamefont {Fedotov}\ \emph {et~al.}(2017)\citenamefont
  {Fedotov}, \citenamefont {Burkert}, \citenamefont {Gothe}, \citenamefont
  {Mokeev},\ and\ \citenamefont {Skorodumina}}]{Fed_an_note:2017}%
  \BibitemOpen
  \bibfield  {author} {\bibinfo {author} {\bibfnamefont {G.~V.}\ \bibnamefont
  {Fedotov}}, \bibinfo {author} {\bibfnamefont {V.~D.}\ \bibnamefont
  {Burkert}}, \bibinfo {author} {\bibfnamefont {R.~W.}\ \bibnamefont {Gothe}},
  \bibinfo {author} {\bibfnamefont {V.~I.}\ \bibnamefont {Mokeev}}, \ and\
  \bibinfo {author} {\bibfnamefont {{\relax Iu}.~A.}\ \bibnamefont
  {Skorodumina}},\ }\href@noop {} {}\bibinfo {howpublished}
  {CLAS-Analysis-2017-101 (CLAS-NOTE-2018-001),
  \url{https://misportal.jlab.org/ul/Physics/Hall-B/clas/viewFile.cfm/2018-001.pdf?documentId=776}}
  (\bibinfo {year} {2017})\BibitemShut {NoStop}%
\bibitem [{\citenamefont {Park}\ \emph {et~al.}(2003)\citenamefont {Park},
  \citenamefont {Burkert},\ and\ \citenamefont {Elouadrhiri}}]{KPark:momcorr}%
  \BibitemOpen
  \bibfield  {author} {\bibinfo {author} {\bibfnamefont {K.}~\bibnamefont
  {Park}}, \bibinfo {author} {\bibfnamefont {V.}~\bibnamefont {Burkert}}, \
  and\ \bibinfo {author} {\bibfnamefont {L.}~\bibnamefont {Elouadrhiri}},\
  }\href@noop {} {}\bibinfo {howpublished} {CLAS-NOTE-2003-012,
  \url{https://www.jlab.org/Hall-B/notes/clas_notes03/03-012.pdf}} (\bibinfo
  {year} {2003})\BibitemShut {NoStop}%
\bibitem [{\citenamefont {Mo}\ and\ \citenamefont {Tsai}(1969)}]{Mo:1968cg}%
  \BibitemOpen
  \bibfield  {author} {\bibinfo {author} {\bibfnamefont {L.~W.}\ \bibnamefont
  {Mo}}\ and\ \bibinfo {author} {\bibfnamefont {Y.-S.}\ \bibnamefont {Tsai}},\
  }\href {\doibase 10.1103/RevModPhys.41.205} {\bibfield  {journal} {\bibinfo
  {journal} {Rev. Mod. Phys.}\ }\textbf {\bibinfo {volume} {41}},\ \bibinfo
  {pages} {205} (\bibinfo {year} {1969})}\BibitemShut {NoStop}%
\bibitem [{\citenamefont {Wu}\ \emph {et~al.}(2005)\citenamefont {Wu},
  \citenamefont {Barth}, \citenamefont {Braun}, \citenamefont {Ernst},
  \citenamefont {Glander}, \citenamefont {Hannappel}, \citenamefont
  {J{\"o}pen}, \citenamefont {Kalinowsky}, \citenamefont {F.~J.~Klein},
  \citenamefont {Klein} \emph {et~al.}}]{Wu:2005wf}%
  \BibitemOpen
  \bibfield  {author} {\bibinfo {author} {\bibfnamefont {C.}~\bibnamefont
  {Wu}}, \bibinfo {author} {\bibfnamefont {J.}~\bibnamefont {Barth}}, \bibinfo
  {author} {\bibfnamefont {W.}~\bibnamefont {Braun}}, \bibinfo {author}
  {\bibfnamefont {J.}~\bibnamefont {Ernst}}, \bibinfo {author} {\bibfnamefont
  {K.-H.}\ \bibnamefont {Glander}}, \bibinfo {author} {\bibfnamefont
  {J.}~\bibnamefont {Hannappel}}, \bibinfo {author} {\bibfnamefont
  {N.}~\bibnamefont {J{\"o}pen}}, \bibinfo {author} {\bibfnamefont
  {H.}~\bibnamefont {Kalinowsky}}, \bibinfo {author} {\bibfnamefont {F.~J.}\
  \bibnamefont {F.~J.~Klein}}, \bibinfo {author} {\bibfnamefont
  {F.}~\bibnamefont {Klein}},  \emph {et~al.},\ }\href {\doibase
  10.1140/epja/i2004-10093-9} {\bibfield  {journal} {\bibinfo  {journal} {Eur.
  Phys. J.}\ }\textbf {\bibinfo {volume} {A23}},\ \bibinfo {pages} {317}
  (\bibinfo {year} {2005})}\BibitemShut {NoStop}%
\bibitem [{\citenamefont {Ripani}\ \emph {et~al.}(2000)\citenamefont {Ripani},
  \citenamefont {Mokeev}, \citenamefont {Anghinolfi}, \citenamefont
  {Battaglieri}, \citenamefont {Fedotov}, \citenamefont {Golovach},
  \citenamefont {Ishkhanov}, \citenamefont {Osipenko}, \citenamefont {Ricco},
  \citenamefont {Sapunenko},\ and\ \citenamefont {Taiuti}}]{Ripani:2000va}%
  \BibitemOpen
  \bibfield  {author} {\bibinfo {author} {\bibfnamefont {M.}~\bibnamefont
  {Ripani}}, \bibinfo {author} {\bibfnamefont {V.}~\bibnamefont {Mokeev}},
  \bibinfo {author} {\bibfnamefont {M.}~\bibnamefont {Anghinolfi}}, \bibinfo
  {author} {\bibfnamefont {M.}~\bibnamefont {Battaglieri}}, \bibinfo {author}
  {\bibfnamefont {G.}~\bibnamefont {Fedotov}}, \bibinfo {author} {\bibfnamefont
  {E.}~\bibnamefont {Golovach}}, \bibinfo {author} {\bibfnamefont
  {B.}~\bibnamefont {Ishkhanov}}, \bibinfo {author} {\bibfnamefont
  {M.}~\bibnamefont {Osipenko}}, \bibinfo {author} {\bibfnamefont
  {G.}~\bibnamefont {Ricco}}, \bibinfo {author} {\bibfnamefont
  {V.}~\bibnamefont {Sapunenko}}, \ and\ \bibinfo {author} {\bibfnamefont
  {M.}~\bibnamefont {Taiuti}},\ }\href {\doibase 10.1016/S0375-9474(99)00853-2}
  {\bibfield  {journal} {\bibinfo  {journal} {Nucl. Phys.}\ }\textbf {\bibinfo
  {volume} {A672}},\ \bibinfo {pages} {220} (\bibinfo {year} {2000})},\ \Eprint
  {http://arxiv.org/abs/hep-ph/0001265} {arXiv:hep-ph/0001265 [hep-ph]}
  \BibitemShut {NoStop}%
\bibitem [{\citenamefont {Aznauryan}\ \emph {et~al.}(2005)\citenamefont
  {Aznauryan}, \citenamefont {Burkert}, \citenamefont {Fedotov}, \citenamefont
  {Ishkhanov},\ and\ \citenamefont {Mokeev}}]{Aznauryan:2005tp}%
  \BibitemOpen
  \bibfield  {author} {\bibinfo {author} {\bibfnamefont {I.~G.}\ \bibnamefont
  {Aznauryan}}, \bibinfo {author} {\bibfnamefont {V.~D.}\ \bibnamefont
  {Burkert}}, \bibinfo {author} {\bibfnamefont {G.~V.}\ \bibnamefont
  {Fedotov}}, \bibinfo {author} {\bibfnamefont {B.~S.}\ \bibnamefont
  {Ishkhanov}}, \ and\ \bibinfo {author} {\bibfnamefont {V.~I.}\ \bibnamefont
  {Mokeev}},\ }\href {\doibase 10.1103/PhysRevC.72.045201} {\bibfield
  {journal} {\bibinfo  {journal} {Phys. Rev.}\ }\textbf {\bibinfo {volume}
  {C72}},\ \bibinfo {pages} {045201} (\bibinfo {year} {2005})},\ \Eprint
  {http://arxiv.org/abs/hep-ph/0508057} {arXiv:hep-ph/0508057 [hep-ph]}
  \BibitemShut {NoStop}%
\bibitem [{\citenamefont {Mokeev}\ \emph {et~al.}()\citenamefont {Mokeev},
  \citenamefont {Burkert}, \citenamefont {Elouadrhiri}, \citenamefont
  {Boluchevsky}, \citenamefont {Fedotov}, \citenamefont {Isupov}, \citenamefont
  {Ishkhanov},\ and\ \citenamefont {V.}}]{Mokeev:2005re}%
  \BibitemOpen
  \bibfield  {author} {\bibinfo {author} {\bibfnamefont {V.~I.}\ \bibnamefont
  {Mokeev}}, \bibinfo {author} {\bibfnamefont {V.~D.}\ \bibnamefont {Burkert}},
  \bibinfo {author} {\bibfnamefont {L.}~\bibnamefont {Elouadrhiri}}, \bibinfo
  {author} {\bibfnamefont {A.~A.}\ \bibnamefont {Boluchevsky}}, \bibinfo
  {author} {\bibfnamefont {G.~V.}\ \bibnamefont {Fedotov}}, \bibinfo {author}
  {\bibfnamefont {E.~L.}\ \bibnamefont {Isupov}}, \bibinfo {author}
  {\bibfnamefont {B.~S.}\ \bibnamefont {Ishkhanov}}, \ and\ \bibinfo {author}
  {\bibfnamefont {S.~N.}\ \bibnamefont {V.}},\ }in\ \href
  {http://www1.jlab.org/Ul/publications/view_pub.cfm?pub_id=6442} {\emph
  {\bibinfo {booktitle} {{Proceedings, (NSTAR 2005)}}}}\ (\bibinfo  {publisher}
  {World Scientific, Hackensack, USA, 2006})\ pp.\ \bibinfo {pages} {47--56},\
  \Eprint {http://arxiv.org/abs/hep-ph/0512164} {arXiv:hep-ph/0512164 [hep-ph]}
  \BibitemShut {NoStop}%
\bibitem [{\citenamefont {Byckling}\ and\ \citenamefont
  {Kajantie}(1971)}]{Byckling:1971vca}%
  \BibitemOpen
  \bibfield  {author} {\bibinfo {author} {\bibfnamefont {E.}~\bibnamefont
  {Byckling}}\ and\ \bibinfo {author} {\bibfnamefont {K.}~\bibnamefont
  {Kajantie}},\ }\href
  {http://www-spires.fnal.gov/spires/find/books/www?cl=QC794.6.K5B99} {\emph
  {\bibinfo {title} {{Particle Kinematics}}}}\ (\bibinfo  {publisher}
  {University of Jyvaskyla},\ \bibinfo {address} {Jyvaskyla, Finland},\
  \bibinfo {year} {1971})\BibitemShut {NoStop}%
\bibitem [{\citenamefont {Skorodumina}\ \emph {et~al.}(2017)\citenamefont
  {Skorodumina}, \citenamefont {Fedotov}, \citenamefont {Burkert},
  \citenamefont {Golovach}, \citenamefont {Gothe},\ and\ \citenamefont
  {Mokeev}}]{Skorodum:EG}%
  \BibitemOpen
  \bibfield  {author} {\bibinfo {author} {\bibfnamefont {{\relax Iu}.~A.}\
  \bibnamefont {Skorodumina}}, \bibinfo {author} {\bibfnamefont {G.~V.}\
  \bibnamefont {Fedotov}}, \bibinfo {author} {\bibfnamefont {V.~D.}\
  \bibnamefont {Burkert}}, \bibinfo {author} {\bibfnamefont {E.~N.}\
  \bibnamefont {Golovach}}, \bibinfo {author} {\bibfnamefont {R.~W.}\
  \bibnamefont {Gothe}}, \ and\ \bibinfo {author} {\bibfnamefont {V.~I.}\
  \bibnamefont {Mokeev}},\ }\href@noop {} {}\bibinfo {howpublished}
  {CLAS12-NOTE-2017-001, arXiv:1703.08081,
  \url{https://misportal.jlab.org/mis/physics/clas12/viewFile.cfm/2017-001.pdf?documentId=36}}
  (\bibinfo {year} {2017})\BibitemShut {NoStop}%
\bibitem [{\citenamefont {Golovach}()}]{Genev:Golova}%
  \BibitemOpen
  \bibfield  {author} {\bibinfo {author} {\bibfnamefont {E.~N.}\ \bibnamefont
  {Golovach}},\ }\href@noop {} {}\bibinfo {howpublished}
  {\url{http://depni.sinp.msu.ru/~golovach/EG/}}\BibitemShut {NoStop}%
\bibitem [{\citenamefont {Golovach}(2018)}]{Golovach:note}%
  \BibitemOpen
  \bibfield  {author} {\bibinfo {author} {\bibfnamefont {E.~N.}\ \bibnamefont
  {Golovach}} (\bibinfo {collaboration} {CLAS Collaboration}),\ }\href@noop {}
  {\bibfield  {journal} {\bibinfo  {journal} {to be published in Phys. Rev. C}\
  } (\bibinfo {year} {2018})},\ \Eprint {http://arxiv.org/abs/1806.01767.}
  {arXiv:1806.01767.} \BibitemShut {NoStop}%
\bibitem [{\citenamefont {{\relax ABBHHM
  Collaboration}}(1968)}]{ABBHHM:1968aa}%
  \BibitemOpen
  \bibfield  {author} {\bibinfo {author} {\bibnamefont {{\relax ABBHHM
  Collaboration}}} (\bibinfo {collaboration}
  {Aachen-Berlin-Bonn-Hamburg-Heidelberg-Munich}),\ }\href {\doibase
  10.1103/PhysRev.175.1669} {\bibfield  {journal} {\bibinfo  {journal} {Phys.
  Rev.}\ }\textbf {\bibinfo {volume} {175}},\ \bibinfo {pages} {1669} (\bibinfo
  {year} {1968})}\BibitemShut {NoStop}%
\bibitem [{\citenamefont {Laforge}\ and\ \citenamefont
  {Schoeffel}(1997)}]{Laforge:1996ts}%
  \BibitemOpen
  \bibfield  {author} {\bibinfo {author} {\bibfnamefont {B.}~\bibnamefont
  {Laforge}}\ and\ \bibinfo {author} {\bibfnamefont {L.}~\bibnamefont
  {Schoeffel}},\ }\href {\doibase 10.1016/S0168-9002(97)00649-9} {\bibfield
  {journal} {\bibinfo  {journal} {Nucl. Instrum. Meth.}\ }\textbf {\bibinfo
  {volume} {A394}},\ \bibinfo {pages} {115} (\bibinfo {year}
  {1997})}\BibitemShut {NoStop}%
\bibitem [{\citenamefont {Bosted}(1995)}]{Bosted:1994tm}%
  \BibitemOpen
  \bibfield  {author} {\bibinfo {author} {\bibfnamefont {P.~E.}\ \bibnamefont
  {Bosted}},\ }\href {\doibase 10.1103/PhysRevC.51.409} {\bibfield  {journal}
  {\bibinfo  {journal} {Phys.Rev.}\ }\textbf {\bibinfo {volume} {C51}},\
  \bibinfo {pages} {409} (\bibinfo {year} {1995})}\BibitemShut {NoStop}%
\bibitem [{Mok()}]{Mok_page}%
  \BibitemOpen
  \href@noop {} {}\bibinfo {howpublished}
  {\url{https://userweb.jlab.org/~mokeev/resonance_electrocouplings/}}\BibitemShut
  {NoStop}%
\bibitem [{\citenamefont {Dugger}\ \emph {et~al.}(2009)\citenamefont {Dugger}
  \emph {et~al.}}]{Dugger:2009pn}%
  \BibitemOpen
  \bibfield  {author} {\bibinfo {author} {\bibfnamefont {M.}~\bibnamefont
  {Dugger}} \emph {et~al.} (\bibinfo {collaboration} {CLAS Collaboration}),\
  }\href {\doibase 10.1103/PhysRevC.79.065206} {\bibfield  {journal} {\bibinfo
  {journal} {Phys. Rev.}\ }\textbf {\bibinfo {volume} {C79}},\ \bibinfo {pages}
  {065206} (\bibinfo {year} {2009})},\ \Eprint {http://arxiv.org/abs/0903.1110}
  {arXiv:0903.1110 [hep-ex]} \BibitemShut {NoStop}%
\bibitem [{\citenamefont {Aznauryan}\ \emph {et~al.}(2009)\citenamefont
  {Aznauryan} \emph {et~al.}}]{Aznauryan:2009mx}%
  \BibitemOpen
  \bibfield  {author} {\bibinfo {author} {\bibfnamefont {I.~G.}\ \bibnamefont
  {Aznauryan}} \emph {et~al.} (\bibinfo {collaboration} {CLAS Collaboration}),\
  }\href {\doibase 10.1103/PhysRevC.80.055203} {\bibfield  {journal} {\bibinfo
  {journal} {Phys. Rev.}\ }\textbf {\bibinfo {volume} {C80}},\ \bibinfo {pages}
  {055203} (\bibinfo {year} {2009})},\ \Eprint {http://arxiv.org/abs/0909.2349}
  {arXiv:0909.2349 [nucl-ex]} \BibitemShut {NoStop}%
\bibitem [{\citenamefont {Dalton}\ \emph {et~al.}(2009)\citenamefont {Dalton},
  \citenamefont {Adams}, \citenamefont {Ahmidouch}, \citenamefont {Angelescu},
  \citenamefont {Arrington}, \citenamefont {Asaturyan}, \citenamefont {Baker},
  \citenamefont {Benmouna}, \citenamefont {Bertoncini}, \citenamefont
  {Boeglinand} \emph {et~al.}}]{Dalton:2008aa}%
  \BibitemOpen
  \bibfield  {author} {\bibinfo {author} {\bibfnamefont {M.~M.}\ \bibnamefont
  {Dalton}}, \bibinfo {author} {\bibfnamefont {G.~S.}\ \bibnamefont {Adams}},
  \bibinfo {author} {\bibfnamefont {A.}~\bibnamefont {Ahmidouch}}, \bibinfo
  {author} {\bibfnamefont {T.}~\bibnamefont {Angelescu}}, \bibinfo {author}
  {\bibfnamefont {J.}~\bibnamefont {Arrington}}, \bibinfo {author}
  {\bibfnamefont {R.}~\bibnamefont {Asaturyan}}, \bibinfo {author}
  {\bibfnamefont {O.~K.}\ \bibnamefont {Baker}}, \bibinfo {author}
  {\bibfnamefont {N.}~\bibnamefont {Benmouna}}, \bibinfo {author}
  {\bibfnamefont {C.}~\bibnamefont {Bertoncini}}, \bibinfo {author}
  {\bibfnamefont {W.~U.}\ \bibnamefont {Boeglinand}},  \emph {et~al.},\ }\href
  {\doibase 10.1103/PhysRevC.80.015205} {\bibfield  {journal} {\bibinfo
  {journal} {Phys. Rev.}\ }\textbf {\bibinfo {volume} {C80}},\ \bibinfo {pages}
  {015205} (\bibinfo {year} {2009})},\ \Eprint {http://arxiv.org/abs/0804.3509}
  {arXiv:0804.3509 [hep-ex]} \BibitemShut {NoStop}%
\bibitem [{\citenamefont {Denizli}\ \emph {et~al.}(2007)\citenamefont {Denizli}
  \emph {et~al.}}]{Denizli:2007tq}%
  \BibitemOpen
  \bibfield  {author} {\bibinfo {author} {\bibfnamefont {H.}~\bibnamefont
  {Denizli}} \emph {et~al.} (\bibinfo {collaboration} {CLAS Collaboration}),\
  }\href {\doibase 10.1103/PhysRevC.76.015204} {\bibfield  {journal} {\bibinfo
  {journal} {Phys. Rev.}\ }\textbf {\bibinfo {volume} {C76}},\ \bibinfo {pages}
  {015204} (\bibinfo {year} {2007})},\ \Eprint {http://arxiv.org/abs/0704.2546}
  {arXiv:0704.2546 [nucl-ex]} \BibitemShut {NoStop}%
\bibitem [{\citenamefont {Thompson}\ \emph {et~al.}(2001)\citenamefont
  {Thompson} \emph {et~al.}}]{Thompson:2000by}%
  \BibitemOpen
  \bibfield  {author} {\bibinfo {author} {\bibfnamefont {R.}~\bibnamefont
  {Thompson}} \emph {et~al.} (\bibinfo {collaboration} {CLAS Collaboration}),\
  }\href {\doibase 10.1103/PhysRevLett.86.1702} {\bibfield  {journal} {\bibinfo
   {journal} {Phys. Rev. Lett.}\ }\textbf {\bibinfo {volume} {86}},\ \bibinfo
  {pages} {1702} (\bibinfo {year} {2001})},\ \Eprint
  {http://arxiv.org/abs/hep-ex/0011029} {arXiv:hep-ex/0011029 [hep-ex]}
  \BibitemShut {NoStop}%
\bibitem [{\citenamefont {Armstrong}\ \emph {et~al.}(1999)\citenamefont
  {Armstrong} \emph {et~al.}}]{Armstrong:1998wg}%
  \BibitemOpen
  \bibfield  {author} {\bibinfo {author} {\bibfnamefont {C.~S.}\ \bibnamefont
  {Armstrong}} \emph {et~al.} (\bibinfo {collaboration} {Jefferson Lab
  E94014}),\ }\href {\doibase 10.1103/PhysRevD.60.052004} {\bibfield  {journal}
  {\bibinfo  {journal} {Phys. Rev.}\ }\textbf {\bibinfo {volume} {D60}},\
  \bibinfo {pages} {052004} (\bibinfo {year} {1999})},\ \Eprint
  {http://arxiv.org/abs/nucl-ex/9811001} {arXiv:nucl-ex/9811001 [nucl-ex]}
  \BibitemShut {NoStop}%
\bibitem [{\citenamefont {Burkert}\ \emph {et~al.}(2003)\citenamefont
  {Burkert}, \citenamefont {De~Vita}, \citenamefont {Battaglieri},
  \citenamefont {Ripani},\ and\ \citenamefont {Mokeev}}]{Burkert:2002zz}%
  \BibitemOpen
  \bibfield  {author} {\bibinfo {author} {\bibfnamefont {V.}~\bibnamefont
  {Burkert}}, \bibinfo {author} {\bibfnamefont {R.}~\bibnamefont {De~Vita}},
  \bibinfo {author} {\bibfnamefont {M.}~\bibnamefont {Battaglieri}}, \bibinfo
  {author} {\bibfnamefont {M.}~\bibnamefont {Ripani}}, \ and\ \bibinfo {author}
  {\bibfnamefont {V.}~\bibnamefont {Mokeev}},\ }\href {\doibase
  10.1103/PhysRevC.67.035204} {\bibfield  {journal} {\bibinfo  {journal}
  {Phys.Rev.}\ }\textbf {\bibinfo {volume} {C67}},\ \bibinfo {pages} {035204}
  (\bibinfo {year} {2003})},\ \Eprint {http://arxiv.org/abs/hep-ph/0212108}
  {arXiv:hep-ph/0212108 [hep-ph]} \BibitemShut {NoStop}%
\bibitem [{\citenamefont {Park}\ \emph {et~al.}(2015)\citenamefont {Park} \emph
  {et~al.}}]{Park:2014yea}%
  \BibitemOpen
  \bibfield  {author} {\bibinfo {author} {\bibfnamefont {K.}~\bibnamefont
  {Park}} \emph {et~al.} (\bibinfo {collaboration} {CLAS Collaboration}),\
  }\href {\doibase 10.1103/PhysRevC.91.045203} {\bibfield  {journal} {\bibinfo
  {journal} {Phys. Rev.}\ }\textbf {\bibinfo {volume} {C91}},\ \bibinfo {pages}
  {045203} (\bibinfo {year} {2015})},\ \Eprint {http://arxiv.org/abs/1412.0274}
  {arXiv:1412.0274 [nucl-ex]} \BibitemShut {NoStop}%
\bibitem [{\citenamefont {Tiator}\ \emph {et~al.}(2011)\citenamefont {Tiator},
  \citenamefont {Drechsel}, \citenamefont {Kamalov},\ and\ \citenamefont
  {Vanderhaeghen}}]{Tiator:2011pw}%
  \BibitemOpen
  \bibfield  {author} {\bibinfo {author} {\bibfnamefont {L.}~\bibnamefont
  {Tiator}}, \bibinfo {author} {\bibfnamefont {D.}~\bibnamefont {Drechsel}},
  \bibinfo {author} {\bibfnamefont {S.~S.}\ \bibnamefont {Kamalov}}, \ and\
  \bibinfo {author} {\bibfnamefont {M.}~\bibnamefont {Vanderhaeghen}},\ }\href
  {\doibase 10.1140/epjst/e2011-01488-9} {\bibfield  {journal} {\bibinfo
  {journal} {Eur. Phys. J. ST}\ }\textbf {\bibinfo {volume} {198}},\ \bibinfo
  {pages} {141} (\bibinfo {year} {2011})},\ \Eprint
  {http://arxiv.org/abs/1109.6745} {arXiv:1109.6745 [nucl-th]} \BibitemShut
  {NoStop}%
\bibitem [{\citenamefont {Olive}\ \emph {et~al.}(2014)\citenamefont {Olive}
  \emph {et~al.}}]{Agashe:2014kda}%
  \BibitemOpen
  \bibfield  {author} {\bibinfo {author} {\bibfnamefont {K.~A.}\ \bibnamefont
  {Olive}} \emph {et~al.} (\bibinfo {collaboration} {Particle Data Group}),\
  }\href {\doibase 10.1088/1674-1137/38/9/090001} {\bibfield  {journal}
  {\bibinfo  {journal} {Chin. Phys.}\ }\textbf {\bibinfo {volume} {C38}},\
  \bibinfo {pages} {090001} (\bibinfo {year} {2014})}\BibitemShut {NoStop}%
\end{thebibliography}%
\bibliographystyle{apsrev4-1}

\end{document}